\documentclass[11pt, a4paper]{article}
\usepackage{amssymb}
\usepackage{amsfonts}
\usepackage{amsmath}
\usepackage{threeparttable}
\usepackage{color}
\usepackage{graphicx}
\usepackage{subfigure}
\usepackage{multicol}
\usepackage{multirow}
\usepackage{float}
\usepackage{caption}
\usepackage{color}
\usepackage{setspace}
\usepackage{booktabs}
\usepackage{CJKutf8}
\usepackage{CJKpunct}
\usepackage{geometry}
\usepackage{changepage}
\usepackage{titlesec}
\usepackage{tcolorbox}
\usepackage{indentfirst}
\usepackage{enumerate}
\usepackage{CJK}
\newtheorem{lemma}{Lemma}

\newenvironment{proof}[1][Proof]{\noindent\textbf{#1.} }{\ \rule{0.5em}{0.5em}}
\begin{document}

	\date{\today}
	\title{\huge{\textrm{Identification and Auto-debiased Machine Learning for Outcome Conditioned Average Structural Derivatives}}}
	\renewcommand{\baselinestretch}{1.15}
	\author{\\\Large{\textrm{Zequn Jin~~~~~~Lihua Lin~~~~~~Zhengyu Zhang\footnote{Corresponding Author; e-mail: zy.zhang@mail.shufe.edu.cn.}}}\\\\\Large{{School of Economics}}\\\\\Large{{Shanghai
				University of Finance and Economics}}} \maketitle
	\renewcommand{\baselinestretch}{1.15}
	\begin{abstract}\normalsize
		This paper proposes
		a new class of heterogeneous causal quantities, named \textit{outcome conditioned} average structural derivatives (OASD) in a general nonseparable model. OASD is the average partial effect of a marginal change in a continuous treatment on the individuals located at different parts of the outcome distribution,  irrespective of individuals' characteristics.
		OASD combines both features of ATE and QTE: it is interpreted as straightforwardly as ATE
		while at the same time more granular than ATE by breaking the entire population up according to
		the rank of the outcome distribution.
		
		One contribution of this paper is that we establish some close relationships between the \textit{outcome conditioned average partial effects} and a class of parameters measuring the effect of counterfactually
		changing the distribution of a single covariate on the unconditional outcome quantiles. By exploiting such relationship, we can obtain root-$n$ consistent estimator and calculate the semi-parametric efficiency bound for these counterfactual effect parameters. We illustrate this point by two examples: equivalence between OASD and the unconditional partial quantile effect (Firpo et al. (2009)), and equivalence between the marginal partial distribution policy effect (Rothe (2012)) and a corresponding outcome conditioned parameter.

		Because identification of OASD is attained
		under a conditional exogeneity assumption, by controlling for a rich information about covariates,  a researcher may ideally use high-dimensional controls in data.
		We propose for OASD a novel automatic debiased machine learning estimator, and present asymptotic statistical
		guarantees for it. We prove our estimator is root-$n$ consistent, asymptotically normal, and semiparametrically efficient.
		We also prove the validity of the bootstrap procedure for uniform inference on the OASD process.  Simulation studies support our theories.

	\end{abstract}

	\baselineskip=20pt


	\thispagestyle{empty}
	
	\renewcommand{\baselinestretch}{1.18}
	\newpage
	\section{Introduction}\normalsize
	
	When determining the causal effect of an intervention or a treatment of interest $D$ on an outcome $Y$, applied researchers more often focus on mean quantities such as average treatment effect (ATE) than distributional parameters such as quantile treatment effect (QTE), because the formers are easier to interpret.
	Since unobserved heterogeneity is so pervasive in microdata,  distributional parameters such as quantile regression (QR) coefficients or QTE still play an important role in summarizing
	heterogeneous impacts
	of variables on different points of an outcome distribution.
	For example, in understanding the effect of a tax credit reform, one might be more interested in the effect of the tax rate change on the lower tail of the labor supply or savings distribution conditional on individual characteristics than the mean effect. This is the question answered exactly by QR. However, to understand heterogenous effects of such reform, one may ask a different question from the above one: what is the average effect of the tax rate change on labor supply or savings for the individuals at the lower tail of the labor supply (or savings) distribution, irrespective of (by integrating out) individual characteristics?
	
	Formally, assume that there is an outcome variable $Y$ (labor supply, savings or consumptions) with a continuous support $\mathcal{S}_{Y}\subset \mathbb{R}$, a continuous treatment variable $D$ (tax rate) and a $K_{X}$-dimensional vector of covariates
	$X$, which are related through a general nonseparable structural model
	\[
	Y=m\big(D,X,U\big)\eqno(1.1)
	\]
	with $U$  an unobservable random vector, capturing omitted factors and all types of unobserved heterogeneity. This model (1.1) is very general as it imposes on $m(\cdot)$ neither additivity structure nor monotonicity with respect to the error term. Such general model has been studied by Imbens and Newey
	(2009), Altonji and Matzkin (2005), and Hoderlein
	and Mammen (2007, 2009), Firpo (2009), Rothe (2010, 2012), etc.

	Let $\partial_{D}$ be the derivative of $m(\cdot)$ with respect to its first argument.
	In this paper, we focus on the following class of parameters
	\[
	\theta(y_{1},y_{2})=E\left(\partial_{D} m\big(D,X,U\big)\bigg|Y\in(y_{1},y_{2})\right)\eqno(1.2)
	\]
	indexed by $(y_{1},y_{2})\subset \mathcal{S}_{Y}$, which are called the \textit{outcome conditioned average structural derivatives} (OASD). OASD combines both features of ATE and QTE: it is interpreted as straightforwardly as ATE
	while at the same time more granular than ATE by breaking the entire population up according to
	the rank of the outcome distribution.
	
	The OASDs defined by (1.2) provide direct answers to a wide range of questions
	in applied economic analysis. To give a simple example, let $Y$ be log wage, $D$ be years of schooling and $X$ be other individual characteristics.  Applied researchers mainly focus on the average partial effect or $E\left[\partial_{D} m(D,X,U)\right]$, which  measures the mean  gain of receiving one more year of
	schooling for all individuals. To explore the heterogenous effect of receiving more education, one may run quantile regression of $Y$ on $D$ and  $X$ (Buchinsky (1994),  Chamberlain (1994),  Angrist, Chernozhukov and Fernandez-Val  (2006)), and the coefficient on $D$ describes the impact of receiving one more year of schooling on the
	conditional quantiles of wage distribution. Sometimes policymakers may be interested in a related, but more straightforward question: what is the average effect of a small increase in schooling for low/middle/high wage individuals ? Let $y_{0.1}, y_{0.2}, \cdots, y_{0.9}$ equal to the $10\%, 20\%, \cdots, 90\%$-th empirical quantiles of the wage distribution,
	the latter question can be answered by estimating $\theta(y_{0.1},y_{0.2})$, $\cdots$, and $\theta(y_{0.8},y_{0.9})$ directly. If the analysis shows that high-earning individuals, on average,
	earn more than low-earning individuals, for instance, $\theta(y_{0.8},y_{0.9})>\theta(y_{0.2},y_{0.3})$, then one can tell that the dispersion in earnings is likely to go up with schooling and vice versa.
	
	Our OASD and the subsequent identification strategy partially build on the insights of Hoderlein
	and Mammen (2007, 2009), who consider identification and estimation in a nonseparable model of the quantity
	\begin{eqnarray*}
		E\left(\partial_{D} m(D,X,U)\bigg|D=d,X=x,Y=y\right),
	\end{eqnarray*}
	which they call the Local Average Structural Derivative (LASD).
	At first glance, OASD makes only one step forward given LASD by integrating $X$ out conditional on $Y\in(y_{1},y_{2})$. However, as we shall demonstrate later, such modification produces several desirable properties and sheds light on some interesting connections between OASD and other distributional parameters. Specifically, we establish two new identification results for OASD. The first one (Proposition 2.1) points out some important connection between OASD and the unconditional quantile partial effect (UQPE) proposed by Firpo et al. (2009). This relationship has two implications:  it offers an alternative economic interpretation to UQPE, which can be understood as a mean causal effect on an identifiable subpopulation located at some level of the outcome distribution. Moreover, many desirable properties of OASD may carry over to the UQPE. For example, the semiparametric efficiency bound of UQPE may be learnt from that of OASD, which we shall derive in Section 4. Our second identification result (Proposition 2.2) derives an orthogonal score for OASD, which paves the way for developing subsequent
	automatic debiased machine learning estimator  and facilitates establishing some desirable properties of the estimator. For example, we prove that our estimator is root-$n$ consistent and semi-parametrically efficient. We also mention that OASDs have some robustness property against censoring data.  In many applications using micro dataset, censoring is a pervasive phenomenon. It is important that the estimator still works under censoring. All these results are new relative to Hoderlein and Mammen (2007, 2009).

	\textbf{Main contributions to the Literature.} This paper makes three contributions to the literature. First, we propose OASDs as novel and parsimonious  quantities to measure  impacts of a continuous treatment that are heterogeneous across the unconditional distribution of an outcome.
	There are two major differences between OASD and the QR approach.  QR aims to estimate impacts of explanatory variables at different points of the outcome distribution \textit{conditional on} a large number of covariates, while OASD measures the mean effect of a given explanatory variable on the subpopulations located at  different parts of the \textit{unconditional} outcome distribution. \footnote{The difference between \textit{conditional} vs \textit{unconditional} outcome distribution can be understood by a simple example relating wages to years of education. The 0.9 quantile of wage distribution
		conditional on education, which is the subpopulation targeted by QR, refers to the high wage workers within
		each education class, who however may not necessarily be high
		earners overall. However, the unconditional
		0.9 quantile of wage distribution, which is the subpopulation targeted by OASD, refers straightforwardly to the high wage workers.} OASDs are the right estimands to consider
	when the ultimate object of interest is  the people located at specific parts (lower or upper tail) of the unconditional outcome distribution whatever individual characteristics (education, race, age) they have.
	In this sense, OASD can be viewed as generalizing the conception of unconditional quantile treatment effect (Firpo (2007), Frolich and Melly (2013)) for a binary treatment to a continuous treatment.
	The second difference between OASD and QR is more technical:  OASDs are indexed by intervals $(y_{1},y_{2})\subset\mathbb{R}$ instead of $y\in \mathcal{S}_{Y}$.
	As will be clear later, this technical subtlety ensures the resulting estimator has several desirable properties such as $\sqrt{n}$ convergence and semiparametric efficiency. We provide two identification results for OASDs in a general nonseparable model (1.1) without monotonicity.  When the treatment is binary, we show that the outcome conditioned average treatment effect is identified with an additional monotonicity condition (see Remark 2.2.1).

	As a second contribution,  we establish some close relationships between two classes of causal quantities which apparently have different interpretation: one is the outcome conditioned average partial effects (including OASD) studied in the current paper, the other
	is parameters measuring the effect of counterfactually
	changing the distribution of a single covariate on the unconditional distribution of the outcome. Examples of the latter class include the unconditional partial quantile effect (UPQE, Firpo et al. (2009)) and the marginal partial policy effect (MPPE, Rothe 2012). We show there is a close connection between OASD and UPQE, also between MPPE and a corresponding outcome conditioned average effect parameter.
	These results have an important implication: many desirable properties of OASD may well carry over to the UQPE.  For example, to the best of our knowledge,  the literature has not obtained any efficiency bound of UQPE.
	In Section 4, we derive the semiparametric efficiency bound of OASD, which can be used to learn about the efficiency bound of UQPE.
	As another example,
	Firpo et al. (2009, Proposition 1, page 959) has shown that the UQPE can be written as a weighted
	average of a family of conditional quantile partial effects (CQPE), under the monotonicity assumption.
	Using the equivalence result between UQPE and OASD, we show this result (Proposition 2.3-iii) still holds under much weaker assumptions without monotonicity. Similar arguments apply to MPPE.
	
	The third contribution is that we
	propose a novel automatic debiased machine learning (ADML) estimator for OASD, by taking advantage of the debiased estimation approach recently developed by Belloni et al. (2017), Chernozhukov et al. (2022 a,b,c).
	For LASD, Hoderlein and Mammen (2009) have introduced a local polynomial kernel estimator and derived its large sample properties. Although this kernel based approach can be taken to estimate OASD,
	it cannot accommodate high dimensional controls and has no robustness against local perturbations in the nuisance functions.
	Since identification of OASD is attained
	under a conditional exogeneity assumption, by controlling for a rich information about covariates,  a researcher may ideally use high-dimensional controls in data.
	Motivated by this, we contribute to the literature by proposing a first orthogonal score based estimator for OASD (and meanwhile for the LASD), that is shown to be root-$n$ consistent and semiparametrically efficient, and allowing for a flexibility in types of preliminary estimators, e.g., kernel, sieve, and Lasso.

	Like QTE,  our estimator for OASD falls under the framework where there are a continuum of finite-dimensional
	parameters of interest, identified via a continuum of moment conditions that involve a continuum
	of nuisance functions.  We prove the uniform Gaussianity of the OASD \textit{process} and the uniform validity of a multiplier bootstrap,  by taking advantage of the general theory for the Lasso and post-Lasso estimators for functional response
	data established by Belloni et al. (2017).

	\textbf{Relationship to the Literature.} This paper is related to two branches of the literature. The first branch is about identifying causal parameters, particularly those measuring heterogenous distributional impacts in nonseparable models. This branch can be broadly divided into two categories according to whether the treatment variable is discrete or continuous. Contributions to the category dealing with a binary treatment variable include Heckman and Vytlacil (2001) about the policy relevant treatment effects, and Heckman and Vytlacil (2005) about the marginal treatment effects, Firpo  (2007) about the unconditional QTE,  Frolich and Melly (2013) about the local QTE for compliers, Donald and Hsu (2014) about the distributional treatment effects, to name only a few.
	
	Our work is more relevant to the second category about nonparametric identification and estimation of causal parameters or counterfactual policy effect of a continuous treatment, particularly without assuming the error term entering monotonically. Contributions to this category include Altonji and Matzkin (2005),  Chernozhukov et al. (2013),
	Florens, Heckman, Meghir, and Vytlacil (2008), Imbens and Newey (2009) and Hoderlein and
	Mammen (2007, 2009), Rothe (2010, 2012), Firpo et al. (2009), and Ai et al. (2022). Less closely related to our
	work are Chesher (2003, 2005), Chernozhukov and Hansen (2005), Chernozhukov, Imbens
	and Newey (2007) who assume that the error term, at least at some stage, enters the model monotonically.
	Among them, Hoderlein and Mammen (2007, 2009), Firpo et al. (2009)
	are perhaps the most relevant to our work. We contribute to this branch of the literature from two aspects: we propose a new class of quantities, namely OASD to measure heterogeneous impacts of a continuous treatment; we provide insights into the relationship between  OASD and a class of counterfactual policy effect parameters; and we obtain the semi-parametric efficiency bound for OASD.

	Our paper is also related to another fast growing branch of the literature on estimation and
	inference of causal or structural parameters based on orthogonal scores.  The seminal paper by Newey (1994) proposes
	orthogonal scores for many semiparametric models and provides forms of adjustment terms to obtain orthogonal scores from
	moment functions. Chernozhukov et al. (2022a)  propose a general procedure for construction of orthogonal scores
	from moment restriction models. We derive the orthogonal score for OASDs following these general prescriptions.
	Usually, the orthogonal score depends on another unknown function, denoted as $\alpha$ ($\alpha=\partial_{D} \ln f(D,X)$ in OASD) in addition to
	the nonparametric components in the original moment. Chernozhukov et al. (2022 a,b,c) develop a Lasso minimum distance learner of  $\alpha$, that is
	automatic in the sense that it depends only on the identifying moment function and not
	on the functional form of $\alpha$. Our proposed method of estimation takes advantages of these knowledge. A technical novelty in our proof is that we estimate CDF derivatives by a high order partial difference approach, with which we do not need to impose substantially more restrictive approximate sparsity conditions than before while preserving the good rates of convergence
	for high dimensional CDF and well as its derivatives.
	
	In work related but independent from ours, Sasaki et al. (2022) propose a doubly robust score for debiased estimation of the UQPE (Firpo et al. 2009).  Their results complement ours, though the motivation of the two papers are different: Sasaki et al. only consider estimation and inference of the UQPE as a measure
	of heterogeneous counterfactual marginal effects while we focus on outcome conditioned partial effect of continuous treatment. In the absence of the connection between UQPE and OASD (see Section 2.2), the parameters considered by the two papers are entirely different.
	Moreover, Sasaki et al. do not consider the  semiparametric efficiency bound, nor derive the uniform Gaussian distribution of their estimator, while we establish the uniform limiting distribution of the OASD process.
	
	\textbf{Organization of the paper.} The rest of this paper is organized as follows. Section 2 presents the setting, the main identification results and discusses the relationship between the OASD and other counterfactual policy effect parameters.
	Sections 3 develops an automatic debiased learning estimator for OASD. Section 4 provides theoretical
	guarantees of the estimator.   Section 5 presents Monte
	Carlo simulation studies.  The paper is summarized in Section 6. The appendix
	contains proofs and additional details that are important but relegated there due to their lengths.
	
	\textbf{Notations.} We work with the i.i.d. data $\{W_{i}\}_{i=1}^{n}$ which is defined on the probability space $\left(\mathcal{W},\mathcal{A}_{\mathcal{W}},P\right)$. We denote by $\mathbb{P}_{n}$ the empirical probability measure that assigns probability $n^{-1}$ to each $W_{i}\in\{W_{i}\}_{i=1}^{n}$. $\mathbb{E}_{n}$ denotes the expectation with respect to the empirical measure, and $\mathbb{G}_{n}$ denotes the empirical process, that is
	\[
	\mathbb{G}_{n}B(W)=\frac{1}{\sqrt{n}}\sum_{i=1}^{n}\bigg[B(W_{i})-EB(W)\bigg]
	\]
	indexed by a measurable class of functions $\mathcal{B}:\mathcal{W}\mapsto\mathbb{R}$. In what follows, we use $\Vert\cdot\Vert_{P,q}$ to denote the $L^{q}(P)$ norm. $\left\Vert{A}\right\Vert_{\infty}=\max_{i,j}|a_{ij}|$,  $\left\Vert{A}\right\Vert_{1}=\sum_{i,j}|a_{ij}|$, and $\Vert{A}\Vert_{0}$ equals the number of nonzero components of $A$ for a matrix $A=[a_{ij}]$.

	\section{Identification}

	\subsection{Identification of OASD}
	
	Suppose that the outcome variable $Y$ is determined by
	\[
	Y=m\big(D,X,U\big),\eqno(2.1.1)
	\]
	where $D$ is a continuous treatment variable, $X$ is a $K_X$-dimensional vector of covariates, $m(\cdot)$ is a
	smooth measurable function, and $U$ is an unobservable random vector that captures omitted factors and all types of unobserved heterogeneity. Given $(y_{1},y_{2})\subset \mathcal{S}_{Y}$, the object of interest is
	\begin{eqnarray*}
		\theta(y_{1},y_{2})=E\left(\partial_{D} m(D,X,U)\bigg|Y\in(y_{1},y_{2})\right)
	\end{eqnarray*}
	which measures the average partial effect of a marginal change in $D$ on the individuals with $Y\in (y_1, y_2)$.

	Let $Q_{Y}(\tau|d,x)$ denote the $\tau$-th quantile of $Y$ conditional on $D=d$, $X=x$. Under the assumption that the random variables $U$ and $D$ are independent conditional on $X$ and some other technical assumptions (See Appendix A), Hoderlein and Mammen (2007, Theorem 2.1, page 1515) have established that for any
	$(d,x)\in\mathcal{S}_{D}\times\mathcal{S}_{X}$, $0<\tau<1$,
	\[
	E\left(\partial_{D} m(D,X,U)\bigg|D=d,X=x,Y=Q_{Y}(\tau|d,x)\right)=\frac{\partial Q_{Y}(\tau|d,x)}{\partial d}. \eqno(2.1.2)
	\]

	\noindent Let $F_{Y}(\cdot|d,x)$ be the CDF of $Y$ conditional on $(D,X)=(d,x)$. (2.1.2) can be equivalently expressed as
	\[
	E\left(\partial_{D} m(D,X,U)\bigg|D=d,X=x,Y=y\right)=\frac{\partial Q_{Y}(u|d,x)}{\partial d}\bigg|_{u=F_{Y}(y|d,x)}. \eqno(2.1.3)
	\]
	
	Thus $\theta(y_{1},y_{2})$ can be identified straightforwardly by
	\[
	\theta(y_{1},y_{2})=\int_{y_{1}}^{y_{2}}\int\frac{\partial Q_{Y}(u|d,x)}{\partial d}\big|_{u=F_{Y}(y|d,x)}f(d,x,y)dddxdy\bigg/ P(y_{1}<Y<y_{2}). \eqno(2.1.4)
	\]
	
	Below we provide an alternative expression of $\theta(y_{1},y_{2})$, under slightly weaker assumptions than Hoderlein and Mammen (2007).  Note that the error term $U$ in (2.1.1) is invariant with respect to realizations of $D$. More generally, we can assume that $U=U_{D}$, that is, $U$ may change across $d$.
	
	\noindent\textbf{Assumption 2.1.} For each $d\in\mathcal{S}_{D}$, $U_{d}$'s are identically distributed across $d$ conditional on $X$.
	
	Assumption 2.1 is called ``rank similarity'' (Chernozhukov and Hansen (2005)). It permits that the realizations of error term may vary with the treatment intensity, but they should have the same distribution conditional on covariates. Assumption 2.1 incorporates  $U_{d}\equiv U$ as a special case.
	
	\noindent\textbf{Assumption 2.2.}  For each $d\in\mathcal{S}_{D}$, $U_{d}$ is independent of $D$ conditional on $X$.
	
	This conditional independence assumption is weaker than full joint independence
	of $U_{d}$ and $(D,X)$. Other examples of identifying counterfactual or causal parameters based on the conditional exogeneity condition are Firpo et al. (2009), Chernozhukov, Fernandez-Val, and Melly
	(2013), Rothe (2010, 2012).
	
	\noindent\textbf{Proposition 2.1.} Under Assumptions 2.1-2.2 and some regularity conditions (listed in Appendix A),
	\begin{eqnarray*}
		\theta(y_{1},y_{2})=\frac{-1}{P(y_{1}<Y<y_{2})}E\left(\partial_{D}\int_{y_{1}}^{y_{2}}F_{Y}(y|D,X)dy\right).
	\end{eqnarray*}
	
	Compared with (2.1.4), Proposition 2.1 provides three novel insights. First, it shows $\theta(y_{1},y_{2})$ can be expressed as an integral of the average derivative of $F_{Y}(y|d,x)$ over $(y_{1},y_{2})$ with some rescaling. Because the influence function of average derivatives has been well studied in the existing literature
	(Newey, 1994), it is possible to derive the orthogonal score for $\theta(y_{1},y_{2})$ (see Proposition 2.2).   Second, it shows that like the distributional treatment effect, OASD is robust against censoring data. For example, let $Y=\max\{0,Y^{*}\}$. Then $\theta(y_{1},y_{2})$ is  identifiable  for any $(y_{1},y_{2})\subset (0,+\infty)\cap \mathcal{S}_{Y^{*}}$. Third, let
	\[
	\theta(y)=E\left(\partial_{D} m(D,X,U)\bigg|Y=y\right).\eqno(2.1.5)
	\]
	Fix $y_{1}=y$ in Proposition 2.1 and let $y_{2}$ approach $y_{1}=y$,
	\begin{eqnarray*}
		\int_{y_{1}}^{y_{2}}F_{Y}(y|D,X)dy\approx F_{Y}(y|D,X)\Delta y, ~~P(y_{1}<Y<y_{2})\approx f_{Y}(y)\Delta y
	\end{eqnarray*}
	with $\Delta y=y_{2}-y_{1}$. It is straightforward to see
	\[
	\theta(y)=\frac{-E\left(\partial_{D}F_{Y}(y|D,X)\right)}{f_{Y}(y)}\eqno(2.1.6)
	\]
	which is equivalent to the unconditional quantile partial effect (UQPE) proposed  in Firpo et al. (2009). For a detailed discussion on the connection between OASD and UQPE, see Section 2.2.
	
	Below we provide another identification result based on an orthogonal score, which is necessary for estimation with high dimensional controls.
	Let $W=(Y,D,X)$. Let $\eta=\eta(w; y_{1},y_{2})$  collect the (possibly infinite-dimensional) nuisance parameters:
	\[
	\eta\big(W;y_1,y_2\big)=\left(P\big(y_1<Y<y_2\big),\int_{y_1}^{y_2}F_Y\left(y\big|D,X\right)dy,\frac{\partial_Df\big(D,X\big)}{f\big(D,X\big)}\right) \eqno(2.1.7)
	\]
	with $f\big(d,x\big)$ the joint density of $(D,X)$. Let
	\[\begin{aligned}
		&\psi\bigg(W,\theta,\eta;y_1,y_2\bigg)=\frac{-1}{P\big(y_1<Y<y_2\big)}\partial_D\int_{y_1}^{y_2}F_Y(y|D,X)dy-\theta\notag\\
		&-\frac{1}{P(y_1<Y<y_2)}\frac{\partial_Df(D,X)}{f(D,X)}\int_{y_1}^{y_2}\left(F_Y\left(y\big|D,X\right)-1\big\{Y<y\big\}\right)dy\\
		&+\frac{E\left(\partial_D\int_{y_1}^{y_2}F_Y(y|D,X)dy\right)}{P^2(y_1<Y<y_2)}\left(1\big\{y_1<Y<y_2\big\}-P\big(y_1<Y<y_2\big)\right).\\
	\end{aligned}\eqno{(2.1.8)}\]
	
	\noindent\textbf{Proposition 2.2.} (\textit{Identification based on orthogonal score}) Under the same assumptions as Proposition 2.1, we have
	(i) $\theta(y_{1},y_{2})$ satisfies $E\psi\bigg(W,\theta(y_{1},y_{2}),\eta;y_1,y_2\bigg)=0$.
	
	\noindent (ii) $\psi\bigg(W,\theta,\eta;y_1,y_2\bigg)$ satisfies the Neyman orthogonality property:
	\begin{eqnarray*}
		\frac{\partial E\psi\bigg(W,\theta(y_{1},y_{2}),\eta+r(\tilde{\eta}-\eta);y_1,y_2\bigg)}{\partial r}\bigg|_{r=0}=0.
	\end{eqnarray*}
	
	Proposition 2.2 has three implications. First, result (ii) means the functional Gateaux derivative of the moment function $\psi(\cdot)$ with respect to the nonparametric component vanishes when evaluated at the true parameters.
	This orthogonality property is crucial to
	establishing good behaviour (such as root $n$-consistency and asymptotically Gaussian distribution) of the subsequent  automatic debiased machine learning estimator for OASD. Second, we have shown (in Appendix A) that the orthogonal score $\psi(\cdot)$ has some double robustness property in the sense that one may still identify $\theta(y_{1},y_{2})$ from the moment condition, even if one (but not all) component of $\eta$ is incorrectly specified.
	Third, we further prove (Theorem 4.2) that the orthogonal score is also semiparametrically efficient.
	
	\noindent\textbf{Remark 2.1.1.} When the treatment is binary, the quantity in parallel with OASD can be written as
	\begin{eqnarray*}
		\vartheta(y_{1},y_{2})=E\left(m\big(1,X,U_1\big)-m\big(0,X,U_0\big)\bigg|Y\in(y_{1},y_{2})\right)
	\end{eqnarray*}
	which measures the average treatment effect of participating in a program on the individuals with outcome variable $Y\in(y_{1},y_{2})$. In Appendix A, we show $\vartheta(y_{1},y_{2})$ is identifiable under Assumptions 2.1-2.2, with a monotonicity condition. Moreover, we also derive an orthogonal score for $\vartheta(y_{1},y_{2})$.

	\subsection{Relationship to Unconditional Quantile Partial Effects}
	
	By definition, OASDs characterize mean impacts on the subpopulations located at different parts of the unconditional distribution of $Y$.
	Another related and well known approach to estimate counterfactual effects that are heterogeneous across the unconditional outcome distribution
	$F_{Y}$ is the unconditional quantile regression (UQR) proposed by Firpo et al. (2009). This subsection establishes the relationship between OASD and UQR.
	
	Let $Y$ again be generated by a general nonseparable model $Y=m(D,X,U_D)$,
	where $D$ is a scalar continuous treatment variable of
	interest and $X$ consists of controls. The causal parameter UQR aims to identify is called unconditional quantile partial effect (UQPE), which measures the marginal effect of counterfactually shifting the distribution of a coordinate of the explanatory variables on unconditional quantiles of $Y$. The counterfactual distribution of $Y$ after shifting the distribution of $D$ infinitesimally while holding $X$ fixed can be written as
	\begin{eqnarray*}
		F_{Y^{\epsilon}}(y)=\int F_{Y}(y|D=d+\epsilon,X=x)dF_{DX}(d,x).
	\end{eqnarray*}
	Let $Q_{Y^{\epsilon}}(\cdot)$ be the inverse of $F_{Y^{\epsilon}}(y)$.
	The $\tau$-th UQPE with respect to $D$ is defined as
	\[
	UQPE(\tau)=\frac{\partial Q_{Y^{\epsilon}}(\tau)}{\partial\epsilon}\bigg|_{\epsilon=0}.\eqno(2.2.1)
	\]
	
	Under the assumption of conditional exogeneity, as shown by Firpo et al.
	(2009, Proposition 1, page 959), the UQPE can be interpreted as the causal effect of changing the distribution of $D$ infinitesimally. Without such an assumption, UQPE may still be of interest as
	a summary statistic of the counterfactual distributional relationship between $Y$ and $D$.

	Let $\theta(y)$ be the average structural derivative of $D$ conditional on $Y=y$, namely,
	\begin{eqnarray*}
		\theta(y)=E\left(\partial_{D} m(D,X,U_D)\bigg|Y=y\right).
	\end{eqnarray*}
	The relationship between $\theta(y_{1},y_{2})$ and $\theta(y)$ is
	\begin{eqnarray*}
		\theta(y_{1})=\lim_{y_{2}\rightarrow y_{1}}\theta(y_{1},y_{2}),~~~\theta(y_{1},y_{2})=\frac{1}{P(y_{1}<Y<y_{2})}\int_{y_{1}}^{y_{2}}\theta(y)f_Y(y)dy.
	\end{eqnarray*}
	
	Let $Q_{Y}\big(\tau\big)$ denote the $\tau$-th quantile of $Y$. Like Firpo et al. (2009, page 959), we define the CQPE, which is the effect of a small change of $D$ on the conditional quantile of $Y$:
	\[
	CQPE_{\tau}(d,x)=\frac{\partial{Q}_{Y}(\tau|d,x)}{\partial{d}}.
	\]
	Let $\zeta_{\tau}(d,x)=F_{Y}\left(Q_{Y}(\tau)|d,x\right)$ be the matching function in Firpo et al. (2009).

	\noindent\textbf{Proposition 2.3.} Under the same assumptions as Proposition 2.1, we have (i)
	\begin{eqnarray*}
		\theta\left(Q_{Y}\big(\tau\big)\right)=UQPE(\tau).
	\end{eqnarray*}
	(ii)
	\begin{eqnarray*}
		\theta\bigg(Q_{Y}\big(\tau_{1}\big),Q_{Y}\big(\tau_{2}\big)\bigg)=\frac{1}{\tau_{2}-\tau_{1}}\int_{\tau_{1}}^{\tau_{2}}UQPE(\tau)d\tau.
	\end{eqnarray*}
	(iii)
	\begin{eqnarray*}
		UQPE(\tau)=E\left[ CQPE_{\zeta_{\tau}(D,X)}(D,X)\frac{f_{Y}\left(Q_{Y}(\tau)|D,X\right)}{f_{Y}\left(Q_{Y}(\tau)\right)}\right].
	\end{eqnarray*}
	
	Results (i)-(ii) indicate some interesting  connections between OASD and UQPE, that is, both quantities actually possess the identical economic interpretation, and such equivalence holds under a very general nonseparable data generating process without monotonicity. Based on this equivalence, we obtain two additional insights. First, UQPE can be alternatively interpreted as a mean causal effect on an identifiable subpopulation located at some part of the outcome distribution. Second, many desirable properties of OASD may carry over to the UQPE. For example, until now, the literature has not obtained any efficiency bound of UQPE. By Proposition 2.3, the semiparametric efficiency bound of UQPE may be learnt from that of OASD, which we shall derive in Section 4. Last but not least, result (iii) has been proved by Firpo et al. (2009, Proposition 1, page 959).
	There, the proof needs two assumptions: (i) joint independence of $(D,X)$ and $U_D$ and (ii) $m(d,x,u)$ is monotonic in $u$.
	Here we show the result still holds under much weaker assumptions, that is, we only need conditional exogeneity and without monotonicity.
	
	\subsection{Marginal Partial Policy Effect and Outcome Conditioned Average Partial Effect}

	The preceding subsection demonstrates there is a close connection between two different classes of quantities: one is the outcome conditioned average partial effect of a continuous treatment (OASD) and the other
	class incorporates parameters measuring the effect of a counterfactual change
	in the distribution of a single covariate on unconditional quantiles of the outcome (UQPE).
	This subsection provides an additional example in support of this insight, that is,  we show another counterfactual policy effect parameter, called the marginal partial policy effects (Rothe 2012) can also be represented as some outcome conditioned average partial effect.
	
	Rothe (2012) proposes a class of quantities to evaluate the effect of a counterfactual change
	in the \textit{unconditional distribution} of a single covariate on the unconditional distribution
	of an outcome variable of interest,  holding everything else, in particular the
	dependence structures of the covariates constant.  Using the notations in this paper,  the parameters in Rothe (2012) can be described as follows.
	An outcome variable $Y$ is related to a continuously distributed covariate $D$, and $K$ dimensional vector of covariates
	$X=(X_{1},\cdots,X_{K})$ through a general nonseparable structural model
	\begin{eqnarray*}
		Y=m(D,X,U_D).
	\end{eqnarray*}
	Let $Q_{0}(\cdot), Q_{1}(\cdot),\cdots, Q_{K}(\cdot)$ be the unconditional quantile function of $D$, $X_{1},\cdots, X_{K}$ respectively.
	Then $(D,X)$ can be equivalently expressed in
	terms of their unconditional quantile
	functions and a rank vector $R=\left(R_{0},R_{1},\cdots, R_{K}\right)$ of standard uniformly distributed latent variables, that is,
	\[
	Y=m\left(Q_{0}\big(R_{0}\big),Q_{1}\big(R_{1}\big),\cdots, Q_{K}\big(R_{K}\big),U_D\right),\eqno(2.3.1)
	\]
	with $R_{k}\sim^{d}\mbox{Uniform} (0,1)$, $k=0,\cdots, K$.
	The joint distribution of $R$, also the copula function of $(D,X)$, measures the
	dependence structure between $D$ and $X$. Because $D$ is continuously distributed, the latent rank
	variable $R_{0}$ constitutes a one-to-one transformation of $D$. Define the outcome
	$Y_{H}$ of the counterfactual experiment in which the unconditional distribution of
	$D$ has been changed to some CDF $H(\cdot)$, but everything else has been held constant,
	\[
	Y_{H}=m\left(H^{-1}\big(R_{0}\big),Q_{1}\big(R_{1}\big),\cdots, Q_{K}\big(R_{K}\big),U_D\right).\eqno(2.3.2)
	\]

	Let $Q_{A}(\cdot)$ denote the quantile function of a generic random variable $A$. It is natural to define the $\tau$-th partial quantile policy effect of changing the marginal distribution of $D$ from $F_{0}(\cdot)=Q_{0}^{-1}(\cdot)$ to $H(\cdot)$ as
	\begin{eqnarray*}
		Q_{Y_{H}}(\tau)-Q_{Y}(\tau).
	\end{eqnarray*}
	In practice, most policies are contracted, expanded or adjusted gradually, thus one may naturally consider the effect of an infinitesimal change of the distribution of $D$ at a certain given direction. Let $G_{0}$ be any fixed CDF,  representing the direction of policy change.
	Let $H=H_{t}$ be an element of a continuum of CDFs indexed by $t\in[0,1]$ such
	that
	\[
	H_{t}=F_{0}+t\left(G_{0}-F_{0}\right).\eqno(2.3.3)
	\]
	Then the $\tau$-th marginal quantile partial policy effect (MQPE) at the direction of $G_{0}$ is given by\footnote{In applications, the policy effect of changing the unconditional distribution of $D$ from $F_{0}$ to any given fixed CDF $H$ can be well approximated by the effect of an appropriately designed infinitesimal change. To see this, let $t$ be a sufficiently small real number, say, $t=0.01$. To learn about the effect of changing $F_{0}$ to a fixed $H$, one may solve $G_{0}$ from  $H=F_{0}+t\left(G_{0}-F_{0}\right)$, that is,
		$G_{0}=\frac{H-(1-t)F_{0}}{t}$. Then $Q_{Y_{H}}(\tau)-Q_{Y}(\tau)\simeq MQPE(\tau,G_{0}) \cdot t$.}
	\[
	MQPE(\tau,G_{0})=\frac{\partial Q_{Y_{H_{t}}}(\tau)}{\partial t}\bigg|_{t=0}.\eqno(2.3.4)
	\]
	
	The MQPE described above is very similar to the UQPE defined by (2.2.1). Given the equivalence between UQPE and OASD,
	an interesting question is whether there exists some outcome conditioned average partial effect quantity, which is equivalent to MQPE? The next proposition answers this question.
	
	\noindent\textbf{Proposition 2.4.} Define
	\begin{eqnarray*}
		\varsigma(y,G_{0})=E\left(\frac{\partial m\left(H_{t}^{-1}(R_{0}),X,U_D\right)}{\partial t}\bigg|_{t=0}\bigg|Y=y\right)
	\end{eqnarray*}
	with $H_{t}$ given by (2.3.3).
	$\varsigma(y,G_{0})$ is called \textit{outcome conditioned average partial policy effect}, which measures the average effect of changing the unconditional distribution of $D$ infinitesimally towards the direction of $G_{0}$ on the individuals with $Y=y$.
	Similarly, we can define
	\begin{eqnarray*}
		\varsigma(y_{1},y_{2},G_{0})=E\left(\frac{\partial m\left(H_{t}^{-1}(R_{0}),X,U_D\right)}{\partial t}\bigg|_{t=0}\bigg|Y\in(y_{1},y_{2})\right).
	\end{eqnarray*}
	Under the same assumptions in Proposition 2.1,   (i)
	\begin{eqnarray*}
		\varsigma(Q_{Y}\big(\tau\big),G_{0})=MQPE(\tau,G_{0})
	\end{eqnarray*}
	(ii)
	\begin{eqnarray*}
		\varsigma\bigg(Q_{Y}\big(\tau_{1}\big),Q_{Y}\big(\tau_{2}\big),G_{0}\bigg)=\frac{1}{\tau_{2}-\tau_{1}}\int_{\tau_{1}}^{\tau_{2}}MQPE\big(\tau,G_{0}\big)d\tau.
	\end{eqnarray*}
	
	Like OASD and UQPE, the above proposition indicates MQPE can be interpreted as a mean effect on an identifiable subpopulation located at some part of the outcome distribution. As Rothe (2012) has shown MQPE is identified under a conditional exogeneity condition, a corollary of Proposition 2.4 is that both  $\varsigma(y,G_{0})$ and $\varsigma(y_{1},y_{2},G_{0})$ are also identifiable under the same conditions. Further investigation of these parameters is beyond the scope of this paper.

	\section{Auto-Debiased Machine Learning Estimator}

	The preceding section shows that OASD is identified
	under a conditional exogeneity assumption, by controlling for a rich information about covariates $X$, thus it is ideal to consider an estimation procedure using high-dimensional controls in data.
	We propose an automatic debiaed/double machine learning (ADML) procedure for estimating OASDs with high dimensional covariates. The procedure is easily implemented and semiparametrically efficient.
	The estimation method consists of three steps:
	
	\noindent(i) Estimate the CDF $F_{Y}\left(y|d,x\right)$, its integral and derivatives  using high-dimensional nonparametric methods with model selection.
	
	\noindent(ii) Using the orthogonal score to estimate $\frac{\partial_Df(D,X)}{f(D,X)}$ automatically.
	
	\noindent (iii) Estimate $\theta(y_{1},y_{2})$ based on the orthogonal score via the plug-in rule.

	\noindent We now describe the estimation procedure  in detail.
	
	\textbf{Step 1.} (\textit{Estimate CDF}) Let $b(d,x)=\{b_{k}(d,x)\}_{k=1}^{p}$ be the basis functions used to approximate $F_{Y}(y|d,x)$, and $\Lambda(\cdot)$ be logistic link function. Then  $F_{Y}(y|d,x)$ can be estimated by
	\[
	\widehat{F}_{Y}(y|d,x)=\Lambda\left(b^{\prime}(d,x)\widehat{\beta}(y)\right).
	\]
	To obtain $\widehat{\beta}(y)$, we first estimate $\widetilde{\beta}(y)$ by the Lasso penalized distribution regression
	\[
	\widetilde{\beta}(y)=\arg\min_{\beta}\frac{1}{n}\sum_{i=1}^{n}\bigg[1\bigg\{Y_{i}\leq{y}\bigg\}\ln\Lambda\bigg(b^{\prime}(D_{i},X_{i})\beta\bigg)+1\bigg\{Y_{i}>{y}\bigg\}\ln\left(1-\Lambda\bigg(b^{\prime}(D_{i},X_{i})\beta\bigg)\right)\bigg]+\frac{\lambda}{n}\left\Vert\widehat{\Psi}_{y}\beta\right\Vert_{1},
	\]
	where $\lambda$ denotes the penalty level to guarantee good theoretical properties of the lasso estimator, and $\widehat{\Psi}_y=\text{diag}(\widehat{\psi}_{y1}^q,\cdots,
	\widehat{\psi}_{yp}^q)$ denotes the diagonal matrix of penalty loadings. According to Belloni et al. (2017), we set the penalty level $\lambda$ as
	\[
	\lambda=1.1\sqrt{n}\Phi^{-1}\left(1-\frac{0.1/\ln(n)}{2p n}\right).
	\]
	The penalty loadings can be obtained by the following algorithm, proposed by Belloni et al. (2017, Algorithm 6.1, page 261)
	\begin{enumerate}[(1)]
		\item  Set $q=0$ and initialize $\widehat{\psi}_{yk}^0$ for each $k=1,\cdots,p$
		\[
		\widehat{\psi}_{yk}^0=\frac{1}{2}\sqrt{\frac{1}{n}\sum_{i=1}^{n}b_k^2(D_i,X_i)},k=1,\cdots,p.
		\]
		\item Calculate the lasso and post-lasso estimators $\tilde{\beta}(y)$ and $\widehat{\beta}(y)$, based on $\widehat{\Psi}_y=\text{diag}(\widehat{\psi}_{y1}^q,\cdots,
		\widehat{\psi}_{yp}^q)$.
		\item Set
		\[
		\widehat{\psi}_{yk}^{q+1}=\sqrt{\frac{1}{n}\sum_{i=1}^{n}b_k^2(D_i,X_i)\left(1\{Y_i\leq y\}-\Lambda\left(b'(D,X)\widehat{\beta}(y)\right)\right)},k=1,\cdots,p,
		\]
		\item If $q>q^*$ for the upper bound on the number of iterations $q^*$ stop; otherwise set $q\leftarrow q+1$ and go to step (2).
	\end{enumerate}
	
	Given $y\in\mathcal{S}_{Y}$, define $\widehat{I}(y)=\text{supp}\left(\widehat{\beta}(y)\right)$. The post-lasso estimator $\widehat{\beta}(y)$ is a solution to
	\[\begin{aligned}
		\widehat{\beta}(y)=&\arg\min_{\beta}\frac{1}{n}\sum_{i=1}^{n}\bigg[1\bigg\{Y_{i}\leq{y}\bigg\}\ln\Lambda\bigg(b^{\prime}(D_{i},X_{i})\beta\bigg)+1\bigg\{Y_{i}>{y}\bigg\}\ln\left(1-\Lambda\bigg(b^{\prime}(D_{i},X_{i})\beta\bigg)\right)\bigg], \\
		&\text{s.t.}\ \beta_{j}=0,\ j\notin\widehat{I}(y).
	\end{aligned}\]
	
	(\textit{Estimate the integral of CDF}) Let $\Delta{y}=\left(y_{2}-y_{1}\right)/J$ for some positive integer $J$. Notice that by definition of integration,
	\[
	IF(y_{1},y_{2},d,x)\doteq\int_{y_{1}}^{y_{2}}F_{Y}(y|d,x)dy=\lim_{J\to\infty}\sum_{j=1}^{J}F_{Y}(y_{1}+j\Delta{y}|d,x)\Delta{y}.
	\]
	Thus, the estimator of the integral of CDF can be constructed by
	\[
	\widehat{IF}(y_{1},y_{2},d,x)=\sum_{j=1}^{J}\widehat{F}_{Y}(y_{1}+j\Delta{y}|d,x)\Delta{y}=\sum_{j=1}^{J}\Lambda\left(b^{\prime}(d,x)\widehat{\beta}\left(y_{1}+j\Delta{y}\right)\right)\Delta{y}. \eqno{(3.1)}
	\]
	
	(\textit{Estimate the derivative of the integral of CDF}) Let
	\begin{eqnarray*}
		DIF(y_{1},y_{2},d,x)\doteq\partial_d\int_{y_1}^{y_2}F_Y(y|d,x)dy.
	\end{eqnarray*}
	We estimate $DIF(y_{1},y_{2},d,x)$ by a high order partial difference approach, by borrowing the idea from Belloni et al. (2019) in dealing with the estimation of conditional density function.
	Let $\ell$ be some positive integer. A partial difference estimator  of $DIF(y_{1},y_{2},d,x)$ with a bias of general order $O\left(h_{n}^{2\ell}\right)$ is given by
	\[
	\widehat{DIF}(y_{1},y_{2},d,x)=\frac{1}{2h_{n}}\sum_{l=1}^{\ell}\eta_{l}\left(\widehat{IF}(y_{1},y_{2},d+lh_{n},x)-\widehat{IF}(y_{1},y_{2},d-lh_{n},x)\right), \eqno{(3.2)}
	\]
	with $h_{n}$ the bandwidth satisfying $h_{n}\to{0}$. The constants $\eta_{l}$ are determined by\footnote{For example, we have $\eta_{1}=1$ for $\ell=1$; $\eta_{1}=4/3$ and $\eta_{2}=-1/6$ for $\ell=2$; $\eta_{1}=3/2$, $\eta_{2}=-3/10$ and $\eta_{3}=1/30$ for $\ell=3$.}
	\[
	\sum_{l=1}^{\ell}l\cdot\eta_{l}=1 \eqno{(3.3)}
	\]
	and for $v=3,5,\dots,2\ell-1$,
	\[
	\sum_{l=1}^{\ell}l^{v}\cdot\eta_{l}=0. \eqno{(3.4)}
	\]
	As will be clear in the next section, the bandwidth should satisfy the following two conditions
	\[
	h_{n}^{4}n\to\infty \quad\text{and}\quad h_{n}^{2\ell}n^{1/4}\to{0}.
	\]
	Thus, we suggest to select the bandwidth $h_{n}=n^{-1/\left(4\ell+2\right)}$ by maximizing the convergence rate of $\widehat{DIF}$. We also compare the finite sample performance of \textit{DIF} in Eq.(3.2) with Sasaki et al. (2022)'s estimator. Simulation results are reported in Appendix B, which show that our estimator performs better in terms of bias ratio.
	
	\noindent\textbf{Remark 3.1.} By Eq.(3.2), we employ an estimator in the spirit of the high order bias reduction kernel smoothing to estimate the CDF derivative instead of directly differentiating the lasso CDF estimator.
	By using a partial difference estimator with a bias of sufficiently high order, we do not need to impose substantially more restrictive approximate sparsity conditions than before while preserving the good rates of convergence
	for both CDF and its derivatives. To understand how estimator (3.2) works, we first take Taylor expansion of $\widehat{IF}$ with respect to $d\pm{lh_{n}}$ at $d$.
	By the construction of (3.3), all the coefficients of first order derivatives sum up to one, while the coefficients of remaining orders sum up to zero by construction of Equation (3.4). Thus, only the first order derivative of CDF and the terms of order $O\left(h_{n}^{2\ell}\right)$ are left.

	\textbf{Step 2.} (\textit{Automatic estimation of $\frac{\partial_Df(D,X)}{f(D,X)}$.}) According to Chernozhukov et al. (2022a,b) and Singh and Sun (2021), we estimate $\frac{\partial_Df(D,X)}{f(D,X)}$ automatically based on the double robustness property of orthogonal score function $\psi$ (defined below Proposition 2.2). For any real function $\delta(d,x)$, notice that
	\[
	\frac{\partial}{\partial\tau}E\left[\psi\bigg(W,\theta,\eta+\tau\delta;y_1,y_2\bigg)\right]\bigg|_{\tau=0}=0.
	\eqno{(3.5)}\]
	Let $\delta(d,x)=\left(0,\widetilde{\delta}(d,x),0\right)^{\prime}$. $(3.5)$ is equal to
	\[
	E\left[\partial_{D}\widetilde{\delta}(D,X)+\frac{\partial_Df(D,X)}{f(D,X)}\widetilde{\delta}(D,X)\right]=0. \eqno{(3.6)}
	\]
	The automatic estimator for $\frac{\partial_Df(D,X)}{f(D,X)}$ is constructed on the basis of  $(3.6)$. Suppose $L(D,X)=\frac{\partial_Df(D,X)}{f(D,X)}$ is replaced by a linear combination $b^{\prime}(D,X)\gamma$ and let $\widetilde{\delta}(D,X)$ be one element $b_{k}(D,X)$ for $k=1,\dots,p$. Then $L(D,X)=\frac{\partial_Df(D,X)}{f(D,X)}$ can be estimated by
	\[
	\widehat{L}(d,x)=b^{\prime}(d,x)\widehat{\gamma}.
	\]
	$\widehat{\gamma}$ is the Lasso estimator which is constructed by\footnote{We can also use post-lasso estimator to replace $\gamma$.}
	\[
	\widehat{\gamma}=\arg\min_{\gamma}-2\widehat{M}^{\prime}\gamma+\gamma^{\prime}\widehat{G}\gamma+2\widetilde{\lambda}\left\Vert\gamma\right\Vert_{1}, \eqno{(3.7)}
	\]
	where $\widetilde{\lambda}>0$ is a positive scalar to control for the degree of penalty, and
	\[\begin{aligned}
		&\widehat{M}=-\frac{1}{n}\sum_{i=1}^{n}\partial_{D_{i}}b(D_{i},X_{i}), \\
		&\widehat{G}=\frac{1}{n}\sum_{i=1}^{n}b(D_{i},X_{i})b^{\prime}(D_{i},X_{i}). \\
	\end{aligned}\]
	We estimate (3.7) by the iterative tuning procedure for data-driven regularization parameter $\tilde{\lambda}$ , proposed by Chernozhukov et al. (2022b, Appendix A, page 1000).
	\\

	\textbf{Step 3.} (\textit{Estimate $\theta(y_{1},y_{2})$ by plug-in})  $P(y_{1},y_{2})=P(y_{1}<Y<y_{2})$ can be directly estimated by
	\[
	\widehat{P}(y_{1},y_{2})=\frac{1}{n}\sum_{i=1}^{n}1\bigg\{y_{1}<Y_{i}<y_{2}\bigg\}. \eqno{(3.8)}
	\]
	Based on the estimator of $
	\eta=\left(P\big(y_1<Y<y_2\big),\int_{y_1}^{y_2}F_Y\left(y\big|D,X\right)dy,\frac{\partial_Df\big(D,X\big)}{f\big(D,X\big)}\right)$,
	$\theta(y_{1},y_{2})$ is straightforwardly estimated via a plug-in rule, such that
	\[
	\frac{1}{n}\sum_{i=1}^{n}\psi\bigg(W_{i},\widehat{\theta},\widehat{\eta};y_{1},y_{2}\bigg)=0.
	\]
	Or
	\[\begin{aligned}
		\widehat{\theta}(y_{1},y_{2})=&-\frac{1}{n}\sum_{i=1}^{n}\bigg[\frac{1}{\widehat{P}(y_{1},y_{2})}\widehat{DIF}(y_{1},y_{2},D_{i},X_{i})\\
		&-\frac{1}{\widehat{P}(y_{1},y_{2})}\widehat{L}(D_{i},X_{i})\bigg(\widehat{IF}(y_{1},y_{2},D_{i},X_{i})-\int_{y_{1}}^{y_{2}}1\bigg\{Y_{i}<y\bigg\}dy\bigg)\\
		&+\left(\frac{1}{n}\sum_{j=1}^{n}\frac{\widehat{DIF}(y_{1},y_{2},D_{j},X_{j})}{\widehat{P}^{2}(y_{1},y_{2})}\right)\left(1\bigg\{y_1<Y_{i}<y_2\bigg\}-\widehat{P}(y_{1},y_{2})\right)\bigg].
	\end{aligned}\eqno{(3.9)}\]
	where
	\[
	\int_{y_{1}}^{y_{2}}1\bigg\{Y_{i}<y\bigg\}dy=1\bigg\{Y_{i}\leq y_{1}\bigg\}(y_{2}-y_{1})+1\bigg\{y_{1}<Y_{i}<y_{2}\bigg\}(y_{2}-Y_{i}).
	\]

	\noindent\textbf{Remark 3.2.} As is standard in the literature (Chernozhukov et al. (2018)), we can  use  various data splitting methods to further relax the entropy condition required in the subsequent asymptotic analysis.
	There is no asymptotic efficiency loss from sample splitting under cross fitting. See Chernozhukov et al. (2018) for more details. For simplicity, we do not describe the estimation procedure using the data splitting method in this paper.
	One reason is that the entropy of the function classes can be easily verified (which can be found in the next section), e.g.,  differentiability of density function. Second, sample splitting facilitates the statistical inference, at the cost of making estimation more involved. For example, researchers need to choose the number of fold $K$, and repeat running Steps 1-3 $K$ times.
	\\

	\noindent\textbf{Algorithm 1.} (\textit{Auto Double Machine Learning Estimator })
	
	\noindent\textbf{Step 1.} Pick a finite set $\mathcal{Y}\subset\mathcal{S}_{Y}$ of grid points of outcome values.
	
	\noindent\textbf{Step 2.} For any $y\in\mathcal{Y}$, compute $\widetilde{\beta}(y)$ from $\ell_{1}$-penalized logistic regression of $1\{Y<y\}$ on $b(D,X)$.
	
	\noindent\textbf{Step 3.} For any $y\in\mathcal{Y}$, compute $\widehat{\beta}(y)$ from logistic regression of $1\{Y<y\}$ on $\left\{b_k(D,X):\widetilde{\beta}_{k}\neq{0}\right\}$.
	
	\noindent\textbf{Step 4.} Estimate the integral of CDFs and its derivative via (3.1) and (3.2).
	
	\noindent\textbf{Step 5.} Compute $\widehat{\gamma}$ from $\ell_{1}$-penalized GMM via (3.7).
	
	\noindent\textbf{Step 6.} Estimate the unconditional probability $P(y_{1},y_{2})$ via (3.8).
	
	\noindent\textbf{Step 7.} For a pair of $(y_{1},y_{2})\in\mathcal{Y}\times\mathcal{Y}$ and $y_{1}<y_{2}$, compute $\widehat{\theta}(y_{1},y_{2})$ via (3.9).

	\section{Asymptotic Properties and Inference}
	
	In this section, we establish the asymptotic properties for the ADML estimator of $\theta(y_{1},y_{2})$. Overall, the ADML estimator, which can be viewed as a stochastic process of $(y_{1},y_{2})$, is proved to be uniformly Gaussian based on some high level conditions. Following, we provide a set of sufficient conditions for these high level conditions, which are convenient to hold in practice. We show the ADML estimator is semiparametrically efficient, that is, it achieves the semiparametric efficiency bound. Finally, we derive the uniform validity of the multiplier bootstrap used to construct uniform confidence bands.
	
	\subsection{Large Sample Properties}
	
	Consider fixed sequence of numbers $\Delta_{n}\to{0^{+}}$ at a speed at most polynomial in $n$ (e.g., $\Delta_{n}\geq{1/n^{\bar{c}}}$ for some $\bar{c}>0$), and positive constants $c$, $C$, $H$ and $T$.
	
	\subsubsection{Limiting Distribution of ADML Estimator}
	
	We introduce a set of assumptions used to prove the uniform Gaussianity of the ADML estimator. Some of these assumptions are high level. Sufficient conditions for these high level conditions are provided in the next subsection.
	
	\noindent\textbf{Assumption 4.1.} The random element $W$ takes values in a compact measure space $(\mathcal{W},\mathcal{A}_{\mathcal{W}})$ and its law is determined by a probability measure $P$. The observed data $\{W_{i}\}_{i=1}^{n}$ consist of $n$ i.i.d. copies of a random element $W=(Y,D,X)\in\mathbb{R}^{2+d_{X}}$.
	
	\noindent\textbf{Assumption 4.2.} Let $u=(y_{1},y_{2})\in\mathcal{U}\subset\mathbb{R}^{2}$ be the index of target parameter $\theta$. $\mathcal{U}$ is a totally bounded metric space equipped with a semi-metric $d_{\mathcal{U}}$.\footnote{Our OASDs are defined on $(y_1,y_2)\in \mathcal{S}_Y \times \mathcal{S}_Y$ with $y_1<y_2$. Let $c_{0}>0$, $c_{1}$ and $c_{2}$ be three constants.  The metric space $\mathcal{U}$ can be defined as $\mathcal{U}=\left\{(y_{1},y_{2}):c_{1}\leq{y_{1}+c_{0}}\leq{y_{2}}\leq{c_{2}}\right\}$, which is a bounded upper triangular.} Denote $Y(u)$ as a measurable transform $t(Y,u)$ of $Y$ and $u$.\footnote{Specifically, $Y(u)\in\left\{\displaystyle\int_{y_{1}}^{y_{2}}1\{Y\leq{y}\}dy,1\{y_{1}\leq{Y}\leq{y_{2}}\}\right\}$ in this paper.} The map $u\mapsto{Y(u)}$ obeys the following uniform continuity property:
	\[
	\lim\limits_{\epsilon\to{0^{+}}} \sup_{d_{\mathcal{U}}(u,\bar{u})\leq\epsilon}\left\Vert Y(u)-Y(\bar{u})\right\Vert_{P,2}=0, \quad E\sup_{u\in\mathcal{U}}|Y(u)|^{2+c}<\infty,
	\]
	where the supremum in the first expression is taken over $u,\bar{u}\in\mathcal{U}$.
	
	Assumption 4.2 defines a valid metric space for OASDs, and restricts the continuity and boundedness of $Y$. According to Assumption 4.2, there exists a positive constant $H$, which ensures that $\mathcal{U}\subset[-H,H]\times[-H,H]$. Denote the space $\mathcal{H}=\{y: |y|\leq{H}\}$.
	
	\noindent\textbf{Assumption 4.3.} Assume the functions $F_{Y}(y|d,x)$ and $L(d,x)$ can be approximated by
	\[
	F_{Y}(y|d,x)=\Lambda\bigg(b(d,x)^{\prime}\beta(y)\bigg)+r_{F}(y,d,x)
	\]
	and
	\[
	L(d,x)=b(d,x)^{\prime}\gamma+r_{L}(d,x),
	\]
    where $r_{F}(y,d,x)$ and $r_{L}(d,x)$ are the approximation errors. Then uniformly over $y\in\mathcal{H}$ and $u\in\mathcal{U}$,
	
	(i) The sparsity condition $\Vert\beta(y)\Vert_{0}\leq{s_{\beta}}$ and $\Vert\gamma\Vert_{0}\leq{s_{\gamma}}$ holds, and the approximation errors satisfy $\Vert r_{F}\Vert_{P,2}=o_{p}\left(h_{n}n^{-1/4}\right)$, $\Vert r_{F}\Vert_{P,\infty}=o_{p}\left(h_{n}\right)$, and $\Vert r_{L}\Vert_{P,2}=o_{p}\left(n^{-1/4}\right)$, $\Vert r_{L}\Vert_{P,\infty}=o_{p}(1)$. The sparsity indices $s_{\beta}$, $s_{\gamma}$ and the number of terms $p$ in the vector $b(d,x)$ obeying $s_{\beta}^{2}\log^{2}(p\vee n)=o\left(h_{n}^{4}n\right)$, and $s_{\gamma}^{2}\log^{2}(p\vee n)=o(n)$. The bandwidth $h_{n}$ satisfies $h_{n}^{2\ell}=o\left(n^{-1/4}\right)$ for some positive integer $\ell$.
	
	(ii) There are estimators $\widehat{\beta}(y)$ and $\widehat{\gamma}$ such that, with probability no less than $1-\Delta_{n}$, the estimation errors satisfy $\left\Vert b(D,X)^{\prime}\left(\widehat{\beta}(y)-\beta(y)\right)\right\Vert_{\mathbb{P}_{n},2}=o_{p}\left(h_{n}n^{-1/4}\right)$, $K_{n}\left\Vert \widehat{\beta}(y)-\beta(y)\right\Vert_{1}=o_{p}\left(h_{n}\right)$, and $\left\Vert b(D,X)^{\prime}\left(\widehat{\gamma}-\gamma\right)\right\Vert_{\mathbb{P}_{n},2}=o_{p}\left(n^{-1/4}\right)$, $K_{n}\left\Vert \widehat{\gamma}-\gamma\right\Vert_{1}=o_{p}(1)$; the estimators are sparse such that $\left\Vert\widehat{\beta}(y)\right\Vert_{0}\leq{s_{\beta}}$ and $\Vert\widehat{\gamma}\Vert_{0}\leq{s_{\gamma}}$.
	
	(iii) The empirical and population norms induced by the Gram matrix formed by $\{b_{j}(d,x)\}_{j=1}^{p}$ are equivalent on sparse subsets, such as
	\[
	\sup_{\Vert\upsilon\Vert_{0}\leq{s\log{n}}}\left|\frac{\Vert b(D,X)^{\prime}\upsilon\Vert_{\mathbb{P}_{n},2}}{\Vert b(D,X)^{\prime}\upsilon\Vert_{P,2}}-1\right|\leq{\epsilon_{n}}.
	\]
	The boundedness conditions hold:
	$\left\Vert\left\Vert\partial^{l}_{D} b(D,X)\right\Vert_{\infty}\right\Vert_{P,\infty}\leq{K_{n}}$ for $l=0,1$, $\Vert Y(u)\Vert_{P,\infty}\leq{C}$ and $\left\Vert \partial^{2}_{D}F(y|D,X)\right\Vert_{P,\infty}\leq{C}$, $\left\Vert\partial_{D}F(y|D,X)\right\Vert_{P,2}\leq{C}$.
	
	(iv) $\partial_{d}F(y|d,x)$ is bounded and $\sigma$-th continuously differentiable with respect to $(d,x)$, and satisfies $2\sigma>\max\{1+d_{X_{c}},4(\ell-1)\}$, where $d_{X_{c}}$ denotes the dimension of continuous components in $X$.
	
	\noindent\textbf{Remark 4.1.} Assumption 4.3-(i) includes conditions on the approximate sparsity of the model and bandwidths used to estimate CDFs, their derivatives and $L(D,X)=\frac{\partial_Df(D,X)}{f(D,X)}$. To ensure that the CDF and its derivative have a desirable convergence rate, i.e., $o_{p}\left(n^{-1/4}\right)$, the growth rate of the approximate sparsity and the order of the approximation error are restricted. The sparsity condition of $s_{\beta}$ and the convergence rate condition of $h_{n}$ together require that $s_{\beta}$ should grow slower than $n^{1/2-1/(4\ell)}$. This growth rate can be improved if the order of bias $2\ell$ increases, namely that the CDF derivative is estimated with a higher order bias, accompanied with additional smoothness assumptions (Assumption 4.3(iv)). Sasaki et al. (2022) estimate the CDF derivative by directly differentiating  the CDF function. In practice, such estimation procedure might not be satisfactory. The uniform convergence rate of the derivative estimator  mainly depends on the level of sparsity, and is usually slower than the CDF estimator. To achieve the faster uniform convergence rate, particularly faster than $n^{-1/4}$, the divergence rate of sparsity should be restricted to grow slower than $n^{1/2-c}$ for some constant $c$, which cannot be improved anymore. As previously discussed, the divergence rate of sparsity $s_{\beta}$ in this paper can be improved to $n^{1/2}$ as close as possible by choosing a partially difference estimator with higher order bias.
	
	\noindent\textbf{Remark 4.2.} Assumption 4.3(ii) imposes some high-level conditions on the estimators of nuisance functions. In the next subsection, we provide a set of regular and sufficient conditions for both (Post-)Lasso and automatic estimators to satisfy the uniform bounds. Assumption 4.3(iii) first presents the equivalence between empirical and population norms. Sufficient conditions and primitive examples of functions admitting sparse approximations are given in Belloni et al. (2014). The boundedness conditions in Assumption 4.3(iii) are made to simplify arguments, and they could be removed at the cost of more regular conditions and complicated proofs. Assumption 4.3(iv) is majorly used to establish the Gaussian process of the ADML estimator, which restricts the set of functions $\left\{\partial_{D}\displaystyle\int_{y_{1}}^{y_{2}}F_{Y}(y|D,X)dy:(y_{1},y_{2})\in\mathcal{U}\right\}$ to be a Donsker class. Furthermore, another composition of the smoothness restriction, such that $\sigma>2(\ell-1)$, is assumed to achieve a partially difference estimator with higher order bias in Section 3.
	
	The next theorem establishes the uniform Gaussianity of the empirical reduced-form process $\widehat{Z}_{n}(u)=\sqrt{n}\left(\widehat{\theta}(u)-\theta(u)\right)$ defined by Equation (3.9).
	
	\noindent\textbf{Theorem 4.1.} \textit{Under Assumptions 2.1-2.2 and 4.1-4.3, the reduced-form empirical process admits a linearization; namely}
	\[
	\widehat{Z}_{n}(u)=\sqrt{n}\left(\widehat{\theta}(u)-\theta(u)\right)=Z_{n}(u)+o_{p}(1) \quad in \quad \mathbb{D}=\ell^{\infty}\left(\mathcal{U}\right)
	\]
	\textit{where $Z_{n}(u)=\mathbb{G}_{n}\psi(W,\theta,\eta;u)$. The process $\widehat{Z}_{n}(u)$ is asymptotically Gaussian, namely}
	\[
	\widehat{Z}_{n}(u)\leadsto Z(u) \quad in \quad \mathbb{D}=\ell^{\infty}\left(\mathcal{U}\right)
	\]
	\textit{where $Z(u)=\mathbb{G}\psi(W,\theta,\eta;u)$ with $\mathbb{G}$ denoting Brownian bridge and with $Z(u)$ having bounded, uniformly continuous paths:}
	\[
	E\sup_{u\in\mathcal{U}}\Vert{Z}(u)\Vert<\infty, \quad \lim_{\epsilon\to{0}^{+}}E\sup_{d_{\mathcal{U}}\left(u,\widetilde{u}\right)\leq{\epsilon}}\Vert{Z}(u)-{Z}\left(\widetilde{u}\right)\Vert=0.
	\]

	\subsubsection{Asymptotic Properties for Estimators of Nuisance Functions}
	
	This subsection describes the asymptotic properties for the estimators of the nuisance functions (like the CDFs, their derivatives and $L(D,X)=\frac{\partial_Df(D,X)}{f(D,X)}$). These asymptotic results can be viewed as
 the sufficient conditions for Assumption 4.3(ii) to hold in practice. We first discuss the results for Lasso and Post-Lasso estimators with function valued outcomes and logistic link. Belloni et al. (2017) establish the general results for both linear and logistic link (e.g., Theorem 6.1 and 6.2). They explore that uniform consistency and convergence rate are mainly depends on the rate of sparsity. We invoke the same assumptions of Assumption 6.2 in Belloni et al. (2017).
	
	\noindent\textbf{Assumption 4.4.} The conditions of Assumption 6.2 in Belloni et al. (2017) hold.

	\noindent\textbf{Lemma 4.1.} \textit{If Assumption 4.4 together with Assumption 4.3(i) hold, $K_{n}^{2}s_{\beta}^{2}\log(p\vee{n})=o\left(h_{n}^{2}n\right)$, then}
	\[
	\sup_{y\in\mathcal{H}}\left\Vert b(D,X)^{\prime}\left(\widehat{\beta}(y)-\beta(y)\right)\right\Vert_{\mathbb{P}_{n},2}=o_{p}\left(h_{n}n^{-1/4}\right)
	\]
	\textit{and}
	\[
	K_{n}\sup_{y\in\mathcal{H}}\left\Vert \widehat{\beta}(y)-\beta(y)\right\Vert_{1}=o_{p}\left(h_{n}\right).
	\]
	\begin{proof}
		According to Theorem 6.2 in Belloni et al. (2017), we have
		\[
		\sup_{y\in\mathcal{H}}\left\Vert b(D,X)^{\prime}\left(\widehat{\beta}(y)-\beta(y)\right)\right\Vert_{\mathbb{P}_{n},2}=O_{p}\left(\sqrt{\frac{s_{\beta}\log(p\vee n)}{n}}\right)
		\]
		and
		\[
		\sup_{y\in\mathcal{H}}\left\Vert \widehat{\beta}(y)-\beta(y)\right\Vert_{1}=O_{p}\left(\sqrt{\frac{s_{\beta}^{2}\log(p\vee n)}{n}}\right).
		\]
		Then the results directly follow from Assumption 4.3(i).
	\end{proof}

	\noindent\textbf{Remark 4.3.} Here we restrict the divergence rate of basis functions $K_{n}$, sparsity $s_{\beta}$ and $p\vee{n}$ to satisfy $K_{n}^{2}s_{\beta}^{2}\log(p\vee{n})=o\left(h_{n}^{2}n\right)$, which is quiet strong than one required in Belloni et al. (2017) for standard (Post-)Lasso estimation, but is consistent with one in Belloni et al. (2019, Condition D(P)). The major reason is that in this paper, as well as Belloni et al. (2019), we need to estimate the derivative of (Post-)Lasso estimator. Thus, a faster convergence rate of the original estimator is required to ensure the estimator of derivative function achieving the convergence rate $o_{p}\left(n^{-1/4}\right)$.
\\

The following lemma establishes the bound rates for the derivative of (Post-)Lasso estimator.
\noindent\textbf{Lemma 4.2.} \textit{If Assumption 4.4 together with Assumption 4.3(i) and (iii) hold, $K_{n}^{2}s_{\beta}^{2}\log(p\vee{n})=o\left(h_{n}^{2}n\right)$, then}
\[
\sup_{u\in\mathcal{U}}\left\Vert\widehat{DIF}(u,D,X)-{DIF}(u,D,X)\right\Vert_{\mathbb{P}_{n},2}=o_{p}\left(n^{-1/4}\right)
\]
\textit{and}
\[
\sup_{u\in\mathcal{U}}\left\Vert\widehat{DIF}(u,D,X)-{DIF}(u,D,X)\right\Vert_{\mathbb{P}_{n},\infty}=o_{p}\left(1\right).
\]
\begin{proof}
	According to Lemma 4.1, Assumption 4.3(i) and (iii), we can directly derive
	\[
	\sup_{u\in\mathcal{U}}\left\Vert\widehat{IF}(u,D,X)-{IF}(u,D,X)\right\Vert_{\mathbb{P}_{n},2}=O_{p}\left(\sqrt{\frac{s_{\beta}\log(p\vee n)}{n}}\right)+o_{p}\left(h_{n}n^{-1/4}\right)=o_{p}\left(h_{n}n^{-1/4}\right)
	\]
	and
	\[
	\sup_{u\in\mathcal{U}}\left\Vert\widehat{IF}(u,D,X)-{IF}(u,D,X)\right\Vert_{\mathbb{P}_{n},\infty}=O_{p}\left(\sqrt{\frac{s_{\beta}^{2}\log(p\vee n)}{n}}\right)+o_{p}\left(h_{n}\right)=o_{p}\left(h_{n}\right),
	\]
	by triangle inequality. Thus, we have
\[
	\begin{aligned}
	&\left\Vert\widehat{DIF}(u,D,X)-{DIF}(u,D,X)\right\Vert_{\mathbb{P}_{n},2}  \\
	\leq&\left\Vert\frac{1}{2h_{n}}\sum_{l=1}^{\ell}\left(\widehat{IF}(u,D+lh_{n},X)-{IF}(u,D+lh_{n},X)\right)\right\Vert_{\mathbb{P}_{n},2} \\
	+&\left\Vert\frac{1}{2h_{n}}\sum_{l=1}^{\ell}\left(\widehat{IF}(u,D-lh_{n},X)-{IF}(u,D-lh_{n},X)\right)\right\Vert_{\mathbb{P}_{n},2} \\
	+&\left\Vert\frac{1}{2h_{n}}\sum_{l=1}^{\ell}\left({IF}(u,D+lh_{n},X)-{IF}(u,D-lh_{n},X)\right)-{DIF}(u,D,X)\right\Vert_{\mathbb{P}_{n},2} \\
	=&O_{p}\left(\frac{1}{h_{n}}\sqrt{\frac{s_{\beta}\log(p\vee n)}{n}}\right)+o_{p}\left(n^{-1/4}\right)+O_{p}\left(h_{n}^{2\ell}\right)=o_{p}\left(n^{-1/4}\right), \\
\end{aligned}
\]
and similarly,
\[
	\left\Vert\widehat{DIF}(u,D,X)-{DIF}(u,D,X)\right\Vert_{\mathbb{P}_{n},\infty}
=O_{p}\left(\frac{1}{h_{n}}\sqrt{\frac{s_{\beta}^{2}\log(p\vee n)}{n}}\right)+o_{p}\left(1\right)+O_{p}\left(h_{n}^{2\ell}\right)=o_{p}\left(1\right).
\]
which completes the proof.
\end{proof}
\\

Now we establish the bounds for automatic estimator. Recall that
\[
\widehat{M}=-\frac{1}{n}\sum_{i=1}^{n}\partial_{D_{i}}b(D_{i},X_{i})
\quad\text{and}\quad \widehat{G}=\frac{1}{n}\sum_{i=1}^{n}b(D_{i},X_{i})b^{\prime}(D_{i},X_{i}).
\]
Also define
\[
M=-E\partial_{D}b(D,X) \quad\text{and}\quad G=Eb(D,X)b^{\prime}(D,X).
\]
The following boundedness condition is assumed to achieve convergence rates for $\widehat{M}$ and $\widehat{G}$.

\noindent\textbf{Assumption 4.5.} There exists $K^{*}>0$ such that, with probability 1, $\left\Vert\partial^{l}_{D} b(D,X)\right\Vert_{\infty}\leq{K^{*}}$ for $l=0,1$.

Assumption 4.5 is quiet different with the boundedness condition provided in Assumption 4.3(iii). The boundary of basis functions  with probability $K^*$ is constant, while the boundary with supremum norm $K_{n}$ can diverge to infinity as $n$ increases. Next lemma investigates the convergence rates for $\widehat{M}$ and $\widehat{G}$.

\noindent\textbf{Lemma 4.3.} \textit{If Assumption 4.5 holds, then}
\[
\left\Vert{\widehat{M}-M}\right\Vert_{\infty}=O_{p}\left(\sqrt{\frac{\ln{p}}{n}}\right), \quad \left\Vert{\widehat{G}-G}\right\Vert_{\infty}=O_{p}\left(\sqrt{\frac{\ln{p}}{n}}\right).
\]
\begin{proof}
	For $j=1,\dots,p$, define
	\[
	\widehat{M}_{ij}^{*}=-\partial_{D_{i}}b_{j}(D_{i},X_{i})+E\partial_{D}b_{j}(D,X), \quad \text{and} \quad \widehat{M}_{j}^{*}=\frac{1}{n}\sum_{i=1}^{n}\widehat{M}_{ij}^{*}.
	\]
	For any constant $C^{*}$,
	\[
	P\left(\left\Vert{\widehat{M}-M}\right\Vert_{\infty}\geq{C^{*}}\sqrt{\frac{\ln{p}}{n}}\right)\leq\sum_{j=1}^{p}P\left(\left|\widehat{M}^{*}_{j}\right|\geq{C^{*}}\sqrt{\frac{\ln{p}}{n}}\right)\leq{p}\max_{j}P\left(\left|\widehat{M}^{*}_{j}\right|\geq{C^{*}}\sqrt{\frac{\ln{p}}{n}}\right).
	\]
	Notice that $E\widehat{M}_{ij}^{*}=0$ and
	\[
	\left\vert\widehat{M}_{ij}^{*}\right\vert\leq\left\vert\partial_{D_{i}}b_{j}(D_{i},X_{i})\right\vert+E\left\vert\partial_{D_{i}}b_{j}(D_{i},X_{i})\right\vert\leq2K^*.
	\]
	Define $\mathcal{K}=2K^*/\sqrt{\ln{2}}\geq\left\Vert\widehat{M}_{ij}^{*}\right\Vert_{\nu_{2}}$, where $\Vert\cdot\Vert_{\nu_{2}}$ denotes the sub-gaussian norm. It then follows by general Hoeffding's inequality (Theorem 2.6.2 of Vershynin (2018)) that there exists a constant $c$ such that
	\[\begin{aligned}
	{p}\max_{j}P\left(\left|\widehat{M}^{*}_{j}\right|\geq{C^{*}}\sqrt{\frac{\ln{p}}{n}}\right)&={p}\max_{j}P\left(\left|n\widehat{M}^{*}_{j}\right|\geq{C^{*}}\sqrt{n\cdot\ln{p}}\right)\\
	&\leq2p\exp\left(-\frac{c\left({C^{*}}\sqrt{n\cdot\ln{p}}\right)^{2}}{n\mathcal{K}^{2}}\right) \\
		&=2\exp\left(\ln{p}\left[1-\frac{c\left(C^{*}\right)^{2}}{\mathcal{K}^{2}}\right]\right)\to{0} \\
	\end{aligned}\]
	for any $C^{*}>\mathcal{K}/\sqrt{c}$ as $p\to\infty$. Thus, for large enough $C^{*}$, we have
	\[
	P\left(\left\Vert{\widehat{M}-M}\right\Vert_{\infty}\geq{C^{*}}\sqrt{\frac{\ln{p}}{n}}\right)\to{0},
	\]
	which implies the conclusion. Second argument can be similarly proved by another application of general Hoeffding's inequality.
\end{proof}

A set of additional assumptions, for example, the quality of approximation and the level of penalty, are assumed to derive the bounds for automatic estimator.

\noindent\textbf{Assumption 4.6.}  (i) There exist $c^{*}>1$, $\xi>0$ such that for all $\bar{s}\leq{c}^{*}\left(\frac{\ln{p}}{n}\right)^{-\frac{1}{1+2\xi}}$. There exists some $\widetilde{\gamma}\in\mathbb{R}^{p}$ with $\left\Vert\widetilde{\gamma}\right\Vert_{1}\leq{c^{*}}$ and $\bar{s}$ nonzero elements such that $\left\Vert L-b^{\prime}\widetilde{\gamma}\right\Vert_{P,2}^{2}\leq{c^{*}}\bar{s}^{-\xi}$.

\noindent(ii) $G$ is nonsingular with largest eigenvalue uniformly bounded in $n$.

\noindent(iii) Denote $\mathcal{S}_{\bar{\bar{\gamma}}}$ as the support of $\bar{\bar{\gamma}}$. There exist $k>3$ such that for $\bar{\bar{\gamma}}\in\left\{\widetilde{\gamma},\bar{\gamma}\right\}$,
\[
RE(k)=\inf_{\upsilon\neq{0},\sum_{j\in\mathcal{S}_{\bar{\bar{\gamma}}}^{c}}|\upsilon_{j}|\leq{k}\sum_{j\in\mathcal{S}_{\bar{\bar{\gamma}}}}|\upsilon_{j}|}\frac{\upsilon^{\prime}G\upsilon}{\sum_{j\in\mathcal{S}_{\bar{\bar{\gamma}}}}\upsilon_{j}^{2}}>0,
\]
where
\[
\bar{\gamma}=\arg\min_{\dot{\gamma}}\left\Vert L-b^{\prime}\dot{\gamma}\right\Vert_{P,2}^{2}+2\widetilde{\lambda}\left\Vert\dot{\gamma}\right\Vert_{1}.
\]

\noindent(iv) $\ln{p}=O(\ln{n})$ and $\widetilde{\lambda}=\kappa_{n}\sqrt{\frac{\ln{p}}{n}}$ for some $\kappa_{n}\to\infty$.

\noindent\textbf{Lemma 4.4.} \textit{If Assumptions 4.5 and 4.6 hold, then}
\[
\left\Vert{b}(D,X)^{\prime}\widehat{\gamma}-L(D,X)\right\Vert_{P,2}=O_{p}\left(\kappa_{n}\left(\frac{\ln{p}}{n}\right)^{\frac{\xi}{1+2\xi}}\right).
\]
\begin{proof}
	It directly follows from Lemma 7.2 in Singh and Sun (2021).
\end{proof}

Lemma 4.4 shows that the convergence rate of automatic estimator is faster than $n^{-1/4}$ if $\xi>1/2$, which requires the sparsity and the approximation error to grow with some restricted rate. Based on the sparsity and approximation error conditions in Assumption 4.3(i), we show that the bounds for the automatic estimator required in Assumption 4.3(ii) are satisfied.

\noindent\textbf{Lemma 4.5.} \textit{Let $\phi_{\min}(k)$ be the minimum sparse eigenvalue. If Assumptions 4.5-4.6 together with 4.3(i) and (iii) hold, $\phi_{\min}(k)$ is bounded from zero and $K_{n}\sqrt{s_{\gamma}}=o(n^{1/4})$, then $\widehat{\gamma}$ is sparse, $\left\Vert\widehat{\gamma}\right\Vert_{0}\leq{s_{\gamma}}$, and the following performance bounds hold:}
\[
\left\Vert b(D,X)^{\prime}\left(\widehat{\gamma}-\gamma\right)\right\Vert_{\mathbb{P}_{n},2}=o_{p}\left(n^{-1/4}\right)
\]
\textit{and}
\[
K_{n}\left\Vert \widehat{\gamma}-\gamma\right\Vert_{1}=o_{p}(1).
\]
\begin{proof}
	The sparsity of $\widehat{\gamma}$ can be derived in a similar way with Theorem 3 of Belloni and Chernozhukov (2013). The detailed proof is omitted here. Based on Assumption 4.3(i), it is easy to verify that the sparsity $s_{\gamma}$ and approximation error $r_{L}$ satisfy Assumption 4.6(i) with $\widetilde{\gamma}=\gamma$ and $\xi>1/2$. By Lemma 4.4, we have
	\[
	\left\Vert{b}(D,X)^{\prime}\widehat{\gamma}-L(D,X)\right\Vert_{P,2}=o_{p}\left(n^{-1/4}\right).
	\]
	By triangle inequality and Assumption 4.3(i),
	\[
	\left\Vert{b}(D,X)^{\prime}\left(\widehat{\gamma}-\gamma\right)\right\Vert_{P,2}\leq\left\Vert{b}(D,X)^{\prime}\widehat{\gamma}-L(D,X)\right\Vert_{P,2}+\left\Vert{L}(D,X)-{b}(D,X)^{\prime}{\gamma}\right\Vert_{P,2}=o_{p}\left(n^{-1/4}\right).
	\]
	By Assumption 4.3(iii), there exists a positive constant $C$,
	\[
	\left\Vert{b}(D,X)^{\prime}\left(\widehat{\gamma}-\gamma\right)\right\Vert_{\mathbb{P}_{n},2}\leq{C}\left\Vert{b}(D,X)^{\prime}\left(\widehat{\gamma}-\gamma\right)\right\Vert_{P,2}=o_{p}\left(n^{-1/4}\right).
	\]
	
	For a sparse vector, we have
	\[
	\left\Vert \widehat{\gamma}-\gamma\right\Vert_{1}\leq\sqrt{s_{\gamma}}\left\Vert b(D,X)^{\prime}\left(\widehat{\gamma}-\gamma\right)\right\Vert_{\mathbb{P}_{n},2}/\phi_{\min}\left(s_{\gamma}\right)=o_{p}\left(\sqrt{s_{\gamma}}n^{-1/4}\right).
	\]
	Consequently, we can conclude that
	\[
	K_{n}\left\Vert \widehat{\gamma}-\gamma\right\Vert_{1}=o_{p}\left(K_{n}\sqrt{s_{\gamma}}n^{-1/4}\right)=o_{p}(1).
	\]
\end{proof}

Condition $K_{n}\sqrt{s_{\gamma}}=o(n^{1/4})$ is also required in a similar way with Belloni et al. (2017) to develop uniform Gaussianity, for example, Assumption 6.1(iv) in Belloni et al. (2017).
	
	\subsubsection{Semiparametric Efficiency Bound of OASD }
	
This next theorem shows that $\psi\left(W,\theta,\eta;y_1,y_2\right)$ in Eq.(2.1.8) is also an efficient score, and thus our ADML estimator achieves the semiparametric efficiency bound.
	
	\noindent\textbf{Theorem 4.2.} Under the assumptions of Proposition 4.2, Assumptions 4.1 and 4.2, for any $(y_1,y_2)\in\mathcal{U}$, the semiparametric efficiency bound of $\theta(y_1,y_2)$ is $E\left[\psi\big(W,\theta_0,\eta;y_1,y_2\big)\right]^2$.

	\subsection{Multiplier Bootstrap}
	
	In practice, inference based on directly estimating the asymptotic variance of the limit process can be overly complicated. In such cases, bootstrap methods can be effectively applied to construct the confidence bands. Let $\{\xi\}_{i=1}^{n}$ be a random sample drawn from the distribution with zero-mean and unit-variance. We then define the estimated multiplier process for $Z(u)$ by
	\[
	\widehat{Z}_{n}^{*}(u)=\sqrt{n}\left(\widehat{\theta}^{*}(u)-\widehat{\theta}(u)\right)=\frac{1}{\sqrt{n}}\sum_{i=1}^{n}\xi_{i}\psi\bigg(W_{i},\widehat{\theta},\widehat{\eta};u\bigg).
	\]
	The main result of this section shows that the bootstrap law of the process $\widehat{Z}_{n}^{*}(u)$ provides a valid approximation to the large sample law of $\widehat{Z}_{n}(u)$. We develop such validity by imposing the following regular assumption.
	
	\noindent\textbf{Assumption 4.7.} A random element $\xi$ with values in a measure space $(\Omega,\mathcal{A}_{\Omega})$ that is independent of $(\mathcal{W},\mathcal{A}_{\mathcal{W}})$, and law determined by a probability measure $P_{\xi}$, with zero-mean and unit-variance. The observed data $\{\xi_{i}\}_{i=1}^{n}$ consist of $n$ i.i.d. copies of a random element $\xi$.
	
	We introduce some useful notations to describe the following results. We define the conditional weak convergence of the bootstrap law in probability, denoted by $\widetilde{Z}_{n}(u)\leadsto_{B}\widetilde{Z}(u)$ in $\ell^{\infty}\left(\mathcal{U}\right)$, by
	\[
	\sup_{T\in{BL_{1}\left(\ell^{\infty}\left(\mathcal{U}\right)\right)}}\left|E_{\xi|P}T\left(\widetilde{Z}_{n}(u)\right)-ET\left(\widetilde{Z}(u)\right)\right|=o_{p}(1),
	\]
	where $BL_{1}\left(\mathbb{D}\right)$ denotes the space of functions mapping $\mathbb{D}$ to [0,1] with Lipschitz norm at most 1, and $E_{\xi|P}$ denote the expectation over the multiplier weights $\{\xi_{i}\}_{i=1}^{n}$ holding the data $\{W_{i}\}_{i=1}^{n}$ fixed.
	
	\noindent\textbf{Theorem 4.3.} \textit{If Assumptions 2.1-2.2, 4.1-4.3 and 4.7 hold, the bootstrap law consistently approximates the large sample law $Z(u)$ of $Z_{n}(u)$, namely}
	\[
	\widehat{Z}_{n}^{*}(u)\leadsto_{B} Z(u) \quad in \quad \mathbb{D}=\ell^{\infty}\left(\mathcal{U}\right)
	\]
	
	One of the most relevant practical applications of this result is to test the null hypothesis of treatment homogeneity across $u$:
	\[
	H_{0}: \theta(u)=\theta(u^{\prime}) \ \ \text{for all}\ \ u, u^{\prime}\in\mathcal{U}.
	\]
	To test this hypothesis, we can construct the test statistic $\sup_{u\in\mathcal{U}}\sqrt{n}\left|\widehat{\theta}(u)-\left(\#\mathcal{U}\right)^{-1}\displaystyle\int_{\mathcal{U}}\widehat{\theta}\left(\widetilde{u}\right)d\widetilde{u}\right|$, and use
	\[
	\sup_{u\in\mathcal{U}}\left|\widehat{Z}_{n}^{*}(u)-\left(\#\mathcal{U}\right)^{-1}\displaystyle\int_{\mathcal{U}}\widehat{Z}_{n}^{*}\left(\widetilde{u}\right)d\widetilde{u}\right|
	\]
	to simulate its asymptotic distribution, in which $\#\mathcal{U}$ denotes the area of space $\mathcal{U}$. For example, suppose one is interested in whether the treatment is homogeneous across any fixed length of intervals, such that $c_{1}\leq{y_1}+c_{0}=y_{2}\leq{c_{2}}$ for some  constants $c_{0}$, $c_{1}$ and $c_{2}$ satisfying $c_{0}>0$ and $c_{2}-c_{1}>0$. In such case, we can define $\mathcal{U}=\left\{(y_{1},y_{2}):c_{1}\leq{y_1}+c_{0}=y_{2}\leq{c_{2}}\right\}$ and then $\#\mathcal{U}=c_{2}-c_{1}$.
	
	\section{Monte Carlo Simulation}
		In this section, we study the finite sample performance  of the naive estimator based on moment condition in Proposition 2.1 and the ADML estimator (based on orthogonal score).
Let $y_\tau$ be the $\tau$-th empirical quantile of $Y$. We consider the estimation of $(y_{0.05},y_{0.15})$, $(y_{0.15},y_{0.25})$, $(y_{0.25},y_{0.35})$, $(y_{0.35},y_{0.45})$,$(y_{0.45},y_{0.55})$, $(y_{0.55},y_{0.65})$, $(y_{0.65},y_{0.75})$, $(y_{0.75},y_{0.85})$, $(y_{0.85},y_{0.95})$.

	The data generating process is
	\begin{eqnarray*}
		&&Y=D+X'\left(c_y\delta_0\right)+DX_1+ U\\
		&&D=X'(c_d\delta_0)+V_1\\
		&&U=U_D=\begin{cases}
			V_2&D\leq F_{D}^{-1}(0.3)\\
			V_3&F_{D}^{-1}(0.3)< D\leq F_{D}^{-1}(0.7)\\
			V_4&D> F_{D}^{-1}(0.7)\\
		\end{cases}
	\end{eqnarray*}
	where $V_1,V_2,V_3,V_4\sim N(0,1)$, $X=(X_1,\cdots,X_{p_x})^\intercal\sim N(0,\Sigma)$ with $\Sigma=(0.5)^{|j-k|}$.
	Note that our DGP allows for the error term $U$ relying on realizations of $D$ and a dependence between error term $U$ and the treatment variable $D$.
	$\delta_0$ is a $p_x\times 1$ vector with elements  $\delta_{0,j}=(1/j)^2$ for $j\in\{1,2,...,p_x\}$.
	$c_d$ and $c_y$ are scalars to control the strength of the relationship between the covarites, the outcome, and the treatment variable.
	We use 16 combinations of $c_d$ and $c_y$, respectively, with $c_d=\sqrt{\frac{(\pi^2/3)R_d^2}{(1-R_d^2)\delta_0'\Sigma\delta_0}}$ and $c_y=\sqrt{\frac{R_y^2}{(1-R_y^2)\delta_0'\Sigma\delta_0}}$.
 $R_d^2\in\{0.1,0.2,0.3,0.4\}$ and $R_y^2\in\{0.1,0.2,0.3,0.4\}$, in which $R_d^2$ reflects the sparsity level of the effect of $X$ on $D$ while $R_y^2$ reflects the sparsity level of the effect of  $X$ on  $Y$.
	
	We consider $N=500$, $p_x=30$  throughout the simulation.
	To form the basis function, we include all first order, second order and interaction terms among $(D,X)$. The basis function can be written as $b(D,X)=\left(D,X_1,\cdots,X_{30},D^2,X_1^2,\cdots,X_{30}^2,DX_1,\cdots,DX_{30},X_1X_2,\cdots,X_{29}X_{30}\right)$ with $p=dim\left(b(D,X)\right)=527>N$.
	
	For each design, we calculate the naive estimator and ADML estimator. We do 500 iterations to compute bias ratio, std, mean square error (MSE) and the probability of 500 estimators lies in nominal 95\% confidence interval (Cvg). These results are reported in Tables 5.1-5.4. We find that ADML estimators outperform naive estimators in terms of MSE and Cvg in all cases.
	\begin{table}[H]\small
		\centering
		\begin{threeparttable}
			\begin{tabular}{llccccccccccc}	\multicolumn{13}{l}{\small{\textbf{Table 5.1 }}\small{Sparsity Design for $R_d^2=0.1$}}\\
				\toprule		
				&     & \multicolumn{2}{c}{Bias Ratio} &     & \multicolumn{2}{c}{Std} &     & \multicolumn{2}{c}{MSE} &     & \multicolumn{2}{c}{Cvg} \\
				\cmidrule{3-4}\cmidrule{6-7}\cmidrule{9-10}\cmidrule{12-13}    $R_x^2$ & Quantile & Naive & ADML &     & Naive & ADML &     & Naive & ADML &     & Naive & ADML \\
				\midrule
				\multirow{9}[2]{*}{0.1} & 5\%-15\% & -.058 & -.001 &     & .197 & .159 &     & .041 & .025 &     & .944 & .952 \\
				& 15\%-25\% & .141 & .049 &     & .123 & .121 &     & .024 & .016 &     & .882 & .938 \\
				& 25\%-35\% & .136 & .047 &     & .124 & .115 &     & .023 & .014 &     & .906 & .946 \\
				& 35\%-45\% & .087 & .029 &     & .139 & .129 &     & .023 & .017 &     & .924 & .952 \\
				& 45\%-55\% & .051 & .026 &     & .133 & .126 &     & .019 & .016 &     & .952 & .960 \\
				& 55\%-65\% & .028 & .019 &     & .138 & .136 &     & .020 & .019 &     & .940 & .942 \\
				& 65\%-75\% & .017 & .013 &     & .166 & .166 &     & .028 & .028 &     & .944 & .944 \\
				& 75\%-85\% & .014 & .010 &     & .211 & .211 &     & .045 & .045 &     & .952 & .952 \\
				& 85\%-95\% & .019 & .012 &     & .243 & .243 &     & .060 & .059 &     & .946 & .956 \\
				\midrule
				&     & \multicolumn{2}{c}{Bias Ratio} &     & \multicolumn{2}{c}{Std} &     & \multicolumn{2}{c}{MSE} &     & \multicolumn{2}{c}{Cvg} \\
				\cmidrule{3-4}\cmidrule{6-7}\cmidrule{9-10}\cmidrule{12-13}    $R_x^2$ & Quantile & Naive & ADML &     & Naive & ADML &     & Naive & ADML &     & Naive & ADML \\
				\midrule
				\multirow{9}[2]{*}{0.2} & 5\%-15\% & -.011 & .026 &     & .166 & .137 &     & .028 & .019 &     & .948 & .940 \\
				& 15\%-25\% & .216 & .088 &     & .133 & .125 &     & .035 & .018 &     & .850 & .926 \\
				& 25\%-35\% & .156 & .070 &     & .138 & .122 &     & .028 & .017 &     & .890 & .932 \\
				& 35\%-45\% & .059 & .031 &     & .137 & .127 &     & .020 & .017 &     & .938 & .958 \\
				& 45\%-55\% & .043 & .037 &     & .141 & .138 &     & .021 & .020 &     & .932 & .934 \\
				& 55\%-65\% & .016 & .012 &     & .138 & .138 &     & .019 & .019 &     & .958 & .958 \\
				& 65\%-75\% & .005 & .001 &     & .158 & .158 &     & .025 & .025 &     & .952 & .958 \\
				& 75\%-85\% & .032 & .027 &     & .203 & .203 &     & .043 & .042 &     & .942 & .942 \\
				& 85\%-95\% & .010 & .004 &     & .255 & .253 &     & .065 & .064 &     & .958 & .954 \\
				\midrule
				&     & \multicolumn{2}{c}{Bias Ratio} &     & \multicolumn{2}{c}{Std} &     & \multicolumn{2}{c}{MSE} &     & \multicolumn{2}{c}{Cvg} \\
				\cmidrule{3-4}\cmidrule{6-7}\cmidrule{9-10}\cmidrule{12-13}    $R_x^2$ & Quantile & Naive & ADML &     & Naive & ADML &     & Naive & ADML &     & Naive & ADML \\
				\midrule
				\multirow{9}[2]{*}{0.3} & 5\%-15\% & -.014 & .025 &     & .177 & .146 &     & .031 & .022 &     & .948 & .940 \\
				& 15\%-25\% & .200 & .069 &     & .140 & .122 &     & .032 & .016 &     & .886 & .936 \\
				& 25\%-35\% & .077 & .031 &     & .130 & .115 &     & .019 & .014 &     & .928 & .942 \\
				& 35\%-45\% & .008 & .001 &     & .112 & .109 &     & .012 & .012 &     & .948 & .948 \\
				& 45\%-55\% & .025 & .023 &     & .125 & .125 &     & .016 & .016 &     & .946 & .948 \\
				& 55\%-65\% & .008 & .005 &     & .143 & .142 &     & .020 & .020 &     & .948 & .952 \\
				& 65\%-75\% & .020 & .015 &     & .167 & .167 &     & .028 & .028 &     & .950 & .948 \\
				& 75\%-85\% & .014 & .009 &     & .213 & .212 &     & .046 & .045 &     & .950 & .956 \\
				& 85\%-95\% & .018 & .012 &     & .246 & .245 &     & .062 & .060 &     & .940 & .938 \\
				\midrule
				&     & \multicolumn{2}{c}{Bias Ratio} &     & \multicolumn{2}{c}{Std} &     & \multicolumn{2}{c}{MSE} &     & \multicolumn{2}{c}{Cvg} \\
				\cmidrule{3-4}\cmidrule{6-7}\cmidrule{9-10}\cmidrule{12-13}    $R_x^2$ & Quantile & Naive & ADML &     & Naive & ADML &     & Naive & ADML &     & Naive & ADML \\
				\midrule
				\multirow{9}[2]{*}{0.4} & 5\%-15\% & .025 & .056 &     & .163 & .133 &     & .027 & .018 &     & .958 & .944 \\
				& 15\%-25\% & .180 & .079 &     & .150 & .124 &     & .031 & .017 &     & .898 & .926 \\
				& 25\%-35\% & .059 & .041 &     & .120 & .112 &     & .015 & .013 &     & .930 & .942 \\
				& 35\%-45\% & .023 & .020 &     & .120 & .119 &     & .015 & .014 &     & .940 & .940 \\
				& 45\%-55\% & .024 & .021 &     & .129 & .129 &     & .017 & .017 &     & .960 & .954 \\
				& 55\%-65\% & .020 & .017 &     & .161 & .161 &     & .026 & .026 &     & .950 & .952 \\
				& 65\%-75\% & .019 & .015 &     & .170 & .170 &     & .029 & .029 &     & .946 & .948 \\
				& 75\%-85\% & .023 & .017 &     & .211 & .211 &     & .046 & .045 &     & .934 & .936 \\
				& 85\%-95\% & .015 & .009 &     & .245 & .242 &     & .061 & .059 &     & .938 & .954 \\
				\bottomrule
			\end{tabular}%
		\end{threeparttable}
		\label{tab:addlabel}%
	\end{table}%
	\begin{table}[H]\small
		\centering
		\begin{threeparttable}
			\begin{tabular}{llccccccccccc}	\multicolumn{13}{l}{\small{\textbf{Table 5.2 }}\small{Sparsity Design for $R_d^2=0.2$}}\\
				\toprule
				
				&     & \multicolumn{2}{c}{Bias Ratio} &     & \multicolumn{2}{c}{Std} &     & \multicolumn{2}{c}{MSE} &     & \multicolumn{2}{c}{Cvg} \\
				\cmidrule{3-4}\cmidrule{6-7}\cmidrule{9-10}\cmidrule{12-13}    $R_x^2$ & Quantile & Naive & ADML &     & Naive & ADML &     & Naive & ADML &     & Naive & ADML \\
				\midrule
				\multirow{9}[2]{*}{0.1} & 5\%-15\% & -.178 & -.043 &     & .170 & .143 &     & .045 & .021 &     & .886 & .952 \\
				& 15\%-25\% & .170 & .056 &     & .121 & .118 &     & .026 & .015 &     & .866 & .932 \\
				& 25\%-35\% & .176 & .059 &     & .126 & .119 &     & .028 & .015 &     & .856 & .936 \\
				& 35\%-45\% & .130 & .038 &     & .127 & .117 &     & .023 & .014 &     & .886 & .944 \\
				& 45\%-55\% & .092 & .041 &     & .134 & .124 &     & .022 & .016 &     & .924 & .948 \\
				& 55\%-65\% & .039 & .018 &     & .140 & .135 &     & .021 & .018 &     & .948 & .954 \\
				& 65\%-75\% & .038 & .027 &     & .164 & .161 &     & .028 & .027 &     & .928 & .932 \\
				& 75\%-85\% & .010 & .003 &     & .203 & .203 &     & .041 & .041 &     & .958 & .960 \\
				& 85\%-95\% & .027 & .019 &     & .241 & .241 &     & .061 & .059 &     & .944 & .952 \\
				\midrule
				&     & \multicolumn{2}{c}{Bias Ratio} &     & \multicolumn{2}{c}{Std} &     & \multicolumn{2}{c}{MSE} &     & \multicolumn{2}{c}{Cvg} \\
				\cmidrule{3-4}\cmidrule{6-7}\cmidrule{9-10}\cmidrule{12-13}    $R_x^2$ & Quantile & Naive & ADML &     & Naive & ADML &     & Naive & ADML &     & Naive & ADML \\
				\midrule
				\multirow{9}[2]{*}{0.2} & 5\%-15\% & -.148 & -.031 &     & .158 & .136 &     & .034 & .019 &     & .904 & .950 \\
				& 15\%-25\% & .245 & .081 &     & .125 & .115 &     & .036 & .015 &     & .808 & .924 \\
				& 25\%-35\% & .210 & .066 &     & .130 & .114 &     & .032 & .014 &     & .856 & .940 \\
				& 35\%-45\% & .104 & .031 &     & .134 & .114 &     & .022 & .013 &     & .920 & .946 \\
				& 45\%-55\% & .059 & .034 &     & .136 & .126 &     & .020 & .016 &     & .930 & .940 \\
				& 55\%-65\% & .034 & .023 &     & .143 & .140 &     & .021 & .020 &     & .948 & .952 \\
				& 65\%-75\% & .033 & .023 &     & .178 & .175 &     & .033 & .031 &     & .938 & .948 \\
				& 75\%-85\% & .023 & .014 &     & .215 & .215 &     & .047 & .046 &     & .954 & .956 \\
				& 85\%-95\% & .031 & .021 &     & .245 & .243 &     & .063 & .060 &     & .946 & .950 \\
				\midrule
				&     & \multicolumn{2}{c}{Bias Ratio} &     & \multicolumn{2}{c}{Std} &     & \multicolumn{2}{c}{MSE} &     & \multicolumn{2}{c}{Cvg} \\
				\cmidrule{3-4}\cmidrule{6-7}\cmidrule{9-10}\cmidrule{12-13}    $R_x^2$ & Quantile & Naive & ADML &     & Naive & ADML &     & Naive & ADML &     & Naive & ADML \\
				\midrule
				\multirow{9}[2]{*}{0.3} & 5\%-15\% & -.115 & -.014 &     & .152 & .135 &     & .027 & .018 &     & .920 & .950 \\
				& 15\%-25\% & .277 & .084 &     & .125 & .110 &     & .038 & .014 &     & .786 & .930 \\
				& 25\%-35\% & .193 & .067 &     & .145 & .118 &     & .033 & .015 &     & .882 & .930 \\
				& 35\%-45\% & .057 & .020 &     & .125 & .112 &     & .017 & .013 &     & .930 & .958 \\
				& 45\%-55\% & .037 & .027 &     & .127 & .124 &     & .017 & .016 &     & .950 & .944 \\
				& 55\%-65\% & .032 & .024 &     & .145 & .144 &     & .022 & .021 &     & .944 & .950 \\
				& 65\%-75\% & .036 & .026 &     & .185 & .182 &     & .035 & .034 &     & .942 & .956 \\
				& 75\%-85\% & .033 & .022 &     & .219 & .218 &     & .050 & .048 &     & .940 & .942 \\
				& 85\%-95\% & .027 & .016 &     & .234 & .232 &     & .057 & .055 &     & .938 & .948 \\
				\midrule
				&     & \multicolumn{2}{c}{Bias Ratio} &     & \multicolumn{2}{c}{Std} &     & \multicolumn{2}{c}{MSE} &     & \multicolumn{2}{c}{Cvg} \\
				\cmidrule{3-4}\cmidrule{6-7}\cmidrule{9-10}\cmidrule{12-13}    $R_x^2$ & Quantile & Naive & ADML &     & Naive & ADML &     & Naive & ADML &     & Naive & ADML \\
				\midrule
				\multirow{9}[2]{*}{0.4} & 5\%-15\% & -.084 & .004 &     & .143 & .133 &     & .022 & .018 &     & .938 & .956 \\
				& 15\%-25\% & .294 & .094 &     & .139 & .115 &     & .041 & .015 &     & .816 & .940 \\
				& 25\%-35\% & .126 & .045 &     & .141 & .114 &     & .024 & .014 &     & .930 & .950 \\
				& 35\%-45\% & .030 & .018 &     & .116 & .112 &     & .014 & .013 &     & .948 & .950 \\
				& 45\%-55\% & .022 & .017 &     & .118 & .117 &     & .014 & .014 &     & .940 & .946 \\
				& 55\%-65\% & .032 & .024 &     & .148 & .147 &     & .022 & .022 &     & .954 & .956 \\
				& 65\%-75\% & .038 & .027 &     & .187 & .184 &     & .037 & .035 &     & .944 & .948 \\
				& 75\%-85\% & .037 & .025 &     & .218 & .217 &     & .050 & .048 &     & .942 & .942 \\
				& 85\%-95\% & .029 & .017 &     & .233 & .230 &     & .057 & .054 &     & .940 & .950 \\
				\bottomrule
			\end{tabular}%
		\end{threeparttable}
		\label{tab:addlabel}%
	\end{table}%
	
	\begin{table}[H]\small
		\centering
		\begin{threeparttable}
			\begin{tabular}{llccccccccccc}	\multicolumn{13}{l}{\small{\textbf{Table 5.3 }}\small{Sparsity Design for $R_d^2=0.3$}}\\
				\toprule
				&     & \multicolumn{2}{c}{Bias Ratio} &     & \multicolumn{2}{c}{Std} &     & \multicolumn{2}{c}{MSE} &     & \multicolumn{2}{c}{Cvg} \\
				\cmidrule{3-4}\cmidrule{6-7}\cmidrule{9-10}\cmidrule{12-13}    $R_x^2$ & Quantile & Naive & ADML &     & Naive & ADML &     & Naive & ADML &     & Naive & ADML \\
				\midrule
				\multirow{9}[2]{*}{0.1} & 5\%-15\% & -.239 & -.047 &     & .144 & .129 &     & .047 & .018 &     & .780 & .956 \\
				& 15\%-25\% & .190 & .078 &     & .123 & .121 &     & .029 & .017 &     & .834 & .924 \\
				& 25\%-35\% & .194 & .074 &     & .125 & .119 &     & .030 & .016 &     & .850 & .918 \\
				& 35\%-45\% & .176 & .065 &     & .126 & .118 &     & .028 & .016 &     & .870 & .932 \\
				& 45\%-55\% & .114 & .042 &     & .143 & .131 &     & .026 & .018 &     & .920 & .950 \\
				& 55\%-65\% & .045 & .014 &     & .133 & .127 &     & .019 & .016 &     & .932 & .950 \\
				& 65\%-75\% & .053 & .038 &     & .165 & .163 &     & .030 & .028 &     & .938 & .940 \\
				& 75\%-85\% & .021 & .011 &     & .201 & .202 &     & .041 & .041 &     & .946 & .946 \\
				& 85\%-95\% & .023 & .014 &     & .251 & .250 &     & .065 & .063 &     & .948 & .954 \\
				\midrule
				&     & \multicolumn{2}{c}{Bias Ratio} &     & \multicolumn{2}{c}{Std} &     & \multicolumn{2}{c}{MSE} &     & \multicolumn{2}{c}{Cvg} \\
				\cmidrule{3-4}\cmidrule{6-7}\cmidrule{9-10}\cmidrule{12-13}    $R_x^2$ & Quantile & Naive & ADML &     & Naive & ADML &     & Naive & ADML &     & Naive & ADML \\
				\midrule
				\multirow{9}[2]{*}{0.2} & 5\%-15\% & -.222 & -.045 &     & .148 & .134 &     & .041 & .019 &     & .848 & .954 \\
				& 15\%-25\% & .245 & .083 &     & .120 & .113 &     & .035 & .015 &     & .780 & .936 \\
				& 25\%-35\% & .247 & .079 &     & .119 & .108 &     & .035 & .014 &     & .774 & .928 \\
				& 35\%-45\% & .190 & .068 &     & .138 & .118 &     & .033 & .016 &     & .870 & .922 \\
				& 45\%-55\% & .089 & .037 &     & .148 & .133 &     & .026 & .018 &     & .922 & .948 \\
				& 55\%-65\% & .062 & .040 &     & .143 & .137 &     & .023 & .020 &     & .938 & .944 \\
				& 65\%-75\% & .038 & .022 &     & .180 & .176 &     & .034 & .031 &     & .950 & .956 \\
				& 75\%-85\% & .024 & .011 &     & .216 & .214 &     & .047 & .046 &     & .940 & .946 \\
				& 85\%-95\% & .040 & .028 &     & .248 & .248 &     & .067 & .064 &     & .940 & .944 \\
				\midrule
				&     & \multicolumn{2}{c}{Bias Ratio} &     & \multicolumn{2}{c}{Std} &     & \multicolumn{2}{c}{MSE} &     & \multicolumn{2}{c}{Cvg} \\
				\cmidrule{3-4}\cmidrule{6-7}\cmidrule{9-10}\cmidrule{12-13}    $R_x^2$ & Quantile & Naive & ADML &     & Naive & ADML &     & Naive & ADML &     & Naive & ADML \\
				\midrule
				\multirow{9}[2]{*}{0.3} & 5\%-15\% & -.194 & -.032 &     & .139 & .134 &     & .032 & .018 &     & .852 & .960 \\
				& 15\%-25\% & .316 & .111 &     & .118 & .109 &     & .043 & .015 &     & .712 & .916 \\
				& 25\%-35\% & .285 & .096 &     & .133 & .112 &     & .042 & .015 &     & .802 & .898 \\
				& 35\%-45\% & .146 & .053 &     & .148 & .119 &     & .029 & .015 &     & .916 & .940 \\
				& 45\%-55\% & .050 & .023 &     & .139 & .130 &     & .020 & .017 &     & .946 & .958 \\
				& 55\%-65\% & .059 & .042 &     & .151 & .147 &     & .025 & .023 &     & .936 & .944 \\
				& 65\%-75\% & .043 & .026 &     & .181 & .177 &     & .035 & .032 &     & .944 & .946 \\
				& 75\%-85\% & .029 & .013 &     & .214 & .212 &     & .047 & .045 &     & .944 & .952 \\
				& 85\%-95\% & .044 & .030 &     & .243 & .242 &     & .067 & .062 &     & .930 & .934 \\
				\midrule
				&     & \multicolumn{2}{c}{Bias Ratio} &     & \multicolumn{2}{c}{Std} &     & \multicolumn{2}{c}{MSE} &     & \multicolumn{2}{c}{Cvg} \\
				\cmidrule{3-4}\cmidrule{6-7}\cmidrule{9-10}\cmidrule{12-13}    $R_x^2$ & Quantile & Naive & ADML &     & Naive & ADML &     & Naive & ADML &     & Naive & ADML \\
				\midrule
				\multirow{9}[2]{*}{0.4} & 5\%-15\% & -.168 & -.016 &     & .134 & .130 &     & .025 & .017 &     & .894 & .954 \\
				& 15\%-25\% & .368 & .132 &     & .125 & .111 &     & .050 & .017 &     & .714 & .908 \\
				& 25\%-35\% & .266 & .091 &     & .146 & .113 &     & .041 & .015 &     & .840 & .920 \\
				& 35\%-45\% & .082 & .032 &     & .135 & .114 &     & .020 & .013 &     & .928 & .948 \\
				& 45\%-55\% & .033 & .019 &     & .127 & .124 &     & .017 & .015 &     & .946 & .952 \\
				& 55\%-65\% & .051 & .036 &     & .152 & .149 &     & .025 & .023 &     & .934 & .938 \\
				& 65\%-75\% & .051 & .034 &     & .193 & .189 &     & .040 & .037 &     & .944 & .944 \\
				& 75\%-85\% & .040 & .023 &     & .220 & .218 &     & .052 & .048 &     & .948 & .950 \\
				& 85\%-95\% & .043 & .028 &     & .239 & .238 &     & .065 & .060 &     & .928 & .934 \\
				\bottomrule
			\end{tabular}%
		\end{threeparttable}
		\label{tab:addlabel}%
	\end{table}%
	
	\begin{table}[H]\small
		\centering
		\begin{threeparttable}
			\begin{tabular}{llccccccccccc}	\multicolumn{13}{l}{\small{\textbf{Table 5.4 }}\small{Sparsity Design for $R_d^2=0.4$}}\\
				\toprule
				&     & \multicolumn{2}{c}{Bias Ratio} &     & \multicolumn{2}{c}{Std} &     & \multicolumn{2}{c}{MSE} &     & \multicolumn{2}{c}{Cvg} \\
				\cmidrule{3-4}\cmidrule{6-7}\cmidrule{9-10}\cmidrule{12-13}    $R_x^2$ & Quantile & Naive & ADML &     & Naive & ADML &     & Naive & ADML &     & Naive & ADML \\
				\midrule
				\multirow{9}[2]{*}{0.1} & 5\%-15\% & -.313 & -.073 &     & .140 & .133 &     & .062 & .020 &     & .672 & .940 \\
				& 15\%-25\% & .168 & .071 &     & .123 & .121 &     & .025 & .016 &     & .866 & .932 \\
				& 25\%-35\% & .176 & .066 &     & .109 & .107 &     & .023 & .013 &     & .844 & .918 \\
				& 35\%-45\% & .170 & .062 &     & .124 & .119 &     & .027 & .016 &     & .874 & .930 \\
				& 45\%-55\% & .125 & .046 &     & .130 & .122 &     & .024 & .016 &     & .912 & .950 \\
				& 55\%-65\% & .070 & .030 &     & .141 & .135 &     & .023 & .019 &     & .940 & .946 \\
				& 65\%-75\% & .062 & .043 &     & .175 & .174 &     & .035 & .032 &     & .922 & .926 \\
				& 75\%-85\% & .033 & .020 &     & .202 & .204 &     & .043 & .042 &     & .952 & .956 \\
				& 85\%-95\% & .017 & .009 &     & .249 & .246 &     & .063 & .060 &     & .942 & .948 \\
				\midrule
				&     & \multicolumn{2}{c}{Bias Ratio} &     & \multicolumn{2}{c}{Std} &     & \multicolumn{2}{c}{MSE} &     & \multicolumn{2}{c}{Cvg} \\
				\cmidrule{3-4}\cmidrule{6-7}\cmidrule{9-10}\cmidrule{12-13}    $R_x^2$ & Quantile & Naive & ADML &     & Naive & ADML &     & Naive & ADML &     & Naive & ADML \\
				\midrule
				\multirow{9}[2]{*}{0.2} & 5\%-15\% & -.294 & -.074 &     & .130 & .124 &     & .049 & .017 &     & .706 & .942 \\
				& 15\%-25\% & .235 & .088 &     & .114 & .109 &     & .031 & .014 &     & .796 & .918 \\
				& 25\%-35\% & .253 & .088 &     & .118 & .109 &     & .035 & .014 &     & .776 & .930 \\
				& 35\%-45\% & .231 & .084 &     & .131 & .116 &     & .037 & .016 &     & .830 & .934 \\
				& 45\%-55\% & .129 & .049 &     & .158 & .138 &     & .032 & .020 &     & .908 & .944 \\
				& 55\%-65\% & .070 & .036 &     & .154 & .145 &     & .027 & .022 &     & .926 & .940 \\
				& 65\%-75\% & .056 & .033 &     & .186 & .181 &     & .038 & .034 &     & .960 & .954 \\
				& 75\%-85\% & .037 & .019 &     & .222 & .221 &     & .052 & .050 &     & .952 & .956 \\
				& 85\%-95\% & .039 & .025 &     & .250 & .251 &     & .068 & .065 &     & .942 & .954 \\
				\midrule
				&     & \multicolumn{2}{c}{Bias Ratio} &     & \multicolumn{2}{c}{Std} &     & \multicolumn{2}{c}{MSE} &     & \multicolumn{2}{c}{Cvg} \\
				\cmidrule{3-4}\cmidrule{6-7}\cmidrule{9-10}\cmidrule{12-13}    $R_x^2$ & Quantile & Naive & ADML &     & Naive & ADML &     & Naive & ADML &     & Naive & ADML \\
				\midrule
				\multirow{9}[2]{*}{0.3} & 5\%-15\% & -.271 & -.063 &     & .129 & .125 &     & .040 & .017 &     & .772 & .944 \\
				& 15\%-25\% & .302 & .114 &     & .114 & .108 &     & .039 & .015 &     & .726 & .916 \\
				& 25\%-35\% & .312 & .106 &     & .122 & .108 &     & .044 & .015 &     & .738 & .912 \\
				& 35\%-45\% & .235 & .088 &     & .144 & .118 &     & .040 & .016 &     & .852 & .930 \\
				& 45\%-55\% & .101 & .043 &     & .156 & .136 &     & .029 & .019 &     & .926 & .946 \\
				& 55\%-65\% & .065 & .036 &     & .159 & .150 &     & .028 & .023 &     & .922 & .940 \\
				& 65\%-75\% & .056 & .032 &     & .193 & .187 &     & .041 & .036 &     & .940 & .950 \\
				& 75\%-85\% & .047 & .027 &     & .225 & .224 &     & .055 & .051 &     & .944 & .952 \\
				& 85\%-95\% & .044 & .028 &     & .242 & .242 &     & .066 & .062 &     & .932 & .944 \\
				\midrule
				&     & \multicolumn{2}{c}{Bias Ratio} &     & \multicolumn{2}{c}{Std} &     & \multicolumn{2}{c}{MSE} &     & \multicolumn{2}{c}{Cvg} \\
				\cmidrule{3-4}\cmidrule{6-7}\cmidrule{9-10}\cmidrule{12-13}    $R_x^2$ & Quantile & Naive & ADML &     & Naive & ADML &     & Naive & ADML &     & Naive & ADML \\
				\midrule
				\multirow{9}[2]{*}{0.4} & 5\%-15\% & -.250 & -.051 &     & .123 & .123 &     & .032 & .016 &     & .816 & .950 \\
				& 15\%-25\% & .371 & .141 &     & .116 & .107 &     & .048 & .016 &     & .674 & .898 \\
				& 25\%-35\% & .347 & .116 &     & .133 & .111 &     & .050 & .016 &     & .740 & .912 \\
				& 35\%-45\% & .195 & .077 &     & .152 & .118 &     & .035 & .016 &     & .888 & .944 \\
				& 45\%-55\% & .070 & .034 &     & .144 & .130 &     & .023 & .017 &     & .934 & .958 \\
				& 55\%-65\% & .071 & .045 &     & .168 & .161 &     & .032 & .027 &     & .936 & .946 \\
				& 65\%-75\% & .062 & .037 &     & .195 & .190 &     & .043 & .038 &     & .938 & .940 \\
				& 75\%-85\% & .053 & .030 &     & .227 & .225 &     & .058 & .053 &     & .932 & .946 \\
				& 85\%-95\% & .046 & .028 &     & .239 & .238 &     & .066 & .060 &     & .924 & .938 \\
				\bottomrule
			\end{tabular}%
		\end{threeparttable}
		\label{tab:addlabel}%
	\end{table}%

	\section{Conclusion}
	
This paper proposes
		a new class of heterogeneous causal quantities, named \textit{outcome conditioned} average structural derivatives (OASD) to measure the average partial effect of a marginal change in a continuous treatment on the individuals located at different parts of the outcome distribution,  irrespective of individuals' characteristics.
		OASD combines both features of ATE and QTE: it is interpreted as straightforwardly as ATE
		while at the same time more granular than ATE by breaking the entire population up according to
		the rank of the outcome distribution.
		
		In addition to providing identification results for OASD, we show there is a close relationship between the \textit{outcome conditioned average partial effects} and a class of parameters measuring the effect of counterfactually
		changing the distribution of a single covariate on the unconditional outcome quantiles. We illustrate this point by two examples: equivalence between OASD and the unconditional partial quantile effect (Firpo et al. (2009)), and equivalence between the marginal partial distribution policy effect (Rothe (2012)) and a corresponding outcome conditioned parameter.

		Because identification of OASD is attained
		under a conditional exogeneity assumption, by controlling for a rich information about covariates,  a researcher may ideally use high-dimensional controls in data.
		We propose for OASD a novel automatic debiased machine learning estimator, and present asymptotic statistical
		guarantees for it. We prove our estimator is root-$n$ consistent, asymptotically normal, and semiparametrically efficient.
		We also prove the validity of the bootstrap procedure for uniform inference on the OASD process.  Simulation studies support our theories.

	\newpage
	\noindent\textbf{Appendix A. Proofs}

	\noindent\textit{{A.1 Notations and Assumptions}}
	
	\noindent{Denote} three nuisance parameters in Proposition 2.2 by $\eta_1(\cdot)$, $\eta_2(\cdot)$, and $\eta_3(\cdot)$, respectively, i.e.,
	\begin{eqnarray*}
		\eta(W;y_1,y_2)&=&\left(P(y_1<Y<y_2),\int_{y_1}^{y_2}F_Y(y|D,X)dy,\frac{\partial_Df(D,X)}{f(D,X)}\right)\\
		&\equiv&\left(\eta_1(y_1,y_2),\eta_2(D,X;y_1,y_2),\eta_3(D,X)\right)
	\end{eqnarray*}
	\noindent{The} following regularity conditions are needed to show Proposition 2.1.
	
	\noindent{\textbf{Assumption A.1.}} Conditional CDF $F_Y(y|d,x)$ is absolutely continuous with respect to the Lebesgue measure for in a neighborhood of $d\in\mathcal{S}_D$ given $x$.  The density $f_Y(y|d,x)$ is continuous at $(y,d)=\left(Q_Y(\tau|d,x),d\right)$ and bounded in $y\in\mathbb{R}$.

\noindent{\textbf{Assumption A.2.}}		$Q_Y(\tau|d,x)$ is partially differentiable with respect to $d$. There exists a measurable function $\Delta$ that satisfies
\begin{eqnarray*}
	P\left(\left|m(d+\delta,x,U_{d})-m(d,x,U_{d})-\delta\Delta(U_{d})\geq \epsilon\delta\right||D=d,X=x\right)=o(\delta)
\end{eqnarray*}
for $\delta\rightarrow 0^+$ and any fixed $\epsilon>0$. We write $\partial_dm(d,x,u)$ for $\Delta(u)$ and $\partial_dm\left(d,x,U_{d}\right)$ for $\Delta(U_{d})$.

\noindent{\textbf{Assumption A.3.}}
The conditional distribution of $\left(Y,\partial_dm\left(d,x,U_{d}\right)\right)$ given $D=d$ and $X=x$ is absolutely continuous with respect to the Lebesgue measure. For the conditional density $f_{Y,\partial_dm\left(d,x,U_{d}\right)|D,X}$ of $(Y;\partial_dm\left(d,x,U_{d}\right))$ given $D$ and $X$, we require that $f_{Y,\partial_dm\left(d,x,U_{d}\right)|D,X}(y,y'|d,x)\leq Cg(y')$, where $C$ is a constant and $g$ is a positive density on $\mathbb{R}$ with finite mean (i.e., $\int |y'|g(y')dy'<\infty$).\\
\\
	\noindent\textit{{A.2 Proofs for Propositions 2.1--2.4}.}\qquad
	
	\noindent{\textbf{Proof of Proposition 2.1.}}
	Note that by definition of $Q_Y(\tau|d,x)$
	\begin{eqnarray*}
		P\left(Y\leq Q_Y(\tau|d,x)|D=d,X=x\right)
		&=&P\left(m(D,X,U_D)\leq Q_Y(\tau|d,x)|D=d,X=x\right)\\
		&=&P\left(m(d,x,U_d)\leq Q_Y(\tau|d,x)|D=d,X=x\right)\\
		&=&\tau
	\end{eqnarray*}
	Similarly, for $\delta>0$
	\begin{eqnarray*}
		P\left(Y\leq Q_Y(\tau|d+\delta,x)|D=d+\delta,X=x\right)&=&P\left(m(D,X,U_D)\leq Q_Y(\tau|d+\delta,x)|D=d+\delta,X=x\right)\\
		&=&P\left(m(d+\delta,x,U_{d+\delta})\leq Q_Y(\tau|d+\delta,x)|D=d+\delta,X=x\right)\\
		&=&\tau
	\end{eqnarray*}
	Thus
	\begin{eqnarray}
		0&=&\tau-\tau\notag\\
		&=&P\left(m(d+\delta,x,U_{d+\delta})\leq Q_Y(\tau|d+\delta,x)|D=d+\delta,X=x\right)\notag\\
		&&-P\left(m(d,x,U_d)\leq Q_Y(\tau|d,x)|D=d,X=x\right)\notag\\
		&=&A_1+A_2+A_3+A_4\label{eq:A.1}
	\end{eqnarray}
	where
	\begin{eqnarray*}
		A_1&=&P\left(m(d+\delta,x,U_{d+\delta})\leq Q_Y(\tau|d+\delta,x)|D=d+\delta,X=x\right)\\
		&&-P\left(m(d+\delta,x,U_{d+\delta})\leq Q_Y(\tau|d,x)|D=d+\delta,X=x\right)\\
		A_2&=&P\left(m(d+\delta,x,U_{d+\delta})\leq Q_Y(\tau|d,x)|D=d+\delta,X=x\right)\\
		&&-P\left(m(d+\delta,x,U_d)\leq Q_Y(\tau|d,x)|D=d+\delta,X=x\right)\\
		A_3&=&P\left(m(d+\delta,x,U_d)\leq Q_Y(\tau|d,x)|D=d+\delta,X=x\right)\\
		&&-P\left(m(d+\delta,x,U_d)\leq Q_Y(\tau|d,x)|D=d,X=x\right)\\
		A_4&=&P\left(m(d+\delta,x,U_d)\leq Q_Y(\tau|d,x)|D=d,X=x\right)\\
		&&-P\left(m(d,x,U_d)\leq Q_Y(\tau|d,x)|D=d,X=x\right)
	\end{eqnarray*}
	For $A_1$ we get, for $\delta\rightarrow0^+$
	\begin{eqnarray}
		A_1&=&P\left(m(d+\delta,x,U_{d+\delta})\leq Q_Y(\tau|d+\delta,x)|D=d+\delta,X=x\right)\notag\\
		&&-P\left(m(d+\delta,x,U_{d+\delta})\leq Q_Y(\tau|d,x)|D=d+\delta,X=x\right)\notag\\
		&=&P\left(Y\leq Q_Y(\tau|d+\delta,x)|D=d+\delta,X=x\right)-P\left(Y\leq Q_Y(\tau|d,x)|D=d+\delta,X=x\right)\notag\\
		&=&\int_{Q_Y(\tau|d,x)}^{Q_Y(\tau|d+\delta,x)}f_Y(y|d+\delta,x)dy\notag\\
		&=&\delta\partial_dQ_Y(\tau|d,x)f_Y\left(Q_Y(\tau|d,x)|d,x\right)+o(\delta)
	\end{eqnarray}
	where the first equality follows from definition of $A_1$, the second from data generating process $Y=m(D,X,U_D)$, the third from simple algebra and Assumptions A1--A2, and the last from Taylor expansion with Peano remainder and $\delta\rightarrow0^+$.
	For $A_2$ we get
	\begin{eqnarray}
		A_2&=&P\left(m(d+\delta,x,U_{d+\delta})\leq Q_Y(\tau|d,x)|D=d+\delta,X=x\right)\notag\\
		&&-P\left(m(d+\delta,x,U_d)\leq Q_Y(\tau|d,x)|D=d+\delta,X=x\right)\notag\\
		&=&P\left(m(d+\delta,x,U_{d+\delta})\leq Q_Y(\tau|d,x)|X=x\right)\notag\\
		&&-P\left(m(d+\delta,x,U_d)\leq Q_Y(\tau|d,x)|X=x\right)\notag\\
		&=&P\left(m(d+\delta,x,U_0)\leq Q_Y(\tau|d,x)|X=x\right)\notag\\
		&&-P\left(m(d+\delta,x,U_0)\leq Q_Y(\tau|d,x)|X=x\right)\notag\\
		&=&0
	\end{eqnarray}
	where the first equality follows from definition of $A_2$, the second from Assumption 2.2, the third from Assumption 2.1, and the last from simple algebra.
	For $A_3$ we get
	\begin{eqnarray}
		A_3&=&P\left(m(d+\delta,x,U_d)\leq Q_Y(\tau|d,x)|D=d+\delta,X=x\right)\notag\\
		&&-P\left(m(d+\delta,x,U_d)\leq Q_Y(\tau|d,x)|D=d,X=x\right)\notag\\
		&=&P\left(m(d+\delta,x,U_d)\leq Q_Y(\tau|d,x)|D=d,X=x\right)\notag\\
		&&-P\left(m(d+\delta,x,U_d)\leq Q_Y(\tau|d,x)|D=d,X=x\right)\notag\\
		&=&0
	\end{eqnarray}
	where the first equality follows from definition of $A_3$, the second from Assumption 2.2, and the last from simple algebra.
	For $A_4$ we get, for $\delta\rightarrow0^+$
	\begin{eqnarray*}
		A_4&=&P\left(m(d+\delta,x,U_d)\leq Q_Y(\tau|d,x)|D=d,X=x\right)\\
		&&-P\left(m(d,x,U_d)\leq Q_Y(\tau|d,x)|D=d,X=x\right)\\
		&=&P\left(m(d+\delta,x,U_d)\leq Q_Y(\tau|d,x)|D=d,X=x\right)\\
		&&-P\left(Y\leq Q_Y(\tau|d,x)|D=d,X=x\right)\\
		&=&P\left(Y\leq Q_Y(\tau|d,x)+Y-m(d+\delta,x,U_d)|D=d,X=x\right)\\
		&&-P\left(Y\leq Q_Y(\tau|d,x)|D=d,X=x\right)\\
		&=&P\left(Q_Y(\tau|d,x)\leq Y\leq Q_Y(\tau|d,x)+Y-m(d+\delta,x,U_d)|D=d,X=x\right)\\
		&&-P\left(Q_Y(\tau|d,x)+Y-m(d+\delta,x,U_d)\leq Y\leq Q_Y(\tau|d,x)|D=d,X=x\right)\\
		&=&P\left(Q_Y(\tau|d,x)\leq Y\leq Q_Y(\tau|d,x)+m(d,x,U_d)-m(d+\delta,x,U_d)|D=d,X=x\right)\\
		&&-P\left(Q_Y(\tau|d,x)+m(d,x,U_d)-m(d+\delta,x,U_d)\leq Y\leq Q_Y(\tau|d,x)|D=d,X=x\right)\\
		&=&P\left(Q_Y(\tau|d,x)\leq Y\leq Q_Y(\tau|d,x)-\delta\partial_dm\left(d,x,U_d\right)|D=d,X=x\right)\\
		&&-P\left(Q_Y(\tau|d,x)-\delta\partial_dm\left(d,x,U_d\right)\leq Y\leq Q_Y(\tau|d,x)|D=d,X=x\right)+o(\delta)\\
		&=&P\left(Y\geq Q_Y(\tau|d,x),\partial_dm\left(d,x,U_d\right)\leq -\frac{Y-Q_Y(\tau|d,x)}{\delta}\bigg|D=d,X=x\right)\\
		&&-P\left(Y\leq Q_Y(\tau|d,x),\partial_dm\left(d,x,U_d\right)\geq -\frac{Y-Q_Y(\tau|d,x)}{\delta}\bigg|D=d,X=x\right)+o(\delta)\\
		&=&\int_{Q_Y(\tau|d,x)}^{+\infty}\int_{-\infty }^{-\frac{y-Q_Y(\tau|d,x)}{\delta}}f_{Y,\partial_dm\left(d,x,U_d\right)|D,X}(y,y'|d,x)dy'dy\\
		&&-\int_{-\infty}^{Q_Y(\tau|d,x)}\int_{-\frac{y-Q_Y(\tau|d,x)}{\delta} }^{+\infty}f_{Y,\partial_dm\left(d,x,U_d\right)|D,X}(y,y'|d,x)dy'dy+o(\delta)\\
	\end{eqnarray*}
	where the first equality follows from definition of $A_4$, the second from data generating process, the third  and fourth from simple algebra, the fifth from data generating process, the sixth from Assumptions A2--A3 and $\delta\rightarrow0^+$, the seventh and last from simple algebra.
	Let $-\frac{y-Q_Y(\tau|d,x)}{\delta}=u$, then $A_4$ can be simplified to
	\begin{eqnarray}
		A_4&=&(-\delta)\int_{0}^{-\infty}\int_{-\infty }^{u}f_{Y,\partial_dm\left(d,x,U_d\right)|D,X}\left(Q_Y(\tau|d,x)-\delta u,y'|d,x\right)dy'du\notag\\
		&&-(-\delta)\int_{+\infty}^{0}\int_{u }^{+\infty}f_{Y,\partial_dm\left(d,x,U_d\right)|D,X}\left(Q_Y(\tau|d,x)-\delta u,y'|d,x\right)dy'du+o(\delta)\notag\\
		&=&\delta\int_{-\infty}^{0}\int_{-\infty }^{u}f_{Y,\partial_dm\left(d,x,U_d\right)|D,X}\left(Q_Y(\tau|d,x)-\delta u,y'|d,x\right)dy'du\notag\\
		&&-\delta\int_{0}^{+\infty}\int_{u }^{+\infty}f_{Y,\partial_dm\left(d,x,U_d\right)|D,X}\left(Q_Y(\tau|d,x)-\delta u,y'|d,x\right)dy'du+o(\delta)\notag\\
		&=&\delta\int_{-\infty}^{0}\int_{y'}^{0}f_{Y,\partial_dm\left(d,x,U_d\right)|D,X}\left(Q_Y(\tau|d,x)-\delta u,y'|d,x\right)dudy'\notag\\
		&&-\delta\int_{0}^{+\infty}\int_{0}^{y'}f_{Y,\partial_dm\left(d,x,U_d\right)|D,X}\left(Q_Y(\tau|d,x)-\delta u,y'|d,x\right)dudy'+o(\delta)\notag\\
		&=&\delta\int_{-\infty}^{0}\int_{y'}^{0}f_{Y,\partial_dm\left(d,x,U_d\right)|D,X}\left(Q_Y(\tau|d,x),y'|d,x\right)dudy'\notag\\
		&&-\delta\int_{0}^{+\infty}\int_{0}^{y'}f_{Y,\partial_dm\left(d,x,U_d\right)|D,X}\left(Q_Y(\tau|d,x),y'|d,x\right)dudy'+o(\delta)\notag\\
		&=&\delta\int_{-\infty}^{0}(-y')f_{Y,\partial_dm\left(d,x,U_d\right)|D,X}\left(Q_Y(\tau|d,x),y'|d,x\right)dy'\notag\\
		&&-\delta\int_{0}^{+\infty}y'f_{Y,\partial_dm\left(d,x,U_d\right)|D,X}\left(Q_Y(\tau|d,x),y'|d,x\right)dy'+o(\delta)\notag\\
		&=&-\delta\int_{-\infty}^{+\infty}y'f_{\partial_dm\left(d,x,U_d\right)|D,X,Y}\left(y'|d,x,Q_Y(\tau|d,x)\right)dy'f_Y\left(Q_Y(\tau|d,x)|d,x\right)+o(\delta)\notag\\
		&=&-\delta E\left(\partial_dm\left(d,x,U_d\right)|Y=Q_Y(\tau|d,x),D=d,X=x\right)f_Y\left(Q_Y(\tau|d,x)|d,x\right)+o(\delta)\notag\\
	\end{eqnarray}
	where the first equality follows from substitution method for definite integral, the second and third from property of the integral, the fourth from $\delta\rightarrow0^+$ and Taylor expansion, the fifth and sixth from simple algebra, and the last from definition of conditional expectation.

	From (1)--(5) we get
	\begin{eqnarray*}
		\partial_dQ_Y(\tau|d,x)=E\left(\partial_dm\left(d,x,U_d\right)|D=d,X=x,Y=Q_Y(\tau|d,x)\right)
	\end{eqnarray*}
	Thus
	\begin{eqnarray}
		\partial_dQ_Y(\tau|d,x)\big|_{\tau=F_Y(y|d,x)}&=&E\left(\partial_dm\left(d,x,U_d\right)|D=d,X=x,Y=y\right)\notag\\
		&=&E\left(\partial_Dm\left(D,X,U_D\right)|D=d,X=x,Y=y\right)
	\end{eqnarray}
	
	The following process provides the identification of $\partial_dQ_Y(\tau|d,x)\big|_{\tau=F(y|d,x)}$ from the perspective of the definition of $\tau$-th conditional quantile
	\begin{eqnarray}
		\int_{-\infty}^{Q_Y(\tau|d,x)}f\left(y|d,x\right)dy=\int_{-\infty}^{Q_Y(\tau|d,x)}\frac{f\left(d,x,y\right)}{f\left(d,x\right)}dy=\tau\notag
	\end{eqnarray}
	Taking derivative with respect to $d$ on both sides
	\begin{eqnarray}
		\partial_d Q_Y(\tau|d,x)\frac{f\left(d,x,Q_Y(\tau|d,x)\right)}{f(d,x)}+\int_{-\infty}^{Q_Y(\tau|d,x)}\frac{\partial_d f\left(d,x,y\right)f(d,x)-\partial_d f(d,x)f(d,x,y)}{f^2\left(d,x\right)}dy=0\notag
	\end{eqnarray}
	By simple algebra
	\begin{eqnarray}
		\partial_d Q_Y(\tau|d,x)=-\frac{\int_{-\infty}^{Q_Y(\tau|d,x)}\partial_d f\left(d,x,y\right)f(d,x)-\partial_df(d,x)f(d,x,y)dy}{f(d,x)f\left(d,x,Q_Y(\tau|d,x)\right)}\notag
	\end{eqnarray}
	Evaluating at $\tau=F_Y(y|d,x)$ on both sides
	\begin{eqnarray}
		\partial_d Q_Y(\tau|d,x)\big|_{\tau=F_Y(y|d,x)}&=&-\frac{\int_{-\infty}^{Q_Y\left(F_Y(y|d,x)|d,x\right)}\partial_d f\left(d,x,y\right)f(d,x)-\partial_df(d,x)f(d,x,y)dy}{f(d,x)f\left(d,x,Q_Y\left(F_Y(y|d,x)|d,x\right)\right)}\notag\\
		&=&-\frac{\int_{-\infty}^{y}\partial_d f\left(d,x,t\right)f(d,x)-\partial_df(d,x)f(d,x,t)dt}{f(d,x)f\left(d,x,y\right)}\notag\\
		&=&-\frac{\partial_d\left( F_Y\left(y|d,x\right)f(d,x)\right)f(d,x)-\partial_df(d,x)f(d,x)F_Y(y|d,x)}{f(d,x)f\left(d,x,y\right)}\notag\\
		&=&-\frac{\partial_dF_Y(y|d,x)}{f_Y(y|d,x)}
	\end{eqnarray}
	Then $\theta(y_{1},y_{2})$ can be identified by
	\begin{eqnarray*}
		\theta(y_{1},y_{2})&=&E\left(\partial_D m(D,X,U_D)\bigg|Y\in(y_{1},y_{2})\right)\\
		&=&\frac{1}{P(y_1<Y<y_2)}E\left(1\{y_1<Y<y_2\}\partial_D m(D,X,U_D)\right)dddxdy\\
		&=&\frac{1}{P(y_1<Y<y_2)}\int_{y_1}^{y_2}E\left(\partial_D m(D,X,U_D)|Y=y\right)f(y)dy\\
		&=&\frac{1}{P(y_1<Y<y_2)}\int_{y_1}^{y_2}E\left(E\left(\partial_D m(D,X,U_D)|D,X,Y=y\right)|Y=y\right)f(y)dy\\
		&=&\frac{1}{P(y_1<Y<y_2)}\int_{y_1}^{y_2}\int\int E\left(\partial_D m(D,X,U_D)|D=d,X=x,Y=y\right)f(d,x,y)dddxdy\\
		&=&\frac{1}{P(y_1<Y<y_2)}\int_{y_1}^{y_2}\int\int\partial_dQ_Y(\tau|d,x)\big|_{\tau=F_Y(y|d,x)}f(d,x,y)dddxdy\\
		&=&\frac{-1}{P(y_1<Y<y_2)}\int_{y_1}^{y_2}\int\int\frac{\partial_dF_Y(y|d,x)}{f\left(y|d,x\right)}f(d,x,y)dddxdy\\
		&=&\frac{-1}{P(y_1<Y<y_2)}\int_{y_1}^{y_2}\int\int\partial_dF_Y(y|d,x)f(d,x)dddxdy\\
		&=&\frac{-1}{P(y_1<Y<y_2)}\int\int\int_{y_1}^{y_2}\partial_dF_Y(y|d,x)dyf(d,x)dddx\\
		&=&\frac{-1}{P(y_1<Y<y_2)}E\left(\int_{y_1}^{y_2}\partial_DF_Y(y|D,X)dy\right)
	\end{eqnarray*}
	where the first equality follows from definition of $\theta_0(y_1,y_2)$, the second from simple algebra, the third from the law of iterated expectation, the fourth from property of conditional expectation and $f(d,x,y)=f(d,x|y)f(y)$, the fifth from equation (6), the sixth from equation (7), and the remaining from simple algebra.$\blacksquare$\\
	
	\noindent{\textbf{Proof of Proposition 2.2.}}
	We rewrite the score function in Proposition 2.2 in terms of $\eta=(\eta_1,\eta_2,\eta_3)$ as
	\begin{eqnarray}
		\psi\left(y_1,y_2,w;\theta(y_1,y_2),\eta\right)&=&-\frac{1}{\eta_1(y_1,y_2)}\partial_D\eta_2(D,X;y_1,y_2)-\theta(y_1,y_2)\notag\\
		&&-\frac{1}{\eta_1(y_1,y_2)}\eta_3(D,X)\left(\eta_2(D,X;y_1,y_2)-\int_{y_1}^{y_2}1\{Y<y\}dy\right)\notag\\
		&&+\frac{E\left(\partial_D\eta_2(D,X;y_1,y_2)\right)}{\eta_1^2(y_1,y_2)}\left(1\{y_1<Y<y_2\}-\eta_1(y_1,y_2)\right)\notag
	\end{eqnarray}
	Note that
	\begin{eqnarray*}
		E\left(	\psi\left(y_1,y_2,W;\theta(y_1,y_2),\eta\right)\right)&=&\underbrace{-\frac{1}{P(y_1<Y<y_2)}E\left(\partial_D\int_{y_1}^{y_2}F(y|D,X)dy\right)-\theta(y_1,y_2)}_0\notag\\
		&&-\frac{1}{P(y_1<Y<y_2)}E\left(\frac{\partial_Df(D,X)}{f(D,X)}\left(\int_{y_1}^{y_2}F_Y(y|D,X)dy-\int_{y_1}^{y_2}1\{Y<y\}dy\right)\right)\notag\\
		&&+\frac{E\left(\partial_D\int_{y_1}^{y_2}F_Y(y|D,X)dy\right)}{P^2(y_1<Y<y_2)}\underbrace{E\left(1\{y_1<Y<y_2\}-P(y_1<Y<y_2)\right)}_0\notag\\
		&=&-\frac{1}{P(y_1<Y<y_2)}\notag\\
		&\cdot&E\left(\frac{\partial_Df(D,X)}{f(D,X)}\left(\int_{y_1}^{y_2}F_Y(y|D,X)dy-E\left(\int_{y_1}^{y_2}1\{Y<y\}dy|D,X\right)\right)\right)\notag\\
		&=&-\frac{1}{P(y_1<Y<y_2)}\notag\\
		&\cdot&E\left(\frac{\partial_Df(D,X)}{f(D,X)}\left(\int_{y_1}^{y_2}F_Y(y|D,X)dy-\int\int_{y_1}^{y_2}1\{t<y\}dyf_Y(t|D,X)dt\right)\right)\notag\\
		&=&-\frac{1}{P(y_1<Y<y_2)}\notag\\
		&\cdot&E\left(\frac{\partial_Df(D,X)}{f(D,X)}\left(\int_{y_1}^{y_2}F_Y(y|D,X)dy-\int_{y_1}^{y_2}\int1\{t<y\}f_Y(t|D,X)dtdy\right)\right)\notag\\
		&=&-\frac{1}{P(y_1<Y<y_2)}\notag\\
		&\cdot&E\left(\frac{\partial_Df(D,X)}{f(D,X)}\left(\int_{y_1}^{y_2}F_Y(y|D,X)dy-\int_{y_1}^{y_2}F_Y(y|D,X)dy\right)\right)\notag\\
		&=&0\notag
	\end{eqnarray*}
	where the first equality follows from definition of $\psi(\cdot)$, Proposition 2.1, and simple algebra, the second from law of iterated expectation, the third from definition of conditional expectation, the fourth from property of double integral, and the remaining from simple algebra.
	
	Thus, $\theta(y_{1},y_{2})$ satisfies $	E\left(	\psi\left(y_1,y_2,W;\theta(y_1,y_2),\eta\right)\right)=0$.

	Note that $\eta$ is the true value of the  nuisance parameter $\tilde{\eta}\in\mathcal{T}$. Then the pathwise (or the Gateaux) derivative against the  nuisance parameters at $0$ is
	\begin{eqnarray*}
		&&\frac{\partial E\psi\bigg(y_1,y_2,W;\theta(y_{1},y_{2}),\eta+r(\tilde{\eta}-\eta)\bigg)}{\partial r}\bigg|_{r=0}\\
		&=&\frac{1}{\left(\eta_1(y_1,y_2)\right)^2}(\tilde{\eta}_1(y_1,y_2)-\eta_1(y_1,y_2))E\left(\partial_D\eta_2(D,X;y_1,y_2)\right)\\
		&&-\frac{1}{\eta_1(y_1,y_2)}E\left(\partial_D\tilde{\eta}_2(D,X;y_1,y_2)-\partial_D\eta_2(D,X;y_1,y_2)\right)\\
		&&+\frac{1}{\left(\eta_1(y_1,y_2)\right)^2}\left(\tilde{\eta}_1(y_1,y_2)-\eta_1(y_1,y_2)\right)E\left(\eta_3(D,X)\left(\eta_2(D,X;y_1,y_2)-\int_{y_1}^{y_1}1\{Y<y\}dy\right)\right)\\
		&&-\frac{1}{\eta_1(y_1,y_2)}E\left(\left(\tilde{\eta}_3(D,X)-\eta_3(D,X)\right)\left(\eta_2(D,X;y_1,y_2)-\int_{y_1}^{y_2}1\{Y<y\}dy\right)\right)\\
		&&-\frac{1}{\eta_1(y_1,y_2)}E\left(\eta_3(D,X)\left(\tilde{\eta}_2(D,X;y_1,y_2)-\eta_2(D,X;y_1,y_2)\right)\right)\\
		&&-2\frac{E\left(\partial_D\eta_2(D,X;y_1,y_2)\right)}{\left(\eta_1(y_1,y_2)\right)^3}\left(\tilde{\eta}_1(y_1,y_2)-\eta_1(y_1,y_2)\right)\underbrace{E\left(1\{y_1<Y<y_2\}-\eta_1(y_1,y_2)\right)}_0\\
		&&+\frac{E\left(\partial_D\tilde{\eta}_2(D,X;y_1,y_2)-\partial_D\eta_2(D,X;y_1,y_2)\right)}{\left(\eta_1(y_1,y_2)\right)^2}\underbrace{E\left(1\{y_1<Y<y_2\}-\eta_1(y_1,y_2)\right)}_0\\
		&&-\frac{E\left(\partial_D\eta_2(D,X;y_1,y_2)\right)}{\left(\eta_1(y_1,y_2)\right)^2}\left(\tilde{\eta}_1(y_1,y_2)-\eta_1(y_1,y_2)\right)\\
		&=&-\frac{1}{\eta_1(y_1,y_2)}E\left(\partial_D\tilde{\eta}_2(D,X;y_1,y_2)-\partial_D\eta_2(D,X;y_1,y_2)\right)\\
		&&+\frac{1}{\left(\eta_1(y_1,y_2)\right)^2}\left(\tilde{\eta}_1(y_1,y_2)-\eta_1(y_1,y_2)\right)E\left(\eta_3(D,X)\underbrace{\left(\eta_2(D,X;y_1,y_2)-E\left(\int_{y_1}^{y_1}1\{Y<y\}dy|D,X\right)\right)}_0\right)\\
		&&-\frac{1}{\eta_1(y_1,y_2)}E\left(\left(\tilde{\eta}_3(D,X)-\eta_3(D,X)\right)\underbrace{\left(\eta_2(D,X;y_1,y_2)-E\left(\int_{y_1}^{y_2}1\{Y<y\}dy|D,X\right)\right)}_0\right)\\
		&&-\frac{1}{\eta_1(y_1,y_2)}E\left(\eta_3(D,X)\left(\tilde{\eta}_2(D,X;y_1,y_2)-\eta_2(D,X;y_1,y_2)\right)\right)
	\end{eqnarray*}
	where the first equality follows from simple algebra, and the second from the law of iterated expectation.
	
	Note that
	\begin{eqnarray*}
		&&E\left(\eta_3(D,X)\left(\tilde{\eta}_2(D,X;y_1,y_2)-\eta_2(D,X;y_1,y_2)\right)\right)\\
		&=&E\left(\frac{\partial_D(D,X)}{f(D,X)}\left(\tilde{\eta}_2(D,X;y_1,y_2)-\eta_2(D,X;y_1,y_2)\right)\right)\\
		&=&\int\int \frac{\partial_df(d,x)}{f(d,x)}\left(\tilde{\eta}_2(d,x;y_1,y_2)-\eta_2(d,x;y_1,y_2)\right)f(d,x)dddx\\
		&=&\int\int \partial_df(d,x)\left(\tilde{\eta}_2(d,x;y_1,y_2)-\eta_2(d,x;y_1,y_2)\right)ddx\\
		&=&-\int\int f(d,x)\partial_d\left(\tilde{\eta}_2(d,x;y_1,y_2)-\eta_2(d,x;y_1,y_2)\right)ddx\\
		&=&E\left(\partial_D\tilde{\eta}_3(D,X;y_1,y_2)-\partial_D\eta_3(D,X;y_1,y_2)\right)
	\end{eqnarray*}
	where the first equality follows from definition of $\eta_3(D,X)$, the second from definition of expectation, the third from simple algebra, the fourth from integration by parts and the regularity condition that $f(d,x)$ is zero on the boundary of $\mathcal{S}_D$, and the last from definition of expectation.
	
	Thus
	\begin{eqnarray*}
		\frac{\partial E\psi\left(y_1,y_2,W;\theta(y_{1},y_{2}),\eta+r(\tilde{\eta}-\eta)\right)}{\partial r}\bigg|_{r=0}=0
	\end{eqnarray*} $\blacksquare$\\
	\\
	\noindent\textbf{Proof of Proposition 2.3.}
	For (i), it follows from (2.1.6) that
	\begin{eqnarray*}
		\theta(y)=\frac{-E\left(\partial_{D}F_{Y}(y|D,X)\right)}{f_{Y}(y)}.
	\end{eqnarray*}
	By Firpo et al. (2009, Corollary 1, page 958), $UQPE(\tau)$ can be expressed as
	\begin{eqnarray*}
		UQPE(\tau)=\frac{-1}{f_{Y}(Q_{Y}(\tau))}\int\frac{\partial F_{Y}\left(Q_{Y}(\tau)\big|d,x\right)}{\partial d}f_{DX}(d,x)dddx
	\end{eqnarray*}
	which is equal to $\theta\left(Q_{Y}\big(\tau\big)\right)$ by replacing $y=Q_{Y}(\tau)$. Result (ii) follows from (i) and
	\begin{eqnarray*}
		\theta(y_{1},y_{2})=\frac{1}{P(y_{1}<Y<y_{2})}\int_{y_{1}}^{y_{2}}\theta(y)f_Y(y)dy.
	\end{eqnarray*}
	For (iii), we only need to show that
	\begin{eqnarray*}
		\theta\left(Q_{Y}\big(\tau\big)\right)=E\left[\frac{\partial Q_{Y}(u|D,X)}{\partial D}\big|_{u=F_{Y}\left(Q_{Y}(\tau)|D,X\right)}\frac{f_{Y}\left(Q_{Y}(\tau)|D,X\right)}{f_{Y}\left(Q_{Y}(\tau)\right)}\right].
	\end{eqnarray*}
	The desired result follows by applying (i). Notice that by (2.1.4),
	\[\begin{aligned}
		\theta\left(y_{1}\right)=&\lim_{y_{2}\to{y_{1}}}\theta(y_{1},y_{2}) \\
		=&\lim_{y_{2}\to{y_{1}}}\int_{y_{1}}^{y_{2}}\int\frac{\partial Q_{Y}(u|d,x)}{\partial d}\big|_{u=F_{Y}(y|d,x)}f_{Y}(y|d,x)f_{DX}(d,x)dddxdy\bigg/ P(y_{1}<Y<y_{2}) \\
		=&\int\frac{\partial Q_{Y}(u|d,x)}{\partial d}\big|_{u=F_{Y}(y_{1}|d,x)}f_{Y}(y_{1}|d,x)f_{DX}(d,x)dddx\bigg/ f_{Y}(y_{1}). \\
	\end{aligned}\]
	Replacing $y_{1}$ by $Q_{Y}(\tau)$ yields
	\[\begin{aligned}
		\theta\left(Q_{Y}\big(\tau\big)\right)=&\int\frac{\partial Q_{Y}(u|d,x)}{\partial d}\big|_{u=F_{Y}\left(Q_{Y}(\tau)|d,x\right)}f_{Y}\left(Q_{Y}(\tau)|d,x\right)f_{DX}(d,x)dddx\bigg/f_{Y}\left(Q_{Y}(\tau)\right)  \\
		=&E\left[\frac{\partial Q_{Y}(u|D,X)}{\partial d}\big|_{u=F_{Y}\left(Q_{Y}(\tau)|D,X\right)}\frac{f_{Y}\left(Q_{Y}(\tau)|D,X\right)}{f_{Y}\left(Q_{Y}(\tau)\right)}\right], \\
	\end{aligned}\]
	which completes the proof. $\blacksquare$\\
	\\
	\noindent\textbf{Proof of Proposition 2.4.}
	
	\noindent{\textbf{Assumption A.4.}}
	The support of $H_t$ is a subset of the support of $D$ conditional on $X$, that is, $\text{supp}(H_t)\subset \text{supp}(D|X=x)$ for all $x\in \text{supp}(X)$.
	
	Under regularity condition Assumption A.4 and Assumptions 2.1--2.2, similar to Rothe (2012, Lemma 1, page 2274), we get
	\begin{eqnarray}
		F_{Y_{H_t}}(y)&=&P\left(Y_{H_t}\leq y\right)\notag\\
		&=&P\left(m(D_{H_t},X,U_D)\leq y\right)\notag\\
		&=&P\left(m\left(H_t^{-1}(R_0),X,U_{F_0^{-1}(R_0)}\right)\leq y\right)\notag\\
		&=&\int\int P\left(m\left(H_t^{-1}(r),x,U_{F_0^{-1}(r)}\right)\leq y|R_0=r,X=x\right)dF_{R_0X}(r,x)\notag\\
		&=&\int\int P\left(m\left(H_t^{-1}(r),x,U_{F_0^{-1}(r)}\right)\leq y|X=x\right)dF_{R_0X}(r,x)\notag\\
		&=&\int\int P\left(m\left(H_t^{-1}(r),x,U_{H_t^{-1}(r)}\right)\leq y|X=x\right)dF_{R_0X}(r,x)\notag\\
		&=&\int\int P\left(m\left(d,x,U_d\right)\leq y|X=x\right)dF_{R_0X}\left(H_t(d),x\right)\notag\\
		&=&\int\int P\left(m\left(d,x,U_d\right)\leq y|D=d,X=x\right)dF_{R_0X}\left(H_t(d),x\right)\notag\\
		&=&\int\int P\left(m\left(D,X,U_D\right)\leq y|D=d,X=x\right)dF_{R_0X}\left(H_t(d),x\right)\notag\\
		&=&\int\int P(Y\leq y|D=d,X=x)dF_{R_0X}\left(H_t(d),x\right)\notag\\
		&=&\int\int P\left(Y\leq y|D=H_t^{-1}(r),X=x\right)dF_{R_0X}\left(r,x\right)\notag\\
		&=&E\left(F_Y\left(y|H_t^{-1}(R_0),X\right)\right)\notag\\
		&=&E\left(F_Y\left(y|H_t^{-1}\left(F_0(D)\right),X\right)\right)
	\end{eqnarray}
	where the first equality follows from definition of unconditional distribution function, the second from data generating process of $Y_{H_t}$, the third from $D_{H_t}=H_t^{-1}(R_0)$, which implies by continuity of $D$, the fourth from the law of iterated expectation, the fifth from Assumption 2.2 and $D=F_0^{-1}(R_0)$, the sixth from Assumption 2.1, the seventh from change of variable and Assumption A.4, a regularity condition, the eighth from Assumption 2.2, the ninth from property of conditional probability, the tenth from data generating process $Y=m(D,X,U_D)$, the eleventh from change of variable and Assumption A.4, the twelfth from definition of expectation, and the last from $R_0=F_0(D)$, which implies by continuity of $D$ and $D=Q_0(R_0)$.
	
	Thus $f_{Y_{H_t}}(\cdot)$, the corresponding PDF of $Y$ under distribution $F_{Y_{H_t}}(\cdot)$, can be written as
	\begin{eqnarray*}
		f_{Y_{H_t}}(y)=E\left(f_Y\left(y|H_t^{-1}\left(F_0(D)\right),X\right)\right)
	\end{eqnarray*}
	Taking the derivative of the both sides with respect to $t$
	\begin{eqnarray*}
		\frac{\partial f_{Y_{H_t}}(y)}{\partial t}=E\left(\frac{\partial f_Y\left(y|H_t^{-1}\left(F_0(D)\right),X\right)}{\partial \left(H_t^{-1}\left(F_0(D)\right)\right)}\frac{\partial H_t^{-1}\left(F_0(D)\right)}{\partial t}\right)
	\end{eqnarray*}
	Evaluating at $t=0$
	\begin{eqnarray}
		\frac{\partial f_{Y_{H_t}}(y)}{\partial t}\bigg|_{t=0}&=&E\left(\partial_Df_Y(y|D,X)\frac{\partial H_t^{-1}\left(F_0(D)\right)}{\partial t}\bigg|_{t=0}\right)\notag\\
		&=&E\left(\partial_Df_Y(y|D,X)\frac{\partial H_t^{-1}(R_0)}{\partial t}\bigg|_{t=0}\right)
	\end{eqnarray}
	where the last equality follows from $R_0=F_0(D)$.
	
	We refer PDFs of continuously distributed random variable $D$ under distribution $F_0$, $G_0$ and $H_t$ as the corresponding lowercase notations $f_0$, $g_0$ and $h_t$, respectively. From $H_t(d)=F_0(d)+t\left(G_0(d)-F_0(d)\right)$, we get $h_t(d)=f_0(d)+t\left(g_0(d)-f_0(d)\right)$.
	
	By definition, we get
	\begin{eqnarray*}
		\int_{-\infty}^{H_t^{-1}(r)}h_t(d)dd=r
	\end{eqnarray*}
	Taking the derivative of both sides with respect to $t$
	\begin{eqnarray*}
		\frac{\partial H_t^{-1}(r)}{\partial t}h_t\left(H_t^{-1}(r)\right)+\int_{-\infty}^{H_t^{-1}(r)}\left(g_0(d)-f_0(d)\right)dd=0
	\end{eqnarray*}
	Evaluating at $t=0$
	\begin{eqnarray*}
		\frac{\partial H_t^{-1}(r)}{\partial t}\big|_{t=0}f_0\left(F_0^{-1}(r)\right)+\int_{-\infty}^{F_0^{-1}(r)}\left(g_0(d)-f_0(d)\right)dd=0
	\end{eqnarray*}
	By simple algebra and property of CDFs
	\begin{eqnarray*}
		\frac{\partial H_t^{-1}(r)}{\partial t}\big|_{t=0}&=&-\frac{\int_{-\infty}^{F_0^{-1}(r)}\left(g_0(d)-f_0(d)\right)dd}{f_0\left(F_0^{-1}(r)\right)}\\
		&=&-\frac{G_0\left(F_0^{-1}(r)\right)-F_0\left(F_0^{-1}(r)\right)}{f_0\left(F_0^{-1}(r)\right)}\\
		&=&-\frac{G_0\left(F_0^{-1}(r)\right)-r}{f_0\left(F_0^{-1}(r)\right)}
	\end{eqnarray*}
	Thus
	\begin{eqnarray}
		\frac{\partial H_t^{-1}(R_0)}{\partial t}\big|_{t=0}\notag
		&=&-\frac{G_0\left(F_0^{-1}(R_0)\right)-R_0}{f_0\left(F_0^{-1}(R_0)\right)}\notag\\
		&=&-\frac{G_0\left(F_0^{-1}\left(F_0(D)\right)\right)-F_0(D)}{f_0\left(F_0^{-1}\left(F_0(D)\right)\right)}\notag\\
		&=&-\frac{G_0(D)-F_0(D)}{f_0(D)}
	\end{eqnarray}
	where the second equality from $R_0=F_0(D)$.
	
	By definition
	\begin{eqnarray*}
		\int_{-\infty}^{Q_{Y_{H_t}}(\tau)}f_{Y_{H_t}}(y)dy=\tau
	\end{eqnarray*}
	Taking the derivative of both sides with respect to $t$
	\begin{eqnarray*}
		\frac{\partial Q_{Y_{H_t}}(\tau)}{\partial t}f_{Y_{H_t}}\left(Q_{Y_{H_t}}(\tau)\right)+\int_{-\infty}^{Q_{Y_{H_t}}(\tau)}\frac{\partial f_{Y_{H_t}}(y)}{\partial t}dy=0
	\end{eqnarray*}
	Evaluating at $t=0$
	\begin{eqnarray*}
		\frac{\partial Q_{Y_{H_t}}(\tau)}{\partial t}\bigg|_{t=0}f_Y\left(Q_Y(\tau)\right)+\int_{-\infty}^{Q_Y(\tau)}\frac{\partial f_{Y_{H_t}}(y)}{\partial t}\bigg|_{t=0}dy=0
	\end{eqnarray*}
	Thus
	\begin{eqnarray}
		MQPE(\tau,G_0)&=&\frac{\partial Q_{Y_{H_t}}(\tau)}{\partial t}\bigg|_{t=0}\notag\\
		&=&-\frac{1}{f_Y\left(Q_Y(\tau)\right)}\int_{-\infty}^{Q_Y(\tau)}\frac{\partial f_{Y_{H_t}}(y)}{\partial t}\bigg|_{t=0}dy\notag\\
		&=&-\frac{1}{f_Y\left(Q_Y(\tau)\right)}\int_{-\infty}^{Q_Y(\tau)}E\left(\partial_Df_Y(y|D,X)\frac{\partial H_t^{-1}(R_0)}{\partial t}\bigg|_{t=0}\right)dy\notag\\
		&=&\frac{1}{f_Y\left(Q_Y(\tau)\right)}\int_{-\infty}^{Q_Y(\tau)}E\left(\partial_Df_Y(y|D,X)\frac{G_0(D)-F_0(D)}{f_0(D)}\right)dy\notag\\
		&=&\frac{1}{f_Y\left(Q_Y(\tau)\right)}E\left(\partial_DF_Y\left(Q_Y(\tau)|D,X\right)\frac{G_0(D)-F_0(D)}{f_0(D)}\right)\notag
	\end{eqnarray}
	where the first equality follows from definition of $MQPE(\tau,G_0)$, the second from simple algebra, the third from equation (9), the fourth from equation (10), and the last from property of CDF.
	
	Moreover, $\varsigma\left(Q_Y(\tau),G_0\right)$, the average treatment effect of changing the unconditional distribution of $D$ infinitesimally towards the direction of $G_0$ on the individuals with $Y=Q_Y(\tau)$, simplifies to
	\begin{eqnarray*}
		\varsigma(Q_Y(\tau),G_{0})&=&E\left(\frac{\partial m\left(H_{t}^{-1}(R_{0}),X,U_D\right)}{\partial t}\big|_{t=0}\bigg|Y=Q_Y(\tau)\right)\\
		&=&E\left(\partial_D m(D,X,U_D)\frac{\partial H_{t}^{-1}(R_{0})}{\partial t}\big|_{t=0}\bigg|Y=Q_Y(\tau)\right)\\
		&=&-E\left(\partial_D m(D,X,U_D)\frac{G_0(D)-F_0(D)}{f_0(D)}\bigg|Y=Q_Y(\tau)\right)\\
		&=&-\int\int E\left(\partial_D m(D,X,U_D)\big|D=d,X=x,Y=Q_Y(\tau)\right)f_{DX}\left(d,x|Q_Y(\tau)\right)\frac{G_0(d)-F_0(d)}{f_0(d)}dddx\\
		&=&\int\int\frac{\partial_dF_Y(Q_Y(\tau)|d,x)}{f_Y(Q_Y(\tau)|d,x)}f_{DX}\left(d,x|Q_Y(\tau)\frac{G_0(d)-F_0(d)}{f_0(d)}\right)dddx\\
		&=&\frac{1}{f_Y\left(Q_Y(\tau)\right)}\int\int\partial_dF_Y(Q_Y(\tau)|d,x)f_{DX}(d,x)\frac{G_0(d)-F_0(d)}{f_0(d)}dddx\\
		&=&\frac{1}{f_Y\left(Q_Y(\tau)\right)}E\left(\partial_DF_Y(Q_Y(\tau)|D,X)\frac{G_0(D)-F_0(D)}{f_0(D)}\right)
	\end{eqnarray*}
	where the first equality follows definition of $\varsigma(\cdot,G_0)$, the second from the chain rule, the third from equation (10), the fourth from the law of iterated expectation, the fifth from equations (6)--(7), which hold under assumption in Proposition 2.1, the sixth from $f\left(d,x,Q_Y(\tau)\right)=f_Y\left(Q_Y(\tau)|d,x\right)f_{DX}(d,x)$, and the last from definition of expectation.
	
	Thus, we get
	\begin{eqnarray*}
		\varsigma\left(Q_Y(\tau),G_0\right)=MQPE(\tau,G_0)
	\end{eqnarray*}
	which completes the proof of (i) in Proposition 2.4.
	
	For (ii),
	\begin{eqnarray*}
		\varsigma\left(Q_Y(\tau_1),Q_Y(\tau_2),G_0\right)&=&\frac{1}{\tau_2-\tau_1}\int_{Q_Y(\tau_1)}^{Q_Y(\tau_2)}\varsigma (y,G_0)f_Y(y)dy\\
		&=&\frac{1}{\tau_2-\tau_1}\int_{\tau_1}^{\tau_2}\varsigma (Q_Y(\tau),G_0)d\tau\\
		&=&\frac{1}{\tau_2-\tau_1}\int_{\tau_1}^{\tau_2}MQPE(\tau,G_0)d\tau
	\end{eqnarray*}
	where the first equality follows from definition of $\varsigma(y_1,y_2,G_0)$ and $\varsigma(y,G_0)$, the second from property of definite integral, and the last from part (i) of Proposition 2.4.
	
	This completes the proof. $\blacksquare$\\
	\\
	\noindent\textit{{A.3 Double Robustness Property}}
	\begin{lemma}(Double robustness property)  $\theta(y_1,y_2)$ can be identified by the orthogonal score given in Proposition 2.2 if either  $\tilde{\eta}_2(D,X;y_1,y_2)=\eta_2(D,X;y_1,y_2)$ or $\tilde{\eta}_3(D,X)=\eta_3(D,X)$, $\tilde{\eta}_2(\cdot)$ and $\tilde{\eta}_3(\cdot)$ being arbitrary models for the true, unknown function $\eta_2(\cdot)$ and $\eta_3(\cdot)$, respectively.
	\end{lemma}
	
	\noindent\textbf{Proof.}
	If $\eta_2(D,X;y_1,y_2)$ is correctly specified and $\eta_3(D,X)$ is specified as $\tilde{\eta}_3(D,X)$, we get
	\begin{eqnarray*}
		&&E\left(	\psi\left(y_1,y_2;W,\theta(y_1,y_2),\eta_1,\eta_2,\tilde{\eta}_3\right)\right)\\
		&=&\underbrace{-\frac{1}{P(y_1<Y<y_2)}E\left(\partial_D\int_{y_1}^{y_2}F(y|D,X)dy\right)-\theta(y_1,y_2)}_0\notag\\
		&&-\frac{1}{P(y_1<Y<y_2)}E\left(\tilde{\eta}_3(D,X)\left(\int_{y_1}^{y_2}F_Y(y|D,X)dy-\int_{y_1}^{y_2}1\{Y<y\}dy\right)\right)\\
		&&+\frac{E\left(\partial_D\int_{y_1}^{y_2}F_Y(y|D,X)dy\right)}{P^2(y_1<Y<y_2)}\underbrace{E\left(1\{y_1<Y<y_2\}-P(y_1<Y<y_2)\right)}_0\\
		&=&-\frac{1}{P(y_1<Y<y_2)}\\
		&\cdot&E\left(\tilde{\eta}_3(D,X)\left(\int_{y_1}^{y_2}F_Y(y|D,X)dy-E\left(\int_{y_1}^{y_2}1\{Y<y\}dy|D,X\right)\right)\right)\\
		&=&-\frac{1}{P(y_1<Y<y_2)}\\
		&\cdot&E\left(\tilde{\eta}_3(D,X)\left(\int_{y_1}^{y_2}F_Y(y|D,X)dy-\int\int_{y_1}^{y_2}1\{t<y\}dyf_Y(t|D,X)dt\right)\right)\\
		&=&-\frac{1}{P(y_1<Y<y_2)}\\
		&\cdot&E\left(\tilde{\eta}_3(D,X)\left(\int_{y_1}^{y_2}F_Y(y|D,X)dy-\int_{y_1}^{y_2}\int1\{t<y\}f_Y(t|D,X)dtdy\right)\right)\\
		&=&-\frac{1}{P(y_1<Y<y_2)}\\
		&\cdot&E\left(\tilde{\eta}_3(D,X)\left(\int_{y_1}^{y_2}F_Y(y|D,X)dy-\int_{y_1}^{y_2}F_Y(y|D,X)dy\right)\right)\\
		&=&0
	\end{eqnarray*}
	where the first equality follows from definition of $\psi(\cdot)$, Proposition 2.1, and simple algebra, the second from law of iterated expectation, the third from definition of conditional expectation, the fourth from property of double integral, and the remaining from simple algebra.
	
	If $\eta_3(D,X)$ is correctly specified and $\eta_2(D,X;y_1,y_2)$ is specified as $\tilde{\eta}_2(D,X;y_1,y_2)$, we get
	\begin{eqnarray*}
		&&E\left(	\psi\left(y_1,y_2,W;\theta(y_1,y_2),\eta_1,\tilde{\eta}_2,\eta_3\right)\right)\\
		&=&-\frac{1}{P(y_1<Y<y_2)}E\left(\partial_D\tilde{\eta}_2(D,X;y_1,y_2)\right)-\theta(y_1,y_2)\\
		&&-\frac{1}{P(y_1<Y<y_2)}E\left(\frac{\partial_Df(D,X)}{f(D,X)}\left(\tilde{\eta}_2(D,X;y_1,y_2)-\int_{y_1}^{y_2}1\{Y<y\}dy\right)\right)\\
		&&+\frac{E\left(\partial_D\tilde{\eta}_2(D,X;y_1,y_2)\right)}{P^2(y_1<Y<y_2)}\underbrace{E\left(1\{y_1<Y<y_2\}-P(y_1<Y<y_2)\right)}_0\\
		&=&-\frac{1}{P(y_1<Y<y_2)}E\left(\partial\tilde{\eta}_2(D,X;y_1,y_2)\right)-\theta(y_1,y_2)\\
		&&-\frac{1}{P(y_1<Y<y_2)}\int\int\partial_df(d,x)\left(\tilde{\eta}_2(d,x;y_1,y_2)-\eta_2(d,x;y_1,y_2)\right)\\
		&=&-\frac{1}{P(y_1<Y<y_2)}E\left(\partial_d\eta_2(D,X;y_1,y_2)\right)-\theta(y_1,y_2)\\
		&=&0
	\end{eqnarray*}
	where the first equality follows from definition of $\psi(\cdot)$ and simple algebra, the second from definition of expectation, the third from integration by parts, and the last from Proposition 2.1.$\blacksquare$
	\\
	\\
	\noindent\textit{{A.4 Binary treatment variable}.}
	
	\noindent When the treatment is binary, the quantity in parallel with OASD is
	\begin{eqnarray*}
		\vartheta(y_{1},y_{2})=E\left(m\left(1,X,U_D\right)-m\left(0,X,U_D\right)|Y\in(y_{1},y_{2})\right)
	\end{eqnarray*}
	\noindent\textbf{Assumption A.5} (Monotonicity condition)	$m(d,x,u)$ is strictly increasing with respect to $u$ for each $d$ and $x$.
	
	\noindent\textbf{Proposition A.1.} Under Assumptions 2.1-2.2 and Assumption A.5,
	\begin{eqnarray*}
		\vartheta(y_1,y_2)&=&\frac{1}{P(y_1<Y<y_2)}\\
		&&\cdot E\left(1\{y_1<Y<y_2\}\left((1-D)Q_Y\left(F_Y(Y|0,X)|1,X\right)-DQ_Y\left(F_Y(Y|1,X)|0,X\right)+(2D-1)Y\right)\right)
	\end{eqnarray*}

	\noindent{\textbf{Proof.}}
	Define
	\begin{eqnarray*}
		\vartheta(d,x,y):= E\left(m(1,X,U_D)-m(0,X,U_D)|D=d,X=x,Y=y\right)
	\end{eqnarray*}
	Thus
	\begin{eqnarray}
		\vartheta(d,x,y)&=&E\left(m(1,X,U_D)-m(0,X,U_D)|D=d,X=x,Y=y\right)\notag\\
		&=&E\left(m(1,x,U_d)-m(0,x,U_d)|D=d,X=x,Y=y\right)\notag\\
		&=&E\left(m(1,x,U_d)-m(0,x,U_d)|D=0,X=x,m(d,x,U_d)=y\right)\notag\\
		&=&E\left(m(1,x,U_d)-m(0,x,U_d)|D=0,X=x,U_d=m^{-1}(d,x,y)\right)\notag\\
		&=&\left(m\left(1,x,U_d\right)-m\left(0,x,U_d\right)\right)\big|_{U_d=m^{-1}(d,x,y)}
	\end{eqnarray}
	where the first equality follows from definition of $\vartheta(d,x,y)$, the second from property of conditional expectation, the third from data generating process, the fourth from Assumption A.5, and the last from property of conditional expectation.

	Denote the distribution of $U_0$ conditional on $X=x$ as $F_{U_0}(\cdot|x):=P(U_0\leq u_0|X=x)$, then $F_{U_0}\left(m^{-1}(d,x,y)|x\right)$ can  be identified for that
	\begin{eqnarray}
		F_{U_0}\left(m^{-1}(d,x,y)|x\right)&=&P\left(U_0\leq m^{-1}(0,x,y)|X=x\right)\notag\\
		&=&P\left(U_d\leq m^{-1}(d,x,y)|X=x\right)\notag\\
		&=&P\left(U_d\leq m^{-1}(d,x,y)|D=d,X=x\right)\notag\\
		&=&P\left(m(d,x,U_d)\leq y|D=d,X=x\right)\notag\\
		&=&P\left(m(D,X,U_D)\leq y|D=d,X=x\right)\notag\\
		&=&\underbrace{P(Y\leq y|D=d,X=x)}_{=F_Y(y|d,x)}
	\end{eqnarray}
	where the first equality follows from the definition of $F_{U_0}(\cdot|x)$, the second from Assumption 2.1, the third from Assumption 2.2, the fourth and fifth from simple algebra, and the last from data generating process and definition of $F_Y(\cdot|d,x)$.
	
	Note that under mild conditions, $F_{U_0}(\cdot|x)$ is strictly increasing with respect to $u$, then we get, from (12)
	\begin{eqnarray}
		m^{-1}(d,x,y)=F_{U_0}^{-1}\left(F_Y(y|d,x)|x\right)
	\end{eqnarray}
	
	Combing the above equality with the identity $m\left(d,x,m^{-1}(d,x,y)\right)\equiv y$, which holds under Assumption A.5, gives
	\begin{eqnarray}
		m\left(d,x,F_{U_0}^{-1}\left(F_Y(y|d,x)|x\right)\right)= y\notag
	\end{eqnarray}
	Evaluating at $y=Q_Y(\tau |d,x)$ gives
	\begin{eqnarray*}
		m\left(d,x,F_{U_0}^{-1}\left(\underbrace{F_Y\left(Q_Y(\tau |d,x)|d,x\right)}_{\equiv \tau}|x\right)\right)= Q_Y(\tau |d,x)\Rightarrow
		m\left(d,x,U_d\right)\big|_{U_d=F_{U_0}^{-1}(\tau |x)}=Q_Y(\tau |d,x)
	\end{eqnarray*}
	Thus
	\begin{eqnarray}
		\left(m\left(1,x,U_d\right)-m\left(0,x,U_d\right)\right)\big|_{U_d=F_{U_0}^{-1}\left(F_Y(y|d,x)|x\right)}=\left(Q_Y(\tau |1,x)-Q_Y(\tau |0,x)\right)\big|_{\tau=F_Y(y|d,x)}
	\end{eqnarray}
	Combining (8), (13) and (14) gives
	\begin{eqnarray}
		\vartheta(d,x,y)=\left(Q_Y(\tau |1,x)-Q_Y(\tau |0,x)\right)\big|_{\tau=F_Y(y|d,x)}\notag
	\end{eqnarray}
	
	This implies that when treatment variable is binary, the relationship shown in (2.1.3) given in the main text still hold under Assumptions 2.1, 2.2 and A.5.
	
	We get
	\begin{eqnarray}
		\vartheta(1,x,y)&=&y-Q_Y(F_Y(y|1,x)|0,x)=y-Q_Y\left(F_Y(y|1,x)|0,x\right)\notag\\
		\vartheta(0,x,y)&=&Q\left(F_Y(y|0,x)|1,x\right)-y=Q_Y\left(F_Y(y|0,x)|1,x\right)-y\notag
	\end{eqnarray}
	Thus, $\vartheta(y_1,y_2)$ can be identified by
	\begin{eqnarray*}
		\vartheta(y_1,y_2)&=&E\left(m(1,X,U_D)-m(0,X,U_D)|Y\in(y_1,y_2)\right)\\
		&=&\frac{1}{P(y_1<Y<y_2)}E\left(1\{y_1<Y<y_2\}\left(m(1,X,U_D)-m(0,X,U_D)\right)\right)\\
		&=&\frac{1}{P(y_1<Y<y_2)}\int_{y_1}^{y_2}E\left(m(1,X,U_D)-m(0,X,U_D)|Y=y\right)f(y)dy\\
		&=&\frac{1}{P(y_1<Y<y_2)}\int_{y_1}^{y_2}E\left(E\left(m(1,X,U_D)-m(0,X,U_D)|D,X,Y=y\right)|Y=y\right)f(y)dy\\
		&=&\frac{1}{P(y_1<Y<y_2)}\\
		&&\cdot\int_{y_1}^{y_2}\int \underbrace{E\left(m(1,X,U_D)-m(0,X,U_D)|D=1,X=x,Y=y\right)}_{\vartheta(1,x,y)}f(1,x|y)dxf(y)dy\\
		&&+\frac{1}{P(y_1<Y<y_2)}\\
		&&\cdot\int_{y_1}^{y_2}\int \underbrace{E\left(m(1,X,U_D)-m(0,X,U_D)|D=0,X=x,Y=y\right)}_{\vartheta(0,x,y)}f(0,x|y)dxf(y)dy\\
		&=&\frac{1}{P(y_1<Y<y_2)}\int_{y_1}^{y_2}\int \left(\vartheta(1,x,y)f(1,x,y)+\vartheta(0,x,y)f(0,x,y)\right)dxdy\\
		&=&\frac{1}{P(y_1<Y<y_2)}E\left(1\{y_1<Y<y_2\}\left(D\vartheta(1,X,Y)+(1-D)\vartheta(0,X,Y)\right)\right)\\
		&=&\frac{-1}{P(y_1<Y<y_2)}E\left(1\{y_1<Y<y_2\}DQ_Y\left(F_Y(Y|1,X)|0,X\right)\right)\\
		&&+\frac{1}{P(y_1<Y<y_2)}E\left(1\{y_1<Y<y_2\}(1-D)Q_Y\left(F_Y(Y|0,X)|1,X\right)\right)\\
		&&+\frac{1}{P(y_1<Y<y_2)}E\left(1\{y_1<Y<y_2\}(2D-1)Y\right)\\
		&=&\frac{1}{P(y_1<Y<y_2)}\\
		&&\cdot E\left(1\{y_1<Y<y_2\}\left((1-D)Q_Y\left(F_Y(Y|0,X)|1,X\right)-DQ_Y\left(F_Y(Y|1,X)|0,X\right)+(2D-1)Y\right)\right)
	\end{eqnarray*}
	where the first equality follows from definition of $\vartheta(y_1,y_2)$, the second and third from law of iterated expectation, and the remaining from simple algebra.
	$\blacksquare$
	\\
	
	To clarify the orthogonal score, we first define the corresponding nuisance parameter
	\begin{eqnarray*}
		&&\Phi(x,y;y_1,y_2)\\
		&=&\bigg(F_Y(y|1,x),F_Y(y|0,x),\frac{f_{DY}(0,y|x)}{f_Y\left(F_y^{-1}\left(F_Y(y|0,x)|1,x\right)|1,x\right)},\frac{f_{DY}(1,y|x)}{f_Y\left(F_Y^{-1}\left(F_Y(y|1,x)|0,x\right)|0,x\right)}\\
		&&,P(D=1|X=x),P\left(y_1<Y<y_2\right)\bigg)
	\end{eqnarray*}
	where $F_Y^{-1}(\cdot|1,x)$ and $F_Y^{-1}(\cdot|0,x)$ are equivalent to $Q_Y(\cdot|1,x)$ and $Q_Y(\cdot|0,x)$, respectively.
	
	Let
	\begin{eqnarray*}
		\Psi(W,\vartheta,\Phi;y_1,y_2)&=&\frac{1}{P(y_1<Y<y_2)}1\{y_1<Y<y_2\}\\
		&&\cdot\left((1-D)F_Y^{-1}\left(F(Y|0,X)|1,X\right)-DF_Y^{-1}\left(F(Y|1,X)|0,X\right)+(2D-1)Y\right)\\
		&&-\vartheta+\phi_1(W;y_1,y_2)+\phi_2(W;y_1,y_2)+\phi_3(y_1,y_2)
	\end{eqnarray*}
	The expressions of $\phi_1(\cdot)$, $\phi_2(\cdot)$, and $\phi_3(\cdot)$ are
	\begin{eqnarray*}
		\phi_1(w;y_1,y_2)&=&\frac{1}{P(y_1<Y<y_2)}\\
		&&\cdot\bigg(\frac{d}{P(D=1|X=x)}\\
		&&\cdot\int_{y_1}^{y_2}\frac{f_{DY}(0,t|x)}{f_Y\left(F_Y^{-1}\left(F_Y(t|0,x)|1,x\right)|1,x\right)} \left( F_Y(t|0,x)-1\{y\leq F_Y^{-1}\left(F_Y(t|0,x)|1,x\right)\}\right)dt\\
		&&+\frac{1-d}{1-P(D=1|X=x)}\int_{y_1}^{y_2}\frac{f_{DY}(0,t|x)}{f_Y\left(F_Y^{-1}\left(F_Y(t|0,x)|1,x\right)|1,x\right)} \left( 1\{y\leq t\}\right)-F_Y(t|0,x)dt\bigg)\\
		\phi_2(w;y_1,y_2)&=&-\frac{1}{P(y_1<Y<y_2)}\\
		&&\cdot\bigg(\frac{d}{P(D=1|X=x)}\int_{y_1}^{y_2}\frac{f_{DY}(1,t|x)}{f_Y\left(F_Y^{-1}\left(F_Y(t|1,x)|0,x\right)|0,x\right)} \left(1\{y\leq t\}-F_Y(t|1,x)\right)dt\\
		&&\frac{1-d}{1-P(D=1|X=x)}\\
		&&\cdot\int_{y_1}^{y_2}\frac{f_{DY}(1,t|x)}{f_Y\left(F_Y^{-1}\left(F_Y(t|1,x)|0,x\right)|0,x\right)} \left(F_Y(t|1,x)- 1\{y\leq F_Y^{-1}\left(F_Y(t|1,x)|0,x\right)\}\right)dt\bigg)\\
		\phi_3(y_1,y_2)&=&\frac{1}{P^2(y_1<Y<y_2)} E\bigg(1\{y_1<Y<y_2\}\big(DF_Y^{-1}\left(F_Y(Y|1,X)|0,X\right)\\
		&&+(1-D)F_Y^{-1}\left(F_Y(Y|0,X)|1,X\right)+(2D-1)Y\big)\bigg)\\
		&&\cdot\left(1\{y_1<Y<y_2\}-P\left(y_1<Y<y_2\right)\right)
	\end{eqnarray*}
	
	\noindent\textbf{Proposition A.2.} (\textit{Identification based on orthogonal score with binary treatment variable}) Under the same assumptions as Proposition A.1, we have
	(i) $\vartheta(y_{1},y_{2})$ satisfies $E\Psi\left(y_1,y_2,W;\vartheta(y_{1},y_{2}),\Phi\right)=0$.
	
	\noindent (ii) $\Psi\left(W,\theta,\eta;y_1,y_2\right)$ satisfies the Neyman orthogonality property:
	\begin{eqnarray*}
		\frac{\partial E\Psi\left(y_1,y_2,W;\theta(y_{1},y_{2}),\Phi+r(\tilde{\Phi}-\Phi)\right)}{\partial r}\big|_{r=0}=0.
	\end{eqnarray*}
	\noindent{\textbf{Proof.}} To simplify the notation, define
	\begin{eqnarray*}
		&&\Phi(y_1,y_2,y,x)\\
		&=&\bigg(F_Y(y|1,x),F_Y(y|0,x),\frac{f_{DY}(0,y|x)}{f_Y\left(F_Y^{-1}\left(F_Y(y|0,x)|1,x\right)|1,x\right)},\frac{f_{DY}(1,y|x)}{f_Y\left(F_Y^{-1}\left(F_Y(y|1,x)|0,x\right)|0,x\right)}\\
		&&,P(D=1|X=x),P\left(y_1<Y<y_2\right)\bigg)\\
		&=&\left(\Phi_1(x,y),\Phi_2(x,y),\Phi_3(x,y),\Phi_4(x,y),\Phi_5(x),\Phi_6(y_1,y_2)\right)
	\end{eqnarray*}
	Rewrite $\Psi(\cdot)$ in terms of $\Phi(\cdot)$
	\begin{eqnarray*}
		\Psi(W,\vartheta,\Phi;y_1,y_2)&=&\frac{1}{\Phi_6(y_1,y_2)}1\{y_1<Y<y_2\}\\
		&&\cdot\left((1-D)\Phi_1^{-1}\left(\Phi_2(Y,X),X\right)-D\Phi_2^{-1}\left(\Phi_1(Y,X),X\right)+(2D-1)Y\right)\\
		&&-\vartheta+\phi_1(W;y_1,y_2)+\phi_2(W;y_1,y_2)+\phi_3(y_1,y_2)
	\end{eqnarray*}
	where $\phi_1^{-1}(\cdot,x)$ and $\phi_2^{-1}(\cdot,x)$ denote the left-inverse transform of $\phi_1(y,x)$ and $\phi_2(y,x)$ with respect to $y$.
	
	Moreover, $\phi_1(\cdot)$, $\phi_2(\cdot)$, and $\phi_3(\cdot)$ can be written as
	\begin{eqnarray*}
		\phi_1(w;y_1,y_2)&=&\frac{1}{\Phi_6(y_1,y_2)}\bigg(\frac{d}{\Phi_5(x)}\int_{y_1}^{y_2}\Phi_3(t,x) \left(\Phi_2(t,x)- 1\{y\leq \Phi_1^{-1}\left(\Phi_2(t,x),x\right)\}\right)dt\\
		&&+\frac{1-d}{1-\Phi_5(x)}\int_{y_1}^{y_2}\Phi_3(t,x) \left(1\{y\leq t\}-\Phi_2(t,x)\right)dt\bigg)\\
		\phi_2(w;y_1,y_2)&=&-\frac{1}{\Phi_6(y_1,y_2)}\bigg(\frac{d}{\Phi_5(x)}\int_{y_1}^{y_2}\Phi_4(t,x) \left(1\{y\leq t\}-\Phi_1(t,x)\right)dt\\
		&&+\frac{1-d}{1-\Phi_5(x)}\int_{y_1}^{y_2}\Phi_4(t,x) \left(\Phi_1(t,x)- 1\{y\leq \Phi_2^{-1}\left(\Phi_1(t,x),x\right)\}\right)dt\bigg)\\
		\phi_3(y_1,y_2)&=&\frac{1}{\left(\Phi_6(y_1,y_2)\right)^2}\\
		&&\cdot E\left(1\{y_1<Y<y_2\}\left((1-D)\Phi_1^{-1}\left(\Phi_2(Y,X),X\right)-D\Phi_2^{-1}\left(\Phi_1(Y,X),X\right)+(2D-1)Y\right)\right)\\
		&&\cdot\left(1\{y_1<Y<y_2\}-\Phi_6(y_1,y_2)\right)
	\end{eqnarray*}
	
	Note that
	\begin{eqnarray*}
		&&E\left(D\int_{y_1}^{y_2}\Phi_3(t,X) \left( \Phi_2(t,X)-1\{Y\leq \Phi_1^{-1}\left(X,\Phi_2(t,X)\right)\}\right)dt|X\right)\\
		&=&E\left(D\int_{y_1}^{y_2}\Phi_3(t,X) \left(F_Y(t|0,X)- 1\{Y\leq F_Y^{-1}\left(F_Y(t|0,X)|1,X\right)\}\right)dt|X\right)\\
		&=&P(D=1|X)E\left(\int_{y_1}^{y_2}\Phi_3(t,X) \left(F_Y(t|0,X)- 1\{Y\leq F_Y^{-1}\left(F_Y(t|0,X)|1,X\right)\}\right)dt|D=1,X\right)\\
		&=&P(D=1|X)\left(\int_{y_1}^{y_2}\Phi_3(t,X) F_Y(t|0,X)dt-\int\int_{y_1}^{y_2}\Phi_3(t,X)  1\{y\leq F_Y^{-1}\left(F_Y(t|0,X)|1,X\right)\}dtf(y|1,X)dy\right)\\
		&=&P(D=1|X)\left(\int_{y_1}^{y_2}\Phi_3(t,X) F_Y(t|0,X)dt-\int_{y_1}^{y_2}\int\Phi_3(t,X)  1\{y\leq F_Y^{-1}\left(F_Y(t|0,X)|1,X\right)\}f(y|1,X)dydt\right)\\
		&=&P(D=1|X)\left(\int_{y_1}^{y_2}\Phi_3(t,X) F_Y(t|0,X)dt-\int_{y_1}^{y_2}\Phi_3(t,X)F_Y(t|0,X)dt\right)\\
		&=&0
	\end{eqnarray*}
	\begin{eqnarray*}
		&&E\left((1-D)\int_{y_1}^{y_2}\Phi_3(t,X) \left(1\{Y\leq t\}-\Phi_2(t,X)\right)dt|X\right)\\
		&=&E\left((1-D)\int_{y_1}^{y_2}\Phi_3(t,X) \left(1\{Y\leq t\}-F_Y(t|0,X)\right)dt|X\right)\\
		&=&\left(1-P(D=1|X)\right)E\left(\int_{y_1}^{y_2}\Phi_3(t,X) \left(1\{Y\leq t\}- F_Y(t|0,X)\right)dt|D=0,X\right)\\
		&=&\left(1-P(D=1|X)\right)\left(\int\int_{y_1}^{y_2}\Phi_3(t,X)  1\{y\leq t\}dtf(y|0,X)dy-\int_{y_1}^{y_2}\Phi_3(t,X) F_Y(t|0,X)dt\right)\\
		&=&\left(1-P(D=1|X)\right)\left(\int_{y_1}^{y_2}\int\Phi_3(t,X)  1\{y\leq t\}f(y|0,X)dydt-\int_{y_1}^{y_2}\Phi_3(t,X) F_Y(t|0,X)dt\right)\\
		&=&\left(1-P(D=1|X)\right)\left(\int_{y_1}^{y_2}\Phi_3(t,X) F_Y(t|0,X)dt-\int_{y_1}^{y_2}\Phi_3(t,X)F_Y(t|0,X)dt\right)\\
		&=&0
	\end{eqnarray*}
	Thus
	\begin{eqnarray}
		E\left(\phi_1(W;y_1,y_2)|X\right)=0
	\end{eqnarray}
	Similarly,
	\begin{eqnarray*}
		&&E\left(D\int_{y_1}^{y_2}\Phi_4(t,X) \left(1\{Y\leq t\}-\Phi_1(t,X)\right)dt|X\right)\\
		&=&E\left(D\int_{y_1}^{y_2}\Phi_4(t,X) \left(1\{Y\leq t\}-F_Y(t|1,X)-\right)dt|X\right)\\
		&=&P(D=1|X)E\left(\int_{y_1}^{y_2}\Phi_4(t,X) \left(1\{Y\leq t\}-F_Y(t|1,X)\right)dt|D=0,X\right)\\
		&=&P(D=1|X)\left(\int\int_{y_1}^{y_2}\Phi_4(t,X)  1\{y\leq t\}dtf(y|1,X)dy-\int_{y_1}^{y_2}\Phi_4(t,X) F_Y(t|1,X)dt\right)\\
		&=&P(D=1|X)\left(\int_{y_1}^{y_2}\int\Phi_4(t,X)  1\{y\leq t\}f(y|1,X)dydt-\int_{y_1}^{y_2}\Phi_4(t,X) F_Y(t|1,X)dt\right)\\
		&=&P(D=1|X)\left(\int_{y_1}^{y_2}\Phi_4(t,X) F_Y(t|1,X)dt-\int_{y_1}^{y_2}\Phi_4(t,X)F_Y(t|1,X)dt\right)\\
		&=&0
	\end{eqnarray*}
	\begin{eqnarray*}
		&&E\left((1-D)\int_{y_1}^{y_2}\Phi_4(t,X) \left(\Phi_1(t,X)- 1\{Y\leq \Phi_2^{-1}\left(X,\Phi_1(t,X)\right)\}\right)dt|X\right)\\
		&=&E\left((1-D)\int_{y_1}^{y_2}\Phi_4(t,X) \left(F_Y(t|1,X)- 1\{Y\leq F_Y^{-1}\left(F_Y(t|1,X)|0,X\right)\}\right)dt|X\right)\\
		&=&\left(1-P(D=1|X)\right)E\left(\int_{y_1}^{y_2}\Phi_4(t,X) \left(F_Y(t|1,X)- 1\{Y\leq F_Y^{-1}\left(F_Y(t|1,X)|0,X\right)\}\right)dt|D=0,X\right)\\
		&=&\left(1-P(D=1|X)\right)\\
		&&\cdot\left(\int_{y_1}^{y_2}\Phi_4(t,X) F_Y(t|1,X)dt-\int\int_{y_1}^{y_2}\Phi_4(t,X)  1\{y\leq F_Y^{-1}\left(F_Y(t|1,X)|0,X\right)\}dtf(y|0,X)dy\right)\\
		&=&\left(1-P(D=1|X)\right)\\
		&&\cdot\left(\int_{y_1}^{y_2}\Phi_4(t,X) F_Y(t|1,X)dt-\int_{y_1}^{y_2}\int\Phi_4(t,X)  1\{y\leq F_Y^{-1}\left(F_Y(t|1,X)|0,X\right)\}f(y|0,X)dydt\right)\\
		&=&\left(1-P(D=1|X)\right)\left(\int_{y_1}^{y_2}\Phi_4(t,X) F_Y(t|1,X)dt-\int_{y_1}^{y_2}\Phi_4(t,X)F_Y(t|1,X)dt\right)\\
		&=&0
	\end{eqnarray*}
	
	Thus
	\begin{eqnarray}
		E\left(\phi_2(W;y_1,y_2)|X\right)=0
	\end{eqnarray}
	From (12) and (16), the proof of (i) is apparent.
	
	Before starting the proof of (ii), we define some useful notations. For $k=1,2,\cdots,6$, $\Phi_{k,r}(\cdot)=\Phi_k(\cdot)+r\left(\tilde{\Phi}_k(\cdot)-\Phi_k(\cdot)\right)$ with $\tilde{\Phi}_k(\cdot)$ being an alternative nuisance function.
	
	Consider that $\Phi_1(y,x)$ is misspecified.
	\begin{eqnarray*}
		&&\frac{d}{dr}E\left(\Psi\left(W,\vartheta(y_1,y_2),\Phi_{1,r},\Phi_2,\Phi_3,\Phi_4,\Phi_5,\Phi_6;y_1,y_2\right)\right)\big|_{r=0}\\
		&=&\underbrace{\frac{1}{\Phi_6(y_1,y_2)}\frac{d}{dr}E\left(1\{y_1<Y<y_2\}(1-D)\Phi_{1,r}^{-1}\left(\Phi_2(Y,X),X\right)\right)\big|_{r=0}}_{\dag_1}\\
		&&\underbrace{-\frac{1}{\Phi_6(y_1,y_2)}\frac{d}{dr}E\left(\frac{D}{\Phi_5(X)}\int_{y_1}^{y_2}\Phi_3(t,X)1\{Y\leq \Phi_{1,r}^{-1}\left(\Phi_2(t,X),X\right)\}dt\right)\big|_{r=0}}_{\dag_2}\\
		&&\underbrace{-\frac{1}{\Phi_6(y_1,y_2)}\frac{d}{dr}E\left(1\{y_1<Y<y_2\}D\Phi_2^{-1}\left(\Phi_{1,r}(Y,X),X\right)\right)\big|_{r=0}}_{\dag_3}\\
		&&\underbrace{+\frac{1}{\Phi_6(y_1,y_2)}\frac{d}{dr}E\left(\frac{D}{\Phi_5(X)}\int_{y_1}^{y_2}\Phi_4(t,X)\Phi_{1,r}(t,X)dt\right)\big|_{r=0}}_{\dag_4}\\
		&&\underbrace{-\frac{1}{\Phi_6(y_1,y_2)}\frac{d}{dr}E\left(\frac{1-D}{1-\Phi_5(X)}\int_{y_1}^{y_2}\Phi_4(t,X)\Phi_{1,r}(t,X)dt\right)\big|_{r=0}}_{\dag_5}\\	&&\underbrace{+\frac{1}{\Phi_6(y_1,y_2)}\frac{d}{dr}E\left(\frac{1-D}{1-\Phi_5(X)}\int_{y_1}^{y_2}\Phi_4(t,X)1\{Y\leq \Phi_2^{-1}\left(\Phi_{1,r}(t,X),X\right)\}dt\right)\big|_{r=0}}_{\dag_6}\\
		&&+\frac{1}{\left(\Phi_6(y_1,y_2)\right)^2}\\
		&&\cdot \frac{d}{dr}E\left(1\{y_1<Y<y_2\}\left((1-D)\Phi_{1,r}^{-1}\left(\Phi_2(Y,X),X\right)-D\Phi_2^{-1}\left(\Phi_{1,r}(Y,X),X\right)+(2D-1)Y\right)\right)\big|_{r=0}\\
		&&\cdot \underbrace{E\left(\left(1\{y_1<Y<y_2\}-\Phi_6(y_1,y_2)\right)\right)}_0
	\end{eqnarray*}
	For $\dag_1$, we get
	\begin{eqnarray*}
		\dag_1&=&\frac{1}{\Phi_6(y_1,y_2)}\frac{d}{dr}E\left(1\{y_1<Y<y_2\}(1-D)\Phi_{1,r}^{-1}\left(\Phi_2(Y,X),X\right)\right)\big|_{r=0}\\
		&=&\frac{1}{\Phi_6(y_1,y_2)}\frac{d}{dr}E\left(1\{y_1<Y<y_2\}(1-D)F_{Y,r}^{-1}\left(F_Y(Y|0,X)|1,X\right)\right)\big|_{r=0}\\
		&=&\frac{1}{\Phi_6(y_1,y_2)}\frac{d}{dr}E\left(E\left(1\{y_1<Y<y_2\}(1-D)F_{Y,r}^{-1}\left(F_Y(Y|0,X)|1,X\right)|X\right)\right)\big|_{r=0}\\
		&=&\frac{1}{\Phi_6(y_1,y_2)}\frac{d}{dr}E\left(\frac{1}{1-P(D=1|X)}E\left(1\{y_1<Y<y_2\}F_{Y,r}^{-1}\left(F_Y(Y|0,X)|1,X\right)|D=0,X\right)\right)\big|_{r=0}\\
		&=&\frac{1}{\Phi_6(y_1,y_2)}\frac{d}{dr}E\left(\frac{1}{1-P(D=1|X)}\int 1\{y_1<y<y_2\}F_{Y,r}^{-1}\left(F_Y(y|0,X)|1,X\right)f_Y(y|0,X)dy\right)\big|_{r=0}\\
		&=&\frac{1}{\Phi_6(y_1,y_2)}\frac{d}{dr}E\left(\frac{1}{1-P(D=1|X)}\int_{y_1}^{y_2}F_{Y,r}^{-1}\left(F_Y(y|0,X)|1,X\right)f_Y(y|0,X)dy\right)\big|_{r=0}\\
		&=&\frac{1}{\Phi_6(y_1,y_2)}E\left(\int_{y_1}^{y_2}\frac{d}{dr}F_{Y,r}^{-1}\left(F_Y(y|0,X)|1,X\right)\big|_{r=0}f_{DY}(0,y|X)dy\right)
	\end{eqnarray*}
	For $\dag_2$, we get
	\begin{eqnarray*}
		\dag_2&=&-\frac{1}{\Phi_6(y_1,y_2)}\frac{d}{dr}E\left(\frac{D}{\Phi_5(X)}\int_{y_1}^{y_2}\Phi_3(t,X)1\{Y\leq \Phi_{1,r}^{-1}\left(\Phi_2(t,X),X\right)\}dt\right)\big|_{r=0}\\
		&=&-\frac{1}{\Phi_6(y_1,y_2)}\frac{d}{dr}E\left(\frac{D}{P(D=1|X)}\int_{y_1}^{y_2}\Phi_3(t,x)1\{Y\leq F_{Y,r}^{-1}\left(F_Y(t|0,X)|1,X\right)\}dt\right)\big|_{r=0}\\
		&=&-\frac{1}{\Phi_6(y_1,y_2)}\frac{d}{dr}E\left(\frac{1}{P(D=1|X)}E\left(D\int_{y_1}^{y_2}\Phi_3(t,x)1\{Y\leq F_{Y,r}^{-1}\left(F_Y(t|0,X)|1,X\right)\}dt|X\right)\right)\big|_{r=0}\\
		&=&-\frac{1}{\Phi_6(y_1,y_2)}\frac{d}{dr}E\left(E\left(\int_{y_1}^{y_2}\Phi_3(t,x)1\{Y\leq F_{Y,r}^{-1}\left(F_Y(t|0,X)|1,X\right)\}dt|D=1,X\right)\right)\big|_{r=0}\\
		&=&-\frac{1}{\Phi_6(y_1,y_2)}\frac{d}{dr}E\left(\int\int_{y_1}^{y_2}\Phi_3(t,x)1\{y\leq F_{Y,r}^{-1}\left(F_Y(t|0,X)|1,X\right)\}f_Y(y|1,x)dtdy\right)\big|_{r=0}\\
		&=&-\frac{1}{\Phi_6(y_1,y_2)}\frac{d}{dr}E\left(\int_{y_1}^{y_2}\int\Phi_3(t,x)1\{y\leq F_{Y,r}^{-1}\left(F_Y(t|0,X)|1,X\right)\}f_Y(y|1,x)dydt\right)\big|_{r=0}\\
		&=&-\frac{1}{\Phi_6(y_1,y_2)}E\left(\int_{y_1}^{y_2}\Phi_3(X,y)\frac{d}{dr}F_Y\left(F_{Y,r}^{-1}\left(F_Y(t|0,X)|1,X\right)|1,X\right)\big|_{r=0}dy\right)\\
		&=&-\frac{1}{\Phi_6(y_1,y_2)}E\left(\int_{y_1}^{y_2}\Phi_3(X,y)\frac{1}{f_Y\left(F_{Y,r}^{-1}\left(F_Y(y|0,X)|1,X\right)|1,X\right)}\frac{d}{dr}F_{Y,r}^{-1}\left(F_Y(t|0,X)|1,X\right)\big|_{r=0}dy\right)\\
		&=&-\frac{1}{\Phi_6(y_1,y_2)}E\left(\int_{y_1}^{y_2}\frac{d}{dr}F_{Y,r}^{-1}\left(F_Y(t|0,X)|1,X\right)\big|_{r=0}f_{DY}(0,y|X)dy\right)
	\end{eqnarray*}
	For $\dag_3$, we get
	\begin{eqnarray*}
		\dag_3&=&-\frac{1}{\Phi_6(y_1,y_2)}\frac{d}{dr}E\left(1\{y_1<Y<y_2\}D\Phi_2^{-1}\left(\Phi_{1,r}(Y,X),X\right)\right)\big|_{r=0}\\
		&=&-\frac{1}{\Phi_6(y_1,y_2)}\frac{d}{dr}E\left(1\{y_1<Y<y_2\}DF_Y^{-1}\left(F_{Y,r}(Y|1,X)|0,X\right)\right)\big|_{r=0}\\
		&=&-\frac{1}{\Phi_6(y_1,y_2)}\frac{d}{dr}E\left(E\left(1\{y_1<Y<y_2\}DF_Y^{-1}\left(F_{Y,r}(Y|1,X)|0,X\right)|X\right)\right)\big|_{r=0}\\
		&=&-\frac{1}{\Phi_6(y_1,y_2)}\frac{d}{dr}E\left(\frac{1}{P(D=1|X)}E\left(1\{y_1<Y<y_2\}F_Y^{-1}\left(F_{Y,r}(Y|1,X)|0,X\right)|D=1,X\right)\right)\big|_{r=0}\\
		&=&-\frac{1}{\Phi_6(y_1,y_2)}\frac{d}{dr}E\left(\frac{1}{P(D=1|X)}\int 1\{y_1<y<y_2\}F_Y^{-1}\left(F_{Y,r}(y|1,X)|0,X\right)f_Y(y|1,X)dy\right)\big|_{r=0}\\
		&=&-\frac{1}{\Phi_6(y_1,y_2)}\frac{d}{dr}E\left(\frac{1}{P(D=1|X)}\int_{y_1}^{y_2}F_Y^{-1}\left(F_{Y,r}(y|1,X)|0,X\right)f_Y(y|1,X)dy\right)\big|_{r=0}\\
		&=&-\frac{1}{\Phi_6(y_1,y_2)}E\left(\frac{1}{P(D=1|X)}\int_{y_1}^{y_2}\frac{d}{dr}F_Y^{-1}\left(F_{Y,r}(y|1,X)|0,X\right)\big|_{r=0}f_Y(y|1,X)dy\right)\\
		&=&-\frac{1}{\Phi_6(y_1,y_2)}E\left(\int_{y_1}^{y_2}\frac{d}{dr}F_Y^{-1}\left(F_{Y,r}(y|1,X)|0,X\right)\big|_{r=0}f_{DY}(1,y|X)dy\right)
	\end{eqnarray*}
	The following identity holds
	\begin{eqnarray*}
		F_Y\left(F_Y^{-1}(\tau|0,x)|0,x\right)=\tau\notag
	\end{eqnarray*}
	Taking derivative with respect to $\tau$ gives
	\begin{eqnarray*}
		f_Y\left(F^{-1}_Y(\tau|0,x)|0,x\right)\frac{d}{d\tau}F_Y^{-1}(\tau|0,x)=1
	\end{eqnarray*}
	By simple algebra
	\begin{eqnarray*}
		\frac{d}{d\tau}F_Y^{-1}(\tau|0,x)=\frac{1}{f_Y\left(F^{-1}_Y(\tau|0,x)|0,x\right)}
	\end{eqnarray*}
	Thus
	\begin{eqnarray*}
		\frac{d}{dr}F_Y^{-1}\left(F_{Y,r}(y|1,X)|0,X\right)\big|_{r=0}=\frac{1}{f_Y\left(F_Y^{-1}\left(F_{Y,r}(y|1,X)|0,X\right)|0,X\right)}\frac{d}{dr}F_{Y,r}(y|1,X)\big|_{r=0}
	\end{eqnarray*}
	We get
	\begin{eqnarray*}
		\dag_3&=&-\frac{1}{\Phi_6(y_1,y_2)}E\left(\int_{y_1}^{y_2}\frac{1}{f\left(F_Y^{-1}\left(F_{Y,r}(y|1,X)|0,X\right)|0,X\right)}\frac{d}{dr}F_{Y,r}(y|1,X)\big|_{r=0}f_{DY}(1,y|X)dy\right)\\
		&=&-\frac{1}{\Phi_6(y_1,y_2)}E\left(\int_{y_1}^{y_2}\Phi_4(X,y)\frac{d}{dr}F_{Y,r}(y|1,X)\big|_{r=0}dy\right)
	\end{eqnarray*}
	For $\dag_4$, we get
	\begin{eqnarray*}
		\dag_4&=&\frac{1}{\Phi_6(y_1,y_2)}\frac{d}{dr}E\left(\frac{D}{\Phi_5(X)}\int_{y_1}^{y_2}\Phi_4(t,X)\Phi_{1,r}(t,X)dt\right)\big|_{r=0}\\
		&=&\frac{1}{\Phi_6(y_1,y_2)}\frac{d}{dr}E\left(\frac{D}{P(D=1|X)}\int_{y_1}^{y_2}\Phi_4(t,X)F_{Y,r}(t|1,X)dt\right)\big|_{r=0}\\
		&=&\frac{1}{\Phi_6(y_1,y_2)}\frac{d}{dr}E\left(\frac{1}{P(D=1|X)}E\left(D\int_{y_1}^{y_2}\Phi_4(t,X)F_{Y,r}(t|1,X)dt|X\right)\right)\big|_{r=0}\\
		&=&\frac{1}{\Phi_6(y_1,y_2)}\frac{d}{dr}E\left(E\left(\int_{y_1}^{y_2}\Phi_4(t,X)F_{Y,r}(t|1,X)dt|D=1,X\right)\right)\big|_{r=0}\\
		&=&\frac{1}{\Phi_6(y_1,y_2)}E\left(\int_{y_1}^{y_2}\Phi_4(X,y)\frac{d}{dr}F_{Y,r}(y|1,X)\big|_{r=0}dy\right)
	\end{eqnarray*}
	For $\dag_5$, we get
	\begin{eqnarray*}
		\dag_5&=&-\frac{1}{\Phi_6(y_1,y_2)}\frac{d}{dr}E\left(\frac{1-D}{1-\Phi_5(X)}\int_{y_1}^{y_2}\Phi_4(t,X)\Phi_{1,r}(t,X)dt\right)\big|_{r=0}\\
		&=&-\frac{1}{\Phi_6(y_1,y_2)}\frac{d}{dr}E\left(\frac{1-D}{1-P(D=1|X)}\int_{y_1}^{y_2}\Phi_4(t,X)F_{Y,r}(t|1,X)dt\right)\big|_{r=0}\\
		&=&-\frac{1}{\Phi_6(y_1,y_2)}\frac{d}{dr}E\left(\frac{1}{1-P(D=1|X)}E\left((1-D)\int_{y_1}^{y_2}\Phi_4(t,X)F_{Y,r}(t|1,X)dt|X\right)\right)\big|_{r=0}\\
		&=&-\frac{1}{\Phi_6(y_1,y_2)}\frac{d}{dr}E\left(E\left(\int_{y_1}^{y_2}\Phi_4(t,X)F_{Y,r}(t|1,X)dt|D=0,X\right)\right)\big|_{r=0}\\
		&=&-\frac{1}{\Phi_6(y_1,y_2)}E\left(\int_{y_1}^{y_2}\Phi_4(t,X)\frac{d}{dr}F_{Y,r}(t|1,X)\big|_{r=0}dt\right)\\
	\end{eqnarray*}
	For $\dag_6$, we get
	\begin{eqnarray*}
		\dag_6&=&\frac{1}{\Phi_6(y_1,y_2)}\frac{d}{dr}E\left(\frac{1-D}{1-\Phi_5(X)}\int_{y_1}^{y_2}\Phi_4(t,X)1\{Y\leq \Phi_2^{-1}\left(\Phi_{1,r}(t,X),X\right)\}dt\right)\big|_{r=0}\\
		&=&\frac{1}{\Phi_6(y_1,y_2)}\frac{d}{dr}E\left(\frac{1}{1-P(D=1|X)}E\left((1-D)\int_{y_1}^{y_2}\Phi_4(t,X)1\{Y\leq F_Y^{-1}\left(F_{Y,r}(t|1,X)|0,X\right)\}dt|X\right)\right)\big|_{r=0}\\
		&=&\frac{1}{\Phi_6(y_1,y_2)}\frac{d}{dr}E\left(E\left(\int_{y_1}^{y_2}\Phi_4(t,X)1\{Y\leq F_Y^{-1}\left(F_{Y,r}(t|1,X)|0,X\right)\}dt|D=0,X\right)\right)\big|_{r=0}\\
		&=&\frac{1}{\Phi_6(y_1,y_2)}\frac{d}{dr}E\left(\int\int_{y_1}^{y_2}\Phi_4(t,X)1\{y\leq F_Y^{-1}\left(F_{Y,r}(t|1,X)|0,X\right)\}f_Y(y|0,X)dtdy\right)\big|_{r=0}\\
		&=&\frac{1}{\Phi_6(y_1,y_2)}E\left(\int_{y_1}^{y_2}\Phi_4(t,X)\frac{d}{dr}F_{Y,r}(t|1,X)\big|_{r=0}dt\right)
	\end{eqnarray*}
	Then
	\begin{eqnarray*}
		\frac{d}{dr}E\left(\Psi\left(W,\vartheta(y_1,y_2),\Phi_{1,r},\Phi_2,\Phi_3,\Phi_4,\Phi_5,\Phi_6;y_1,y_2\right)\right)\big|_{r=0}=0
	\end{eqnarray*}
	Consider that $\Phi_2(y,x)$ is misspecified.
	\begin{eqnarray*}
		&&\frac{d}{dr}E\left(\Psi\left(W,\vartheta(y_1,y_2),\Phi_1,\Phi_{2,r},\Phi_3,\Phi_4,\Phi_5,\Phi_6;y_1,y_2\right)\right)\big|_{r=0}\\
		&=&\underbrace{\frac{1}{\Phi_6(y_1,y_2)}\frac{d}{dr}E\left(1\{y_1<Y<y_2\}(1-D)\Phi_1^{-1}\left(\Phi_{2,r}(Y,X),X\right)\right)\big|_{r=0}}_{\ddagger_1}\\
		&&\underbrace{+\frac{1}{\Phi_6(y_1,y_2)}\frac{d}{dr}E\left(\frac{D}{\Phi_5(X)}\int_{y_1}^{y_2}\Phi_3(t,X)\Phi_{2,r}(t,X)dt\right)\big|_{r=0}}_{\ddagger_2}\\
		&&\underbrace{-\frac{1}{\Phi_6(y_1,y_2)}\frac{d}{dr}E\left(\frac{D}{\Phi_5(X)}\int_{y_1}^{y_2}\Phi_3(t,X)1\{y\leq \Phi_1^{-1}\left(\Phi_{2,r}(t,X),X\right)\}dt\right)\big|_{r=0}}_{\ddagger_3}\\
		&&\underbrace{-\frac{1}{\Phi_6(y_1,y_2)}\frac{d}{dr}E\left(\frac{1-D}{1-\Phi_5(X)}\int_{y_1}^{y_2}\Phi_3(t,X)\Phi_{2,r}(t,X)dt\right)\big|_{r=0}}_{\ddagger_4}\\
		&&\underbrace{-\frac{1}{\Phi_6(y_1,y_2)}\frac{d}{dr}E\left(1\{y_1<Y<y_2\}D\Phi_{2,r}^{-1}\left(\Phi_1(Y,X),X\right)\right)\big|_{r=0}}_{\ddagger_5}\\
		&&\underbrace{+\frac{1}{\Phi_6(y_1,y_2)}\frac{d}{dr}E\left(\frac{1-D}{1-\Phi_5(X)}\int_{y_1}^{y_2}\Phi_4(t,X)1\{Y\leq \Phi_{2,r}^{-1}\left(\Phi_1(t,X),X\right)\}dt\right)\big|_{r=0}}_{\ddagger_6}\\
		&&+\frac{1}{\left(\Phi_6(y_1,y_2)\right)^2}\\
		&&\cdot \frac{d}{dr}E\left(1\{y_1<Y<y_2\}\left((1-D)\Phi_1^{-1}\left(\Phi_{2,r}(Y,X),X\right)-D\Phi_{2,r}^{-1}\left(\Phi_1(Y,X),X\right)+(2D-1)Y\right)\right)\big|_{r=0}\\
		&&\cdot \underbrace{E\left(\left(1\{y_1<Y<y_2\}-\Phi_6(y_1,y_2)\right)\right)}_0
	\end{eqnarray*}
	For $\ddagger_1$, we get
	\begin{eqnarray*}
		\ddagger_1&=&\frac{1}{\Phi_6(y_1,y_2)}\frac{d}{dr}E\left(1\{y_1<Y<y_2\}(1-D)\Phi_1^{-1}\left(\Phi_{2,r}(Y,X),X\right)\right)\big|_{r=0}\\
		&=&\frac{1}{\Phi_6(y_1,y_2)}\frac{d}{dr}E\left(1\{y_1<Y<y_2\}(1-D)F_Y^{-1}\left(F_{Y,r}(Y|0,X)|1,X\right)\right)\big|_{r=0}\\
		&=&\frac{1}{\Phi_6(y_1,y_2)}\frac{d}{dr}E\left(E\left(1\{y_1<Y<y_2\}(1-D)F_Y^{-1}\left(F_{Y,r}(Y|0,X)|1,X\right)|X\right)\right)\big|_{r=0}\\
		&=&\frac{1}{\Phi_6(y_1,y_2)}\frac{d}{dr}E\left(\frac{1}{1-P(D=1|X)}E\left(1\{y_1<Y<y_2\}F_Y^{-1}\left(F_{Y,r}(Y|0,X)|1,X\right)|D=0,X\right)\right)\big|_{r=0}\\
		&=&\frac{1}{\Phi_6(y_1,y_2)}\frac{d}{dr}E\left(\frac{1}{1-P(D=1|X)}\int 1\{y_1<y<y_2\}F_Y^{-1}\left(F_{Y,r}(y|0,X)|1,X\right)f_Y(y|0,X)dy\right)\big|_{r=0}\\
		&=&\frac{1}{\Phi_6(y_1,y_2)}\frac{d}{dr}E\left(\frac{1}{1-P(D=1|X)}\int_{y_1}^{y_2}F_Y^{-1}\left(F_{Y,r}(y|0,X)|1,X\right)f_Y(y|0,X)dy\right)\big|_{r=0}\\
		&=&\frac{1}{\Phi_6(y_1,y_2)}E\left(\int_{y_1}^{y_2}\frac{d}{dr}F_Y^{-1}\left(F_{Y,r}(y|0,X)|1,X\right)\big|_{r=0}f_{DY}(0,y|X)dy\right)
	\end{eqnarray*}
	Similar to the simplification of $\dag_3$, we get
	\begin{eqnarray*}
		\ddagger_1&=&\frac{1}{\Phi_6(y_1,y_2)}E\left(\int_{y_1}^{y_2}\frac{1}{f\left(F_Y\left(F_Y(y|0,X)|1,X\right)|1,X\right)}\frac{d}{dr}F_{Y,r}(y|0,X)\big|_{r=0}f_{DY}(0,y|X)dy\right)\\
		&=&\frac{1}{\Phi_6(y_1,y_2)}E\left(\int_{y_1}^{y_2}\Phi_3(t,X)\frac{d}{dr}F_{Y,r}(t|0,X)\big|_{r=0}dt\right)
	\end{eqnarray*}
	For $\ddagger_2$, we get
	\begin{eqnarray*}
		\ddagger_2&=&\frac{1}{\Phi_6(y_1,y_2)}\frac{d}{dr}E\left(\frac{D}{\Phi_5(X)}\int_{y_1}^{y_2}\Phi_3(t,X)\Phi_{2,r}(t,X)dt\right)\big|_{r=0}\\
		&=&\frac{1}{\Phi_6(y_1,y_2)}\frac{d}{dr}E\left(\frac{D}{P(D=1|X)}\int_{y_1}^{y_2}\Phi_3(t,X)F_{Y,r}(t|0,X)dt\right)\big|_{r=0}\\
		&=&\frac{1}{\Phi_6(y_1,y_2)}\frac{d}{dr}E\left(\frac{1}{P(D=1|X)}E\left(D\int_{y_1}^{y_2}\Phi_3(t,X)F_{Y,r}(t|0,X)dt|X\right)\right)\big|_{r=0}\\
		&=&\frac{1}{\Phi_6(y_1,y_2)}\frac{d}{dr}E\left(E\left(\int_{y_1}^{y_2}\Phi_3(t,X)F_{Y,r}(t|0,X)dt|D=1,X\right)\right)\big|_{r=0}\\
		&=&\frac{1}{\Phi_6(y_1,y_2)}E\left(\int_{y_1}^{y_2}\Phi_3(t,X)\frac{d}{dr}F_{Y,r}(t|0,X)\big|_{r=0}dt\right)
	\end{eqnarray*}
	For $\ddagger_3$, we get
	\begin{eqnarray*}
		\ddagger_3&=&-\frac{1}{\Phi_6(y_1,y_2)}\frac{d}{dr}E\left(\frac{D}{\Phi_5(X)}\int_{y_1}^{y_2}\Phi_3(t,X)1\{Y\leq \Phi_1^{-1}\left(\Phi_{2,r}(t,X),X\right)\}dt\right)\big|_{r=0}\\
		&=&-\frac{1}{\Phi_6(y_1,y_2)}\frac{d}{dr}E\left(\frac{D}{P(D=1|X)}\int_{y_1}^{y_2}\Phi_3(t,x)1\{Y\leq F_Y^{-1}\left(F_{Y,r}(t|0,X)|1,X\right)\}dt\right)\big|_{r=0}\\
		&=&-\frac{1}{\Phi_6(y_1,y_2)}\frac{d}{dr}E\left(\frac{1}{P(D=1|X)}E\left(D\int_{y_1}^{y_2}\Phi_3(t,x)1\{Y\leq F_Y^{-1}\left(F_{Y,r}(t|0,X)|1,X\right)\}dt|X\right)\right)\big|_{r=0}\\
		&=&-\frac{1}{\Phi_6(y_1,y_2)}\frac{d}{dr}E\left(E\left(\int_{y_1}^{y_2}\Phi_3(t,x)1\{Y\leq F_Y^{-1}\left(F_{Y,r}(t|0,X)|1,X\right)\}dt|D=1,X\right)\right)\big|_{r=0}\\
		&=&-\frac{1}{\Phi_6(y_1,y_2)}\frac{d}{dr}E\left(\int\int_{y_1}^{y_2}\Phi_3(t,x)1\{y\leq F_Y^{-1}\left(F_{Y,r}(t|0,X)|1,X\right)\}f_Y(y|1,x)dtdy\right)\big|_{r=0}\\
		&=&-\frac{1}{\Phi_6(y_1,y_2)}\frac{d}{dr}E\left(\int_{y_1}^{y_2}\int\Phi_3(t,x)1\{y\leq F_Y^{-1}\left(F_{Y,r}(t|0,X)|1,X\right)\}f_Y(y|1,x)dydt\right)\big|_{r=0}\\
		&=&-\frac{1}{\Phi_6(y_1,y_2)}E\left(\int_{y_1}^{y_2}\Phi_3(t,x)\frac{d}{dr}F_{Y,r}(t|0,X)\big|_{r=0}dt\right)
	\end{eqnarray*}
	For $\ddagger_4$, we get
	\begin{eqnarray*}
		\ddagger_4&=&-\frac{1}{\Phi_6(y_1,y_2)}\frac{d}{dr}E\left(\frac{1-D}{1-\Phi_5(X)}\int_{y_1}^{y_2}\Phi_3(t,X)\Phi_{2,r}(t,X)dt\right)\big|_{r=0}\\
		&=&-\frac{1}{\Phi_6(y_1,y_2)}\frac{d}{dr}E\left(\frac{1-D}{1-P(D=1|X)}\int_{y_1}^{y_2}\Phi_3(t,X)F_{Y,r}(t|0,X)dt\right)\big|_{r=0}\\
		&=&-\frac{1}{\Phi_6(y_1,y_2)}\frac{d}{dr}E\left(\frac{1}{1-P(D=1|X)}E\left((1-D)\int_{y_1}^{y_2}\Phi_3(t,X)F_{Y,r}(t|0,X)dt|X\right)\right)\big|_{r=0}\\
		&=&-\frac{1}{\Phi_6(y_1,y_2)}\frac{d}{dr}E\left(E\left(\int_{y_1}^{y_2}\Phi_3(t,X)F_{Y,r}(t|0,X)dt|D=1,X\right)\right)\big|_{r=0}\\
		&=&-\frac{1}{\Phi_6(y_1,y_2)}E\left(\int_{y_1}^{y_2}\Phi_3(t,X)\frac{d}{dr}F_{Y,r}(t|0,X)\big|_{r=0}dt\right)
	\end{eqnarray*}
	For $\ddagger_5$, we get
	\begin{eqnarray*}
		\ddagger_5&=&-\frac{1}{\Phi_6(y_1,y_2)}\frac{d}{dr}E\left(1\{y_1<Y<y_2\}D\Phi_{2,r}^{-1}\left(\Phi_1(Y,X),X\right)\right)\big|_{r=0}\\
		&=&-\frac{1}{\Phi_6(y_1,y_2)}\frac{d}{dr}E\left(1\{y_1<Y<y_2\}DF_{Y,r}^{-1}\left(F_Y(Y|1,X)|0,X\right)\right)\big|_{r=0}\\
		&=&-\frac{1}{\Phi_6(y_1,y_2)}\frac{d}{dr}E\left(E\left(1\{y_1<Y<y_2\}DF_{Y,r}^{-1}\left(F_Y(Y|1,X)|0,X\right)|X\right)\right)\big|_{r=0}\\
		&=&-\frac{1}{\Phi_6(y_1,y_2)}\frac{d}{dr}E\left(\frac{1}{P(D=1|X)}E\left(1\{y_1<Y<y_2\}F_{Y,r}^{-1}\left(F_Y(Y|1,X)|0,X\right)|D=1,X\right)\right)\big|_{r=0}\\
		&=&-\frac{1}{\Phi_6(y_1,y_2)}\frac{d}{dr}E\left(\frac{1}{P(D=1|X)}\int 1\{y_1<y<y_2\}F_{Y,r}^{-1}\left(F_Y(y|1,X)|0,X\right)f_Y(y|1,X)dy\right)\big|_{r=0}\\
		&=&-\frac{1}{\Phi_6(y_1,y_2)}\frac{d}{dr}E\left(\frac{1}{P(D=1|X)}\int_{y_1}^{y_2}F_{Y,r}^{-1}\left(F_Y(y|1,X)|0,X\right)f_Y(y|1,X)dy\right)\big|_{r=0}\\
		&=&-\frac{1}{\Phi_6(y_1,y_2)}E\left(\frac{1}{P(D=1|X)}\int_{y_1}^{y_2}\frac{d}{dr}F_{Y,r}^{-1}\left(F_Y(y|1,X)|0,X\right)\big|_{r=0}f_Y(y|1,X)dy\right)\\
		&=&-\frac{1}{\Phi_6(y_1,y_2)}E\left(\int_{y_1}^{y_2}\frac{d}{dr}F_{Y,r}^{-1}\left(F_Y(y|1,X)|0,X\right)\big|_{r=0}f_{DY}(1,y|X)dy\right)
	\end{eqnarray*}
	
	For $\ddagger_6$, we get
	\begin{eqnarray*}
		\ddagger_6&=&\frac{1}{\Phi_6(y_1,y_2)}\frac{d}{dr}E\left(\frac{1-D}{1-\Phi_5(X)}\int_{y_1}^{y_2}\Phi_4(t,X)1\{Y\leq \Phi_{2,r}^{-1}\left(\Phi_1(t,x),x\right)\}dt\right)\big|_{r=0}\\
		&=&\frac{1}{\Phi_6(y_1,y_2)}\frac{d}{dr}E\left(\frac{1}{1-P(D=1|X)}E\left((1-D)\int_{y_1}^{y_2}\Phi_4(t,X)1\{Y\leq F_{Y,r}^{-1}\left(F_Y(t|1,X)|0,X\right)\}dt|X\right)\right)\big|_{r=0}\\
		&=&\frac{1}{\Phi_6(y_1,y_2)}\frac{d}{dr}E\left(E\left(\int_{y_1}^{y_2}\Phi_4(t,X)1\{Y\leq F_{Y,r}^{-1}\left(F_Y(t|1,X)|0,X\right)\}dt|D=0,X\right)\right)\big|_{r=0}\\
		&=&\frac{1}{\Phi_6(y_1,y_2)}\frac{d}{dr}E\left(\int\int_{y_1}^{y_2}\Phi_4(t,X)1\{y\leq F_{Y,r}^{-1}\left(F_Y(t|1,X)|0,X\right)\}f_Y(y|0,X)dtdy\right)\big|_{r=0}\\
		&=&\frac{1}{\Phi_6(y_1,y_2)}E\left(\int_{y_1}^{y_2}\Phi_4(t,X)\frac{d}{dr}F_Y\left(F_{Y,r}^{-1}\left(F_Y(t|1,X)|0,X\right)|0,X\right)\big|_{r=0}dt\right)\\
		&=&\frac{1}{\Phi_6(y_1,y_2)}E\left(\int_{y_1}^{y_2}\Phi_4(t,X)\frac{1}{f_Y\left(F_{Y,r}^{-1}\left(F_Y(t|1,X)|0,X\right)|0,X\right)}\frac{d}{dr}F_{Y,r}^{-1}\left(F_Y(t|1,X)|0,X\right)\big|_{r=0}dt\right)\\
		&=&\frac{1}{\Phi_6(y_1,y_2)}E\left(\int_{y_1}^{y_2}\frac{d}{dr}F_{Y,r}^{-1}\left(F_Y(y|1,X)|0,X\right)\big|_{r=0}f_{DY}(1,y|X)dy\right)
	\end{eqnarray*}
	Then
	\begin{eqnarray*}
		\frac{d}{dr}E\left(\Psi\left(W,\vartheta(y_1,y_2),\Phi_1,\Phi_{2,r},\Phi_3,\Phi_4,\Phi_5,\Phi_6;y_1,y_2\right)\right)\big|_{r=0}=0
	\end{eqnarray*}
	From (15)--(16), we get
	\begin{eqnarray*}
		&&\frac{d}{dr}E\left(\Psi\left(W,\vartheta(y_1,y_2),\Phi_1,\Phi_2,\Phi_{3,r},\Phi_4,\Phi_5,\Phi_6;y_1,y_2\right)\right)\big|_{r=0}=0\\
		&&\frac{d}{dr}E\left(\Psi\left(W,\vartheta(y_1,y_2),\Phi_1,\Phi_2,\Phi_3,\Phi_{4,r},\Phi_5,\Phi_6;y_1,y_2\right)\right)\big|_{r=0}=0\\
		&&\frac{d}{dr}E\left(\Psi\left(W,\vartheta(y_1,y_2),\Phi_1,\Phi_2,\Phi_3,\Phi_4,\Phi_{5,r},\Phi_6;y_1,y_2\right)\right)\big|_{r=0}=0
	\end{eqnarray*}
	Moreover, by (15) and (16), we get
	\begin{eqnarray*}
		&&\frac{d}{dr}E\left(\Psi\left(W,\vartheta(y_1,y_2),\Phi_3,\Phi_2,\Phi_3,\Phi_4,\Phi_5,\Phi_{6,r};y_1,y_2\right)\right)\big|_{r=0}\\
		&=&\frac{1}{\left(\Phi_6(y_1,y_2)\right)^2}\frac{d}{dr}\Phi_{6,r}(y_1,y_2)\big|_{r=0}\\
		&&\cdot E\left(1\{y_1<Y<y_2\}\left((1-D)\Phi_1^{-1}\left(\Phi_2(Y,X),X\right)-D\Phi_2^{-1}\left(\Phi_1(Y,X),X\right)+(2D-1)Y\right)\right)\\
		&&-\frac{2}{\left(\Phi_6(y_1,y_2)\right)^3}\frac{d}{dr}\Phi_{6,r}(y_1,y_2)\big|_{r=0}\\
		&&\cdot E\left(1\{y_1<Y<y_2\}\left((1-D)\Phi_1^{-1}\left(\Phi_2(Y,X),X\right)-D\Phi_2^{-1}\left(\Phi_1(Y,X),X\right)+(2D-1)Y\right)\right)\\
		&&\cdot\underbrace{E\left(1\{y_1<Y<y_2\}-\Phi_6(y_1,y_2)\right)}_0\\
		&&-\frac{1}{\left(\Phi_6(y_1,y_2)\right)^2}\frac{d}{dr}\Phi_{6,r}(y_1,y_2)\big|_{r=0}\\
		&&\cdot E\left(1\{y_1<Y<y_2\}\left((1-D)\Phi_1^{-1}\left(\Phi_2(Y,X),X\right)-D\Phi_2^{-1}\left(\Phi_1(Y,X),X\right)+(2D-1)Y\right)\right)\\
		&=&0
	\end{eqnarray*}
	Then, we can conclude that
	\begin{eqnarray*}
		\frac{\partial E\Psi\left(W,\theta(y_{1},y_{2}),\Phi+r(\tilde{\Phi}-\Phi);y_1,y_2\right)}{\partial r}\big|_{r=0}=0.
	\end{eqnarray*}
	$\blacksquare$\\

	\noindent\textbf{Proof of Theorem 4.1.} In the proof main theorems, $a\lesssim{b}$ means that $a\leq{Ab}$, where the constant $A$ depends on the constants in Assumptions 4.1-4.3 only, but not on $n$. We suppress the claim ``uniformly over $u\in\mathcal{U}$" throughout the proof.
	
	\noindent\textbf{Step 1. (Linearization)} In this step, we establish the claim that the pre-estimator has no first order effects, namely
	\[
	\sqrt{n}\left(\widehat{\theta}(u)-\theta(u)\right)=Z_{n}(u)+o_{p}(1) \quad in \quad \mathbb{D}=\ell^{\infty}\left(\mathcal{U}\right),
	\]
	where $Z_{n}(u)=\mathbb{G}_{n}\psi(W,\theta,\eta;u)$.
	
	Define the following spaces of functions:
	\[
	\mathcal{F}_{1}=\left\{
	\begin{array}{l}
		(d,x,u)\mapsto\displaystyle\int_{y_{1}}^{y_{2}}\Lambda\bigg(b(d,x)^{\prime}\widetilde{\beta}(y)\bigg)dy: \left\Vert\widetilde{\beta}(y)\right\Vert_{0}\leq{Cs} \\
		\left\Vert \displaystyle\int_{y_{1}}^{y_{2}}\Lambda\bigg(b(D,X)^{\prime}\widetilde{\beta}(y)\bigg)dy-\displaystyle\int_{y_{1}}^{y_{2}}F_{Y}(y|D,X)dy\right\Vert_{P,2}=o_{p}\left(n^{-1/4}\right) \\
		\left\Vert \displaystyle\int_{y_{1}}^{y_{2}}\Lambda\bigg(b(D,X)^{\prime}\widetilde{\beta}(y)\bigg)dy-\displaystyle\int_{y_{1}}^{y_{2}}F_{Y}(y|D,X)dy\right\Vert_{P,\infty}=o_{p}(1) \\
	\end{array}
	\right\},
	\]
	\[
	\mathcal{F}_{2}=\left\{
	\begin{array}{l}
		(d,x,u)\mapsto \sum\limits_{l=1}^{\ell}\eta_{l}\left(\widetilde{F}(u,d+lh_{n},x)-\widetilde{F}(u,d-lh_{n},x)\right)\big/2h_{n}: \widetilde{F}\in\mathcal{F}_{1} \\
		\left\Vert \sum\limits_{l=1}^{\ell}\eta_{l}\left(\widetilde{F}(u,d+lh_{n},x)-\widetilde{F}(u,d-lh_{n},x)\right)\big/2h_{n}-\partial_{D}\displaystyle\int_{y_{1}}^{y_{2}}F_{Y}(y|D,X)dy\right\Vert_{P,2}=o_{p}\left(n^{-1/4}\right) \\
		\left\Vert \sum\limits_{l=1}^{\ell}\eta_{l}\left(\widetilde{F}(u,d+lh_{n},x)-\widetilde{F}(u,d-lh_{n},x)\right)\big/2h_{n}-\partial_{D}\displaystyle\int_{y_{1}}^{y_{2}}F_{Y}(y|D,X)dy\right\Vert_{P,\infty}=o_{p}(1) \\
	\end{array}
	\right\},
	\]
	\[
	\mathcal{F}_{3}=\left\{
	\begin{array}{l}
		u\mapsto \mathbb{E}_{n}\widetilde{F}(u,D_{i},X_{i}): \widetilde{F}\in\mathcal{F}_{2} \\
		\left\Vert \mathbb{E}_{n}\widetilde{F}(u,D_{i},X_{i})-E\left[\partial_{D}\displaystyle\int_{y_{1}}^{y_{2}}F_{Y}(y|D,X)dy\right]\right\Vert_{P,2}=o_{p}\left(n^{-1/4}\right) \\
		\left\Vert \mathbb{E}_{n}\widetilde{F}(u,D_{i},X_{i})-E\left[\partial_{D}\displaystyle\int_{y_{1}}^{y_{2}}F_{Y}(y|D,X)dy\right]\right\Vert_{P,\infty}=o_{p}(1) \\
	\end{array}
	\right\},
	\]
	\[
	\mathcal{L}=\left\{
	\begin{array}{l}
		(d,x)\mapsto b(d,x)^{\prime}\widetilde{\gamma}: \left\Vert\widetilde{\gamma}\right\Vert_{0}\leq{Cs} \\
		\left\Vert b(D,X)^{\prime}\widetilde{\gamma}-L(D,X)\right\Vert_{P,2}=o_{p}\left(n^{-1/4}\right) \\
		\left\Vert b(D,X)^{\prime}\widetilde{\gamma}-L(D,X)\right\Vert_{P,\infty}=o_{p}(1) \\
	\end{array}
	\right\}.
	\]
	We observe that with probability no less than $1-\Delta_{n}$,
	\[
	\widehat{IF}(u,d,x)\in\mathcal{F}_{1}, \quad \widehat{DIF}(u,d,x)\in\mathcal{F}_{2}, \quad \mathbb{E}_{n}\widehat{DIF}(u,D_{i},X_{i})\in\mathcal{F}_{3} \quad \text{and} \quad \widehat{L}(d,x)\in\mathcal{L}.
	\]
	To see this, note that under Assumption 4.2 and 4.3,
	\[\begin{aligned}
		&\left\Vert \displaystyle\int_{y_{1}}^{y_{2}}\Lambda\bigg(b(D,X)^{\prime}\widetilde{\beta}(y)\bigg)dy-\displaystyle\int_{y_{1}}^{y_{2}}F_{Y}(y|D,X)dy\right\Vert_{P,2}\\
		\leq&\displaystyle\int_{y_{1}}^{y_{2}}\left\Vert \Lambda\bigg(b(D,X)^{\prime}\widetilde{\beta}(y)\bigg)-F_{Y}(y|D,X)\right\Vert_{P,2}dy\\
		\leq&\displaystyle\int_{y_{1}}^{y_{2}}\left[\left\Vert \Lambda\bigg(b(D,X)^{\prime}\widetilde{\beta}(y)\bigg)-\Lambda\bigg(b(D,X)^{\prime}\beta(y)\bigg)\right\Vert_{P,2}+\left\Vert r_{F}(y,D,X)\right\Vert_{P,2}\right]dy\\
		\lesssim&\displaystyle\int_{y_{1}}^{y_{2}}\left[\Vert\partial\Lambda\Vert_{\infty}\left\Vert b(D,X)^{\prime}\left(\widetilde{\beta}(y)-\beta(y)\right)\right\Vert_{P,2}+\left\Vert r_{F}(y,D,X)\right\Vert_{P,2}\right]dy\\
		\lesssim&\displaystyle\int_{y_{1}}^{y_{2}}\left[\Vert\partial\Lambda\Vert_{\infty}\left\Vert b(D,X)^{\prime}\left(\widetilde{\beta}(y)-\beta(y)\right)\right\Vert_{\mathbb{P}_{n},2}+\left\Vert r_{F}(y,D,X)\right\Vert_{P,2}\right]dy\\
		=&\displaystyle\int_{y_{1}}^{y_{2}}\left[\Vert\partial\Lambda\Vert_{\infty}\times o_{p}\left(h_{n}n^{-1/4}\right)+o_{p}\left(h_{n}n^{-1/4}\right)\right]dy=o_{p}\left(h_{n}n^{-1/4}\right)=o_{p}\left(n^{-1/4}\right),\\
	\end{aligned}\]
	and
	\[\begin{aligned}
		&\left\Vert \displaystyle\int_{y_{1}}^{y_{2}}\Lambda\bigg(b(D,X)^{\prime}\widetilde{\beta}(y)\bigg)dy-\displaystyle\int_{y_{1}}^{y_{2}}F_{Y}(y|D,X)dy\right\Vert_{P,\infty}\\
		\leq&\displaystyle\int_{y_{1}}^{y_{2}}\left\Vert \Lambda\bigg(b(D,X)^{\prime}\widetilde{\beta}(y)\bigg)-F_{Y}(y|D,X)\right\Vert_{P,\infty}dy\\
		\leq&\displaystyle\int_{y_{1}}^{y_{2}}\left[\left\Vert \Lambda\bigg(b(D,X)^{\prime}\widetilde{\beta}(y)\bigg)-\Lambda\bigg(b(D,X)^{\prime}\beta(y)\bigg)\right\Vert_{P,\infty}+\left\Vert r_{F}(y,D,X)\right\Vert_{P,\infty}\right]dy\\
		\leq&\displaystyle\int_{y_{1}}^{y_{2}}\left[\Vert\partial\Lambda\Vert_{\infty}\left\Vert b(D,X)^{\prime}\left(\widetilde{\beta}(y)-\beta(y)\right)\right\Vert_{P,\infty}+\left\Vert r_{F}(y,D,X)\right\Vert_{P,\infty}\right]dy\\
		\lesssim&\displaystyle\int_{y_{1}}^{y_{2}}\left[K_{n}\left\Vert \widetilde{\beta}(y)-\beta(y)\right\Vert_{1}+\left\Vert r_{F}(y,D,X)\right\Vert_{P,\infty}\right]dy=o_{p}\left(h_{n}\right)=o_{p}(1),\\
	\end{aligned}\]
	for $\widetilde{\beta}(y)=\widehat{\beta}(y)$, with evaluation after computing the norms, and for $\Vert\partial\Lambda\Vert_{\infty}$ denoting $\sup_{k\in\mathbb{R}}|\partial\Lambda(k)|$ here and below. Similarly,
	\[\begin{aligned}
		&\left\Vert
		\sum\limits_{l=1}^{\ell}\eta_{l}\displaystyle\int_{y_{1}}^{y_{2}}\left(\Lambda\bigg(b(D+lh_{n},X)^{\prime}\widetilde{\beta}(y)\bigg)-\Lambda\bigg(b(D-lh_{n},X)^{\prime}\widetilde{\beta}(y)\bigg)\right)dy\big/2h_{n} -\partial_{D}\displaystyle\int_{y_{1}}^{y_{2}}F_{Y}(y|D,X)dy\right\Vert_{P,2}\\
		\leq&\left\Vert
		\sum\limits_{l=1}^{\ell}\eta_{l}\displaystyle\int_{y_{1}}^{y_{2}}\left(\Lambda\bigg(b(D+lh_{n},X)^{\prime}\widetilde{\beta}(y)\bigg)-F_{Y}(y|D+lh_{n},X)\right)dy\big/2h_{n}\right\Vert_{P,2}\\
		&+\left\Vert
		\sum\limits_{l=1}^{\ell}\eta_{l}\displaystyle\int_{y_{1}}^{y_{2}}\left(\Lambda\bigg(b(D-lh_{n},X)^{\prime}\widetilde{\beta}(y)\bigg)-F_{Y}(y|D-lh_{n},X)\right)dy\big/2h_{n}\right\Vert_{P,2}\\
		&+\left\Vert
		\sum\limits_{l=1}^{\ell}\eta_{l}\displaystyle\int_{y_{1}}^{y_{2}}\bigg(F_{Y}(y|D+lh_{n},X)-F_{Y}(y|D-lh_{n},X)\bigg)dy\big/2h_{n}-\partial_{D}\displaystyle\int_{y_{1}}^{y_{2}}F_{Y}(y|D,X)dy\right\Vert_{P,2} \\
		=&o_{p}\left(h_{n}n^{-1/4}\right)/h_{n}+O_{p}\left(h_{n}^{2\ell}\right)=o_{p}\left(n^{-1/4}\right), \\
	\end{aligned}\]
	and
	\[\begin{aligned}
		&\left\Vert
		\sum\limits_{l=1}^{\ell}\eta_{l}\displaystyle\int_{y_{1}}^{y_{2}}\left(\Lambda\bigg(b(D+lh_{n},X)^{\prime}\widetilde{\beta}(y)\bigg)-\Lambda\bigg(b(D-lh_{n},X)^{\prime}\widetilde{\beta}(y)\bigg)\right)dy\big/2h_{n} -\partial_{D}\displaystyle\int_{y_{1}}^{y_{2}}F_{Y}(y|D,X)dy\right\Vert_{P,\infty}\\
		\leq&\left\Vert
		\sum\limits_{l=1}^{\ell}\eta_{l}\displaystyle\int_{y_{1}}^{y_{2}}\left(\Lambda\bigg(b(D+lh_{n},X)^{\prime}\widetilde{\beta}(y)\bigg)-F_{Y}(y|D+lh_{n},X)\right)dy\big/2h_{n}\right\Vert_{P,\infty}\\
		&+\left\Vert
		\sum\limits_{l=1}^{\ell}\eta_{l}\displaystyle\int_{y_{1}}^{y_{2}}\left(\Lambda\bigg(b(D-lh_{n},X)^{\prime}\widetilde{\beta}(y)\bigg)-F_{Y}(y|D-lh_{n},X)\right)dy\big/2h_{n}\right\Vert_{P,\infty} \\
		&+\left\Vert
		\sum\limits_{l=1}^{\ell}\eta_{l}\displaystyle\int_{y_{1}}^{y_{2}}\bigg(F_{Y}(y|D+lh_{n},X)-F_{Y}(y|D-lh_{n},X)\bigg)dy\big/2h_{n}-\partial_{D}\displaystyle\int_{y_{1}}^{y_{2}}F_{Y}(y|D,X)dy\right\Vert_{P,\infty} \\
		=&o_{p}\left(h_{n}\right)/h_{n}+O_{p}\left(h_{n}^{2\ell}\right)=o_{p}(1),\\
	\end{aligned}\]
	for $\widetilde{\beta}(y)=\widehat{\beta}(y)$, with evaluation after computing the norms. The verification of the remaining terms are identical and omitted. Moreover, let
	\[
	\mathcal{P}=\left\{
	\begin{array}{l}
		u\mapsto \mathbb{E}_{n}\widetilde{P}(Y_{i};u):\\
		\left\Vert \mathbb{E}_{n}\widetilde{P}(Y_{i};u)-P(u)\right\Vert_{P,2}=o_{p}\left(n^{-1/4}\right) \\
		\left\Vert \mathbb{E}_{n}\widetilde{P}(Y_{i};u)-P(u)\right\Vert_{P,\infty}=o_{p}\left(1\right) \\
	\end{array}
	\right\}.
	\]
	Obviously, with probability no less than $1-\Delta_{n}$, we can observe that $\widehat{P}(u)\in\mathcal{P}$.
	
	We have that
	\[\begin{aligned}
		\sqrt{n}\left(\widehat{\theta}(u)-\theta(u)\right)
		=&\underbrace{\mathbb{G}_{n}\psi(W,\theta,\eta;u)}_{\dagger_{4.1.1}}+\underbrace{\mathbb{G}_{n}\bigg[\psi\left(W,\theta,\widetilde{\eta};u\right)-\psi(W,\theta,\eta;u)\bigg]}_{\dagger_{4.1.2}} \\
		&+\underbrace{\sqrt{n}E\bigg[\psi\left(W,\theta,\widetilde{\eta};u\right)-\psi\left(W,\theta,{\eta};u\right)\bigg]}_{\dagger_{4.1.3}},
	\end{aligned}\]
	with $\widetilde{\eta}$ evaluated at $\widetilde{\eta}=\widehat{\eta}$.
	
	Firstly, we consider $\dagger_{4.1.3}$. Note that for
	\[
	\Delta_{\widetilde{\eta}}=\left(\Delta_{\widetilde{\eta}}^{1},\Delta_{\widetilde{\eta}}^{2},\Delta_{\widetilde{\eta}}^{3}\right)^{\prime}=\widetilde{\eta}-\eta.
	\]
	For all $k=\{k_{j}\}_{j=1}^{3}\in\mathbb{N}^{3}$: $0\leq|k|\leq3$, $|k|=\sum_{j=1}^{3}k_{j}$ and $\partial_{\widetilde{\eta}}^{k}=\partial_{\widetilde{\eta}_{1}}^{k_{1}}\partial_{\widetilde{\eta}_{2}}^{k_{2}}\partial_{\widetilde{\eta}_{3}}^{k_{3}}$. After applying Taylor expansion,
	\[\begin{aligned}
		\dagger_{4.1.3}=&\sqrt{n}\sum_{|k|=1}E\bigg[\partial_{\widetilde{\eta}}^{k}{\psi}\left(W,\theta,{\eta};u\right)\Delta_{\widetilde{\eta}}^{k}\bigg] \\
		&+\sqrt{n}\sum_{|k|=2}\frac{1}{2}E\bigg[\partial_{\widetilde{\eta}}^{k}{\psi}\left(W,\theta,{\eta};u\right)\Delta_{\widetilde{\eta}}^{k}\bigg] \\
		&+\sqrt{n}\sum_{|k|=3}\frac{1}{6}\int_{0}^{1}E\bigg[\partial_{\widetilde{\eta}}^{k}{\psi}\left(W,\theta,{\eta}+\lambda\Delta_{\widetilde{\eta}};u\right)\Delta_{\widetilde{\eta}}^{k}\bigg]d\lambda \\
		=&\dagger_{4.1.3a}+\dagger_{4.1.3b}+\dagger_{4.1.3c},
	\end{aligned}\]
	with $\widetilde{\eta}$ evaluated at $\widetilde{\eta}=\widehat{\eta}$ after computing the expectations. By the law of iterated expectations and the orthogonality property of the moment function for $\eta$, $\forall k\in\mathbb{N}^{3}:|k|=1$,
	\[
	E\bigg[\partial_{\widetilde{\eta}}^{k}{\psi}\left(W,\theta,{\eta};u\right)\Delta_{\widetilde{\eta}}^{k}\bigg]=0,
	\]
	and thus $\dagger_{4.1.3a}=0$. Moreover, uniformly over $\widetilde{\eta}\in\mathcal{R}=\mathcal{P}\times\left(\mathcal{F}_{1}\cup\mathcal{F}_{2}\cup\mathcal{F}_{3}\right)\times\mathcal{L}$, we have
	\[\begin{aligned}
		&|\dagger_{4.1.3b}|\lesssim\sqrt{n}\Vert\widetilde{\eta}-\eta\Vert_{P,2}^{2}=\sqrt{n}o_{p}\left(n^{-1/2}\right)=o_{p}(1), \\
		&|\dagger_{4.1.3c}|\lesssim\sqrt{n}\Vert\widetilde{\eta}-\eta\Vert_{P,2}^{2}\Vert\widetilde{\eta}-\eta\Vert_{P,\infty}=\sqrt{n}o_{p}\left(n^{-1/2}\right)\cdot{o_{p}(1)}=o_{p}(1).
	\end{aligned}\]
	Since $\widehat{\eta}\in\mathcal{R}$, with probability $1-\Delta_{n}$,
	\[
	P\bigg(\left|\dagger_{4.1.3}\right|\lesssim\delta_{n}\bigg)\geq1-\Delta_{n}.
	\]
	
	Then we consider $\dagger_{4.1.2}$. With probability $1-\Delta_{n}$,
	\[\begin{aligned}
		|E\dagger_{4.1.2}^{2}|\leq&\sup_{\widetilde{\eta}\in\mathcal{R}}\left|E\left\{\mathbb{G}_{n}\bigg[\psi\left(W,\theta,\widetilde{\eta};u\right)-\psi\left(W,\theta,{\eta};u\right)\bigg]\right\}^{2}\right| \\
		=&\sup_{\widetilde{\eta}\in\mathcal{R}}\left|E\bigg[\psi\left(W,\theta,\widetilde{\eta};u\right)-\psi\left(W,\theta,{\eta};u\right)\bigg]^{2}\right|. \\
	\end{aligned}\]
	Applying similar arguments as the proceeding one, uniformly over $\widetilde{\eta}\in\mathcal{R}$, we have
	\[
	E\bigg[\psi\left(W,\theta,\widetilde{\eta};u\right)-\psi\left(W,\theta,{\eta};u\right)\bigg]^{2}\lesssim\Vert\widetilde{\eta}-\eta\Vert_{P,2}^{2}+\Vert\widetilde{\eta}-\eta\Vert_{P,2}^{2}\Vert\widetilde{\eta}-\eta\Vert_{P,\infty}^{2}=o_{p}\left(n^{-1/2}\right).
	\]
	Thus, with probability $1-\Delta_{n}$,
	\[
	P\bigg(\left|\dagger_{4.1.2}\right|\lesssim\delta_{n}n^{-1/4}\bigg)\geq1-\Delta_{n}.
	\]
	
	\noindent\textbf{Step 2. (Uniform Donskerness)} Here we claim that Assumptions 4.1-4.3 imply that the set of functions $\{\psi\left(W,\theta,{\eta};u\right)\}_{u\in\mathcal{U}}$ is $P$-Donsker, namely
	\[
	Z_{n}(u)\leadsto Z(u) \quad in \quad \mathbb{D}=\ell^{\infty}\left(\mathcal{U}\right),
	\]
	where $Z(u)=\mathbb{G}\psi\left(W,\theta,{\eta};u\right)$.
	
	We apply Theorem B.1 in Belloni et al. (2017) to verify this claim. The classes of functions
	\[
	\mathcal{V}_{1}=\left\{\displaystyle\int_{y_{1}}^{y_{2}}1\{Y\leq{y}\}dy:(y_{1},y_{2})\in\mathcal{U}\right\}
	\]
	and
	\[
	\mathcal{V}_{2}=\bigg\{1\{y_{1}\leq{Y}\leq{y_{2}}\}:(y_{1},y_{2})\in\mathcal{U}\bigg\}
	\]
	viewed as maps from the sample space $\mathcal{W}$ to the real line, are bounded by constant envelops and have finite VC dimensions. According to Theorem 2.6.7 in van der Vaart and Wellner (1996), we can deduce that
	\[\begin{aligned}
		&\sup_{Q}\log N\left(\epsilon,\mathcal{V}_{1},\Vert\cdot\Vert_{Q,2}\right)\lesssim\log(1/\epsilon)\vee{0},\\
		&\sup_{Q}\log N\left(\epsilon,\mathcal{V}_{2},\Vert\cdot\Vert_{Q,2}\right)\lesssim\log(1/\epsilon)\vee{0},\\
	\end{aligned}\]
	with the supremum taken over all finitely discrete probability measures $Q$ on $(\mathcal{W},\mathcal{A}_{\mathcal{W}})$. According to Lemma L.2 in Belloni et al. (2017), the following class of functions
	\[
	\mathcal{V}_{3}=\left\{\displaystyle\int_{y_{1}}^{y_{2}}F_{Y}(y|D,X)dy:(y_{1},y_{2})\in\mathcal{U}\right\}
	\]
	is bounded by a constant envelop and obeys
	\[
	\sup_{Q}\log N\left(\epsilon,\mathcal{V}_{3},\Vert\cdot\Vert_{Q,2}\right)\lesssim\log(1/\epsilon)\vee{0}.\\
	\]
	The class of functions
	\[
	\mathcal{V}_{4}=\left\{\partial_{D}\displaystyle\int_{y_{1}}^{y_{2}}F_{Y}(y|D,X)dy:(y_{1},y_{2})\in\mathcal{U}\right\}
	\]
	is bounded by a measurable envelop $T\geq\sup_{(y_{1},y_{2})\in\mathcal{U}}\left|\partial_{D}\displaystyle\int_{y_{1}}^{y_{2}}F_{Y}(y|D,X)dy\right|$ with $\Vert{T}\Vert_{P,2}\leq{\infty}$ and obeys
	\[
	\sup_{Q}\log N\left(\epsilon\Vert{T}\Vert_{Q,2},\mathcal{V}_{4},\Vert\cdot\Vert_{Q,2}\right)\lesssim(1/\epsilon)^{(1+d_{X_{c}})/\sigma} \\
	\]
	by Corollary 2.7.2 in van der Vaart and Wellner (1996) and the relationship between covering numbers and bracketing numbers. The classes of functions
	\[
	\mathcal{V}_{5}=\bigg\{P\bigg(y_{1}\leq{Y}\leq{y_{2}}\bigg):(y_{1},y_{2})\in\mathcal{U}\bigg\}
	\]
	and
	\[
	\mathcal{V}_{6}=\left\{E\left[\partial_{D}\displaystyle\int_{y_{1}}^{y_{2}}F_{Y}(y|D,X)dy\right]:(y_{1},y_{2})\in\mathcal{U}\right\}
	\]
	are bounded by constant envelops and obey
	\[\begin{aligned}
		&{\sup_{Q}\log N\left(\epsilon,\mathcal{V}_{5},\Vert\cdot\Vert_{Q,2}\right)\lesssim\log(1/\epsilon)\vee{0},}\\
		&{\sup_{Q}\log N\left(\epsilon,\mathcal{V}_{6},\Vert\cdot\Vert_{Q,2}\right)\lesssim\log(1/\epsilon)\vee{0},}\\
	\end{aligned}\]
	which holds by Lemma L.2 in Belloni et al. (2017) or Lemma A.2 in Ghosal et al. (2000). Moreover, the VC dimension of the measurable function set
	\[
	\mathcal{V}_{7}=\left\{g:(D,X)\mapsto\frac{\partial_{D}f(D,X)}{f(D,X)}\right\}
	\]
	is finite according to Lemma 2.6.15 in van der Vaart and Wellner (1996). Thus, $\mathcal{V}_{7}$ is bounded by a measurable envelop $T^{*}(W)=\left|\dfrac{\partial_{D}f(D,X)}{f(D,X)}\right|$ with $\Vert{T^{*}}\Vert_{P,2}\leq{\infty}$ and obeys
	\[
	\sup_{Q}\log N\left(\epsilon\Vert{T^{*}}\Vert_{Q,2},\mathcal{V}_{7},\Vert\cdot\Vert_{Q,2}\right)\lesssim\log(1/\epsilon)\vee{0}.
	\]
	
	By a slight abuse of notation, let $\eta$ also contains $\displaystyle\int_{y_{1}}^{y_{2}}1\{Y\leq{y}\}dy$ and $1\{y_{1}\leq{Y}\leq{y_{2}}\}$, such that
	\[
	\eta(W;u)=\left(\displaystyle\int_{y_{1}}^{y_{2}}1\{Y\leq{y}\}dy,1\{y_{1}\leq{Y}\leq{y_{2}}\},P\big(y_1<Y<y_2\big),\int_{y_1}^{y_2}F_Y\left(y\big|D,X\right)dy,\frac{\partial_Df\big(D,X\big)}{f\big(D,X\big)}\right).
	\]
	Note that
	\[
	\mathcal{Z}=\big\{
	\psi\left(W,\theta,{\eta};u\right):u\in\mathcal{U}
	\big\}
	\]
	is formed as a uniform Lipschitz transform of the function sets $\mathcal{V}_{1}$, $\mathcal{V}_{2}$, $\mathcal{V}_{3}$, $\mathcal{V}_{4}$, $\mathcal{V}_{5}$, $\mathcal{V}_{6}$ and $\mathcal{V}_{7}$, and is bounded by a measurable envelop $\bar{T}$ with $\left\Vert\bar{T}\right\Vert_{P,2}\leq\infty$.\footnote{The envelop $\bar{T}$ of $\mathcal{Z}$ can be constructed as $\bar{T}=2\left(\sum_{k=1}^{5}L_{k}^{2}R_{k}^{2}\right)^{1/2}$, where $R_{k}$ denotes the envelop of $\mathcal{V}_{k}$, and $L_{k}$ satisfies
		\[
		|\psi\left(W,\theta,\widetilde{\eta};u\right)-\psi\left(W,\theta,\bar{\eta};u\right)|^{2}\leq\sum_{k=1}^{5}L_{k}^{2}(W;u)|\widetilde{\eta}_{k}(W;u)-\bar{\eta}_{k}(W;u)|^{2},
		\]
		for all $\widetilde{\eta}$ and $\bar{\eta}$ in $\mathcal{V}_{1}\times\mathcal{V}_{2}\times\mathcal{V}_{5}\times\left(\mathcal{V}_{3}\cup\mathcal{V}_{4}\cup\mathcal{V}_{6}\right)\times\mathcal{V}_{7}$.
	} According to Theorem 2.10.20 of van der Vaart and Wellner (1996), the class of functions $\mathcal{Z}$ obeys
	\[
	\sup_{Q}\log N\left(\epsilon\left\Vert{\bar{T}}\right\Vert_{Q,2},\mathcal{Z},\Vert\cdot\Vert_{Q,2}\right)\lesssim(1/\epsilon)^{(1+d_{X_{c}})/\sigma}.
	\]
	Since
	\[
	\lim_{\delta\to{0}}\int_{0}^{\delta}\sqrt{(1/\epsilon)^{(1+d_{X_{c}})/\sigma}} d\epsilon\to{0},
	\]
	by Assumption 4.3(iv), the entropy condition (B.2) in Theorem B.1 of Belloni et al. (2017) holds.
	
	The first condition in (B.1) is trivially satisfied. We demonstrate the second condition in (B.1). Consider a sequence of positive constants $\epsilon$ approaching zero, and it suffice to verify that
	\[
	\lim_{\epsilon\to{0}^{+}}\sup_{d_{\mathcal{U}}\left(u,\widetilde{u}\right)\leq{\epsilon}}\left\Vert\psi\left(W,\theta,{\eta};u\right)-\psi\left(W,\theta,{\eta};\widetilde{u}\right)\right\Vert_{P,2}=0.
	\]
	Notice that
	\[\begin{aligned}
		&\left\Vert\psi\left(W,\theta,{\eta};u\right)-\psi\left(W,\theta,{\eta};\widetilde{u}\right)\right\Vert_{P,2}\\
		\lesssim&\underbrace{\left\Vert{\displaystyle\int_{y_{1}}^{y_{2}}1\{Y\leq{y}\}dy-\displaystyle\int_{\widetilde{y}_{1}}^{\widetilde{y}_{2}}1\{Y\leq{y}\}dy}\right\Vert_{P,2}}_{\dagger_{4.1.4}}
		+\underbrace{\left\Vert1\bigg\{y_{1}\leq{Y}\leq{y_{2}}\bigg\}-1\bigg\{\widetilde{y}_{1}\leq{Y}\leq{\widetilde{y}_{2}}\bigg\}\right\Vert_{P,2}}_{\dagger_{4.1.5}} \\
		+&\underbrace{\left\Vert{\displaystyle\int_{y_{1}}^{y_{2}}F_{Y}(y|D,X)dy-\displaystyle\int_{\widetilde{y}_{1}}^{\widetilde{y}_{2}}F_{Y}(y|D,X)dy}\right\Vert_{P,2}}_{\dagger_{4.1.6}}
		+\underbrace{\left\Vert{\displaystyle\int_{y_{1}}^{y_{2}}\partial_{D}F_{Y}(y|D,X)dy-\displaystyle\int_{\widetilde{y}_{1}}^{\widetilde{y}_{2}}\partial_{D}F_{Y}(y|D,X)dy}\right\Vert_{P,2}}_{\dagger_{4.1.7}} \\
		+&\underbrace{{E\displaystyle\int_{y_{1}}^{y_{2}}\partial_{D}F_{Y}(y|D,X)dy-E\displaystyle\int_{\widetilde{y}_{1}}^{\widetilde{y}_{2}}\partial_{D}F_{Y}(y|D,X)dy}}_{\dagger_{4.1.8}}\\
		+&\underbrace{ E1\bigg\{y_{1}\leq{Y}\leq{y_{2}}\bigg\}-E1\bigg\{\widetilde{y}_{1}\leq{Y}\leq{\widetilde{y}_{2}}\bigg\}}_{\dagger_{4.1.9}}. \\
	\end{aligned}\]
	Under Assumption 4.2, $\dagger_{4.1.4}$ and $\dagger_{4.1.5}$ converges to $0$ as $d_{\mathcal{U}}\left(u,\widetilde{u}\right)\to{0}$. Note that
	\[
	\displaystyle\int_{y_{1}}^{y_{2}}F_{Y}(y|D,X)dy=E\left[\displaystyle\int_{y_{1}}^{y_{2}}1\{Y\leq{y}\}dy\bigg|D,X\right]
	\]
	By the contradiction property of the conditional expectation,
	\[
	\dagger_{4.1.6}\leq\dagger_{4.1.4}\to{0}
	\]
	as $d_{\mathcal{U}}\left(u,\widetilde{u}\right)\to{0}$. For $\dagger_{4.1.7}$,
	\[\begin{aligned}
		\dagger_{4.1.7}=&\left\Vert{\displaystyle\int_{\widetilde{y}_{2}}^{y_{2}}\partial_{D}F_{Y}(y|D,X)dy-\displaystyle\int_{\widetilde{y}_{1}}^{y_{1}}\partial_{D}F_{Y}(y|D,X)dy}\right\Vert_{P,2} \\
		\leq&\left\Vert\displaystyle\int_{\widetilde{y}_{2}}^{y_{2}}\partial_{D}F_{Y}(y|D,X)dy\right\Vert_{P,2}+\left\Vert\displaystyle\int_{\widetilde{y}_{1}}^{y_{1}}\partial_{D}F_{Y}(y|D,X)dy\right\Vert_{P,2} \\
		\leq&|y_{2}-\widetilde{y}_{2}|\sup_{y\in\mathcal{M}}\left\Vert\partial_{D}F_{Y}(y|D,X)\right\Vert_{P,2}+|y_{1}-\widetilde{y}_{1}|\sup_{y\in\mathcal{M}}\left\Vert\partial_{D}F_{Y}(y|D,X)\right\Vert_{P,2}
		\to{0} \\
	\end{aligned}\]
	as $d_{\mathcal{U}}\left(u,\widetilde{u}\right)\to{0}$ by Assumption 4.3(iii). Consequently, we have that $\dagger_{4.1.8}\to{0}$ and $\dagger_{4.1.9}\to{0}$ as $d_{\mathcal{U}}\left(u,\widetilde{u}\right)\to{0}$.
	$\blacksquare$\\
	
	\noindent\textbf{Proof of Theorem 4.2.} The joint density of the observed variables $W=(Y,D,X)$ can be written as
	\begin{eqnarray}
		f(y,d,x)=f_Y(y|d,x)f_D(d|x)f_{X}(x)\notag
	\end{eqnarray}
	Consider a regular parametric submodel indexed by $\epsilon$ with $\epsilon_0$ corresponding to the true model:\\$f(y,d,x;\epsilon_0)\equiv f(y,d,x)$. The density of $f(y,d,x;\epsilon)$ can be written as
	\begin{eqnarray}
		f(y,d,x;\epsilon)=f_Y(y|d,x;\epsilon)f_D(d|x;\epsilon)f_X(x;\epsilon)\notag
	\end{eqnarray}
	We will assume that all terms of the previous equation admit an interchange of the order of integration and differention, which will hold under sufficient condition given by Theorem 1.3.2 of Amemiya (1985) such that
	\[
	\int\frac{\partial f(y,d,x;\epsilon)}{\partial\epsilon}dxdddy=\frac{\partial}{\partial\epsilon}\underbrace{\int f(y,d,x;\epsilon)dxdddy}_1=0 \eqno{(A.4.2.1)}
	\]
	The corresponding score of $f(y,d,x;\epsilon)$ is
	\[
	s(y,d,x;\epsilon)=\frac{\partial \ln f(y,d,x;\epsilon)}{\partial\epsilon}=\check{f}_Y(y|d,x;\epsilon)+\check{f}_D(d|x;\epsilon)+\check{f}_{X}(x;\epsilon)
	\]
	where $\check{f}$ defines a derivative of the log, that is, $\check{f}_Y(y|d,x;\epsilon)\equiv\partial \ln f_Y(y|d,x;\epsilon)/\partial\epsilon$, $\check{f}_D(d|x;\epsilon)\equiv\partial \ln f_D(d|x;\epsilon)/\partial\epsilon$ and $\check{f}_{X}(x;\epsilon)\equiv\partial \ln f_{X}(x;\epsilon)/\partial\epsilon$. Notice that the expectation of the score is zero if $\epsilon$ is evaluated at the true value $\epsilon_{0}$. According to Proposition 2.1, we have
	\[
	\theta(u)=-\dfrac{\displaystyle\int\bigg(\int1\{y_{1}<y<y_{2}\}1\{t\leq y\}\partial_df_Y(t|d,x)dtdy\bigg)f_{D}(d|x)f_{X}(x)dddx}{\displaystyle\int1\{y_{1}<y<y_{2}\}f(y,d,x)dxdddy}.
	\]
	Therefore, the parameter $\theta(u;\epsilon)$ induced by the submodel $f(d,y,x;\epsilon)$ satisfies
	\[
	\theta(u;\epsilon)=-\dfrac{\displaystyle\int\bigg(\int1\{y_{1}<y<y_{2}\}1\{t\leq y\}\partial_df_Y(t|d,x;\epsilon)dtdy\bigg)f_{D}(d|x;\epsilon)f_{X}(x;\epsilon)dddx}{\displaystyle\int1\{y_{1}<y<y_{2}\}f(y,d,x;\epsilon)dxdddy}.
	\]
	
	The tangent space of the model is the set of functions that are mean zero and satisfy the additive structure of the score:
	\begin{eqnarray}
		\Im=\{s_y(y|d,x)+s_d(d|x)+s_x(x)\}\notag
	\end{eqnarray}
	for any functions $s_y$, $s_{d}$ and $s_{x}$ satisfying the mean zero property
	\[
	E[s_y(Y|D,X)|D,X]=E[s_d(D|X)|X]=Es_x(X)=0.
	\]
	Then the semiparametric variance bound of $\theta(u)$ is the variance of the projection on $\Im$ of a function $\Gamma(W;u)$(with $E\Gamma(\cdot;u)=0$ and $E\left\|\Gamma^2(\cdot;u)\right\|<\infty$ for any $u\in\mathcal{U}$) that satisfies for all regular parametric submodels
	\begin{eqnarray}
		\frac{\partial\theta(u;\epsilon)}{\partial\epsilon}\bigg|_{\epsilon=\epsilon_0}=E\left[\Gamma(W;u)\cdot s(W;\epsilon_0)\right]\notag
	\end{eqnarray}
	If $\Gamma(W;u)$ itself already lies in the tangent space, the variance bound is given by $E\Gamma^2(W;u)$ for any $u\in\mathcal{U}$.
	
	We first calculate $\dfrac{\partial\theta(u;\epsilon)}{\partial\epsilon}\bigg|_{\epsilon=\epsilon_0}$. By calculation,
	\[\begin{aligned}
		&\frac{\partial\theta(u;\epsilon)}{\partial\epsilon}\bigg|_{\epsilon=\epsilon_0}\\
		=&\frac{E\left(\partial_D\displaystyle\int_{y_1}^{y_2}F(y|D,X)dy\right)}{P^{2}(u)}\bigg(\int1\{y_{1}<y<y_{2}\}\frac{\partial f_Y(y|d,x;\epsilon)}{\partial\epsilon}\bigg|_{\epsilon=\epsilon_0}f_D(d|x)f_{X}(x)dxdddy\\
		+&\int1\{y_{1}<y<y_{2}\}  f_Y(y|d,x)\frac{\partial f_D(d|x;\epsilon)}{\partial\epsilon}\bigg|_{\epsilon=\epsilon_0}f_{X}(x)dxdddy\\
		+&\int1\{y_{1}<y<y_{2}\}  f_Y(y|d,x)f_D(d|x)\frac{\partial f_{X}(x;\epsilon)}{\partial\epsilon}\bigg|_{\epsilon=\epsilon_0}dxdddy\bigg)\\
		-&\frac{1}{P\left(u\right)}\displaystyle\int\left(\int1\{y_{1}<y<y_{2}\}1\{t\leq y\}\partial_d\left(\frac{\partial f_Y(t|d,x;\epsilon)}{\partial\epsilon}\bigg|_{\epsilon=\epsilon_0}\right)dtdy\right)f_{D}(d|x)f_{X}(x)dddx \\
		-&\frac{1}{P\left(u\right)}\displaystyle\int\bigg(\int1\{y_{1}<y<y_{2}\}1\{t\leq y\}\partial_df_Y(t|d,x)dtdy\bigg)\frac{\partial f_D(d|x;\epsilon)}{\partial\epsilon}\bigg|_{\epsilon=\epsilon_0}f_{X}(x)dddx \\
		-&\frac{1}{P\left(u\right)}\displaystyle\int\bigg(\int1\{y_{1}<y<y_{2}\}1\{t\leq y\}\partial_df_Y(t|d,x)dtdy\bigg)f_{D}(d|x)\frac{\partial f_{X}(x;\epsilon)}{\partial\epsilon}\bigg|_{\epsilon=\epsilon_0}dddx. \\
	\end{aligned}\]
	Recall that the Neyman-orthogonal score is
	\[\begin{aligned}
		\psi\left(W,\theta,\eta;u\right)=&-\frac{1}{P(u)}\partial_D\int_{y_1}^{y_2}F_Y(y|D,X)dy-\theta(u)\\
		&-\frac{1}{P(u)}\frac{\partial_Df(D,X)}{f(D,X)}\int_{y_1}^{y_2}\bigg(F_Y\left(y\big|D,X\right)-1\big\{Y<y\big\}\bigg)dy\\
		&+\frac{E\left(\partial_D\displaystyle\int_{y_1}^{y_2}F_Y(y|D,X)dy\right)}{P^2(u)}\bigg(1\big\{y_1<Y<y_2\big\}-P(u)\bigg).
	\end{aligned}\]	
	To justify our theorem, it suffice to show that (i)
	\[
	\dfrac{\partial\theta(u;\epsilon)}{\partial\epsilon}\bigg|_{\epsilon=\epsilon_0}=E\left[\psi\left(W,\theta,\eta;u\right)\cdot s(W;\epsilon_0)\right]
	\]
	and (ii) $\psi\left(W,\theta,\eta;u\right)$ lies in the tangent space $\Im$ for any $u\in\mathcal{U}$.

	The second argument can be easily verified. For (i), substituting the representation $\psi\left(W,\theta,\eta;u\right)$ into $E\left[\psi\left(W,\theta,\eta;u\right)\cdot s(W;\epsilon_0)\right]$ implies
	\[
	E\left[\psi\left(W,\theta,\eta;u\right)\cdot s(W;\epsilon_0)\right]=-\frac{1}{P(u)}\bigg(\dagger_{4.2.1}+\dagger_{4.2.2}\bigg)+\frac{E\left(\partial_D\displaystyle\int_{y_1}^{y_2}F_Y(y|D,X)dy\right)}{P^2(u)}\dagger_{4.2.3},
	\]
	where
	\[
	\dagger_{4.2.1}=E\left[\left(\partial_D\int_{y_1}^{y_2}F_Y(y|D,X)dy+\theta(u)\right)\cdot\bigg(\check{f}_Y(Y|D,X;\epsilon_{0})+\check{f}_D(D|X;\epsilon_{0})+\check{f}_{X}(X;\epsilon_{0})\bigg)\right],
	\]
	\[
	\dagger_{4.2.2}=E\left[\frac{\partial_Df(D,X)}{f(D,X)}\int_{y_1}^{y_2}\bigg(F_Y\left(y\big|D,X\right)-1\big\{Y<y\big\}\bigg)dy\cdot\bigg(\check{f}_Y(Y|D,X;\epsilon_{0})+\check{f}_D(D|X;\epsilon_{0})+\check{f}_{X}(X;\epsilon_{0})\bigg)\right],
	\]
	and
	\[
	\dagger_{4.2.3}=E\left[\bigg(1\big\{y_1<Y<y_2\big\}-P(u)\bigg)\cdot\bigg(\check{f}_Y(Y|D,X;\epsilon_{0})+\check{f}_D(D|X;\epsilon_{0})+\check{f}_{X}(X;\epsilon_{0})\bigg)\right].
	\]
	By $(A.4.2.1)$, we have
	\[
	E\left[\check{f}_Y(Y|D,X;\epsilon_{0})|D,X\right]=E\left[\check{f}_D(D|X;\epsilon_{0})|X\right]=E\check{f}_{X}(X;\epsilon_{0})=0.
	\]
	For $\dagger_{4.2.1}$, we get
	\[\begin{aligned}
		\dagger_{4.2.1}=&E\left[\partial_D\int_{y_1}^{y_2}F_Y(y|D,X)dy\cdot\bigg(\check{f}_D(D|X;\epsilon_{0})+\check{f}_{X}(X;\epsilon_{0})\bigg)\right] \\
		=&\int\partial_d\int_{y_1}^{y_2}F_Y(y|d,x)dy\left(\frac{\partial f_D(d|x;\epsilon)}{\partial\epsilon}\bigg|_{\epsilon=\epsilon_0}f_{X}(x)+f_{D}(d|x)\frac{\partial f_{X}(x;\epsilon)}{\partial\epsilon}\bigg|_{\epsilon=\epsilon_0}\right)dddx \\
		=&\displaystyle\int\bigg(\int1\{y_{1}<y<y_{2}\}1\{t\leq y\}\partial_df_Y(t|d,x)dtdy\bigg)\frac{\partial f_D(d|x;\epsilon)}{\partial\epsilon}\bigg|_{\epsilon=\epsilon_0}f_{X}(x)dddx \\
		&+\displaystyle\int\bigg(\int1\{y_{1}<y<y_{2}\}1\{t\leq y\}\partial_df_Y(t|d,x)dtdy\bigg)f_{D}(d|x)\frac{\partial f_{X}(x;\epsilon)}{\partial\epsilon}\bigg|_{\epsilon=\epsilon_0}dddx. \\
	\end{aligned}\]
Similarly, we can derive
	\[\begin{aligned}
	\dagger_{4.2.2}=&E\left[\frac{\partial_Df(D,X)}{f(D,X)}\int_{y_1}^{y_2}\bigg(F_Y\left(y\big|D,X\right)-1\big\{Y<y\big\}\bigg)dy\cdot\check{f}_Y(Y|D,X;\epsilon_{0})\right] \\
	=&-E\left[\frac{\partial_Df(D,X)}{f(D,X)}\int_{y_1}^{y_2}1\big\{Y<y\big\}dy\cdot\check{f}_Y(Y|D,X;\epsilon_{0})\right] \\
	=&-\int\left[\int1\{y_{1}<y<y_{2}\}1\{t<y\}dy\right]\partial_df(d,x)\cdot\frac{\partial}{\partial\epsilon}f_{Y}(t|d,x;\epsilon)\bigg|_{\epsilon=\epsilon_{0}}dtdddx \\
	=&\int\left[\int1\{y_{1}<y<y_{2}\}1\{t<y\}dy\right]f(d,x)\cdot\partial_d\left(\frac{\partial}{\partial\epsilon}f_{Y}(t|d,x;\epsilon)\bigg|_{\epsilon=\epsilon_{0}}\right)dtdddx \\
	=&\displaystyle\int\left(\int1\{y_{1}<y<y_{2}\}1\{t\leq y\}\partial_d\left(\frac{\partial f_Y(t|d,x;\epsilon)}{\partial\epsilon}\bigg|_{\epsilon=\epsilon_0}\right)dtdy\right)f_{D}(d|x)f_{X}(x)dddx, \\
	\end{aligned}\]
and
\[\begin{aligned}
	\dagger_{4.2.3}=&E\left[1\big\{y_1<Y<y_2\big\}\cdot\bigg(\check{f}_Y(Y|D,X;\epsilon_{0})+\check{f}_D(D|X;\epsilon_{0})+\check{f}_{X}(X;\epsilon_{0})\bigg)\right] \\
	=&\int1\big\{y_1<y<y_2\big\}\cdot\bigg(\check{f}_Y(y|d,x;\epsilon_{0})+\check{f}_D(d|x;\epsilon_{0})+\check{f}_{X}(x;\epsilon_{0})\bigg)f(y,d,x)dxdddy \\
	=&\int1\{y_{1}<y<y_{2}\}\frac{\partial f_Y(y|d,x;\epsilon)}{\partial\epsilon}\bigg|_{\epsilon=\epsilon_0}f_D(d|x)f_{X}(x)dxdddy\\
	+&\int1\{y_{1}<y<y_{2}\}  f_Y(y|d,x)\frac{\partial f_D(d|x;\epsilon)}{\partial\epsilon}\bigg|_{\epsilon=\epsilon_0}f_{X}(x)dxdddy\\
	+&\int1\{y_{1}<y<y_{2}\}  f_Y(y|d,x)f_D(d|x)\frac{\partial f_{X}(x;\epsilon)}{\partial\epsilon}\bigg|_{\epsilon=\epsilon_0}dxdddy, \\
\end{aligned}\]
which completes the proof. $\blacksquare$
\\

	\noindent\textbf{Proof of Theorem 4.3.} Let $P^{*}=P\times{P}_{\xi}$. Then the operator $E_{P^{*}}$ then denotes the expectation with respect to $P^{*}=P\times{P_{\xi}}$ and $\mathbb{G}_{n}$ denotes the corresponding empirical process, that is
	\[
	\mathbb{G}_{n}B(\xi,W)=\frac{1}{\sqrt{n}}\sum_{i=1}^{n}\bigg[B(\xi_{i},W_{i})-E_{P^{*}}B(\xi,W)\bigg].
	\]
	
	Recall that we define the bootstrap draw as
	\[
	\widehat{Z}_{n}^{*}(u)=\sqrt{n}\left(\widehat{\theta}^{*}(u)-\widehat{\theta}(u)\right)=\frac{1}{\sqrt{n}}\sum_{i=1}^{n}\xi_{i}\psi\bigg(W_{i},\widehat{\theta},\widehat{\eta};u\bigg)=\mathbb{G}_{n}\xi\psi\bigg(W,\widehat{\theta},\widehat{\eta};u\bigg),
	\]
	since $E_{P^{*}}\left[\xi\psi\bigg(W,\widehat{\theta},\widehat{\eta};u\bigg)\right]=0$ because $\xi$ is independent of $W$ and has zero mean. The proof also consists of two steps.
	
	\noindent\textbf{Step 1.} In this step, we establish that
	\[
	\widehat{Z}_{n}^{*}(u)={Z}_{n}^{*}(u)+o_{p^{*}}(1), \quad in \quad \mathbb{D}=\ell^{\infty}\left(\mathcal{U}\right),
	\]
	where $Z_{n}^{*}(u)=\mathbb{G}_{n}\xi\psi(W,\theta,\eta;u)$.
	
	Define $\widetilde{\psi}(W,\eta;u)=\psi(W,\theta,\eta;u)+\theta$. We then have the representation
	\[\begin{aligned}
		&\sqrt{n}\left(\widehat{\theta}^{*}(u)-\widehat{\theta}(u)\right)\\
		=&\mathbb{G}_{n}\xi\psi(W,\theta,\eta;u)-\left(\widehat{\theta}(u)-{\theta}(u)\right)\mathbb{G}_{n}\xi+\underbrace{\mathbb{G}_{n}\bigg[\xi\widetilde{\psi}\left(W,\widehat{\eta};u\right)-\xi\widetilde{\psi}\left(W,\eta;u\right)\bigg]}_{\dagger_{4.3.1}}. \\
	\end{aligned}\]
	According to Theorem 4.1, $\left(\widehat{\theta}(u)-{\theta}(u)\right)\mathbb{G}_{n}\xi=O_{p^{*}}\left(n^{-1/2}\right)=o_{p^{*}}(1)$. With probability $1-\Delta_{n}$,
	\[\begin{aligned}
		\left|E_{P^{*}}\left(\dagger_{4.3.1}\right)^{2}\right|\leq&\sup_{\widetilde{\eta}\in\mathcal{R}}\left|E_{P^{*}}\left\{\mathbb{G}_{n}\bigg[\xi\widetilde{\psi}\left(W,\widetilde{\eta};u\right)-\xi\widetilde{\psi}\left(W,\eta;u\right)\bigg]\right\}^{2}\right| \\
		=&\sup_{\widetilde{\eta}\in\mathcal{R}}\left|E_{P^{*}}\bigg[\xi\widetilde{\psi}\left(W,\widetilde{\eta};u\right)-\xi\widetilde{\psi}\left(W,\eta;u\right)\bigg]^{2}\right| \\
		=&\sup_{\widetilde{\eta}\in\mathcal{R}}\left|E\bigg[\widetilde{\psi}\left(W,\widetilde{\eta};u\right)-\widetilde{\psi}\left(W,\eta;u\right)\bigg]^{2}\right|. \\
	\end{aligned}\]
	Similarly, uniformly over $\widetilde{\eta}\in\mathcal{R}$,
	\[
	E\bigg[\widetilde{\psi}\left(W,\widetilde{\eta};u\right)-\widetilde{\psi}\left(W,\eta;u\right)\bigg]^{2}\lesssim\Vert\widetilde{\eta}-\eta\Vert_{P,2}^{2}+\Vert\widetilde{\eta}-\eta\Vert_{P,2}^{2}\Vert\widetilde{\eta}-\eta\Vert_{P,\infty}^{2}=o_{p}\left(n^{-1/2}\right).
	\]
	Thus, we can conclude that with probability $1-\Delta_{n}$, $\dagger_{4.3.1}=o_{p^{*}}(1)$.
	
	\noindent\textbf{Step 2.} Here we claim that
	\[
	\widehat{Z}_{n}^{*}(u)\leadsto_{B} Z(u) \quad in \quad \mathbb{D}=\ell^{\infty}\left(\mathcal{U}\right).
	\]
	Applying Theorem B.2 in Belloni et al. (2017) or equivalently, Theorem 2 in Kosorok (2003), we have ${Z}_{n}^{*}(u)\leadsto_{B} Z(u)$ in $\mathbb{D}=\ell^{\infty}\left(\mathcal{U}\right)$. Then by Lemma 2 in Chiang et al. (2019) and the result in Step 1, we have $\widehat{Z}_{n}^{*}(u)\leadsto_{B} Z(u)$ in $\mathbb{D}=\ell^{\infty}\left(\mathcal{U}\right)$. $\blacksquare$\\
	\\
		\noindent\textbf{Appendix B. Monte Carlo Simulations for CDF derivative}

	In this appendix, we compare the performance of CDF derivative based on our estimator and Sasaki et al. (2022).
	We consider the same data-generating process as Sasaki et al. (2022).
	The outcome variable is generated according to the partial linear high-dimensional modem
	\begin{eqnarray*}
		Y|(D,X)\sim N\left(g(D)+\sum_{j=1}^{p}\alpha_jX_j,1\right)
	\end{eqnarray*}
	where the function $g(d)$ is defined in the following three ways
	\begin{eqnarray*}
		g(d)=\begin{cases}
			d&\text{in GDP 1}\\
			d-0.1d^2&\text{in GDP 2}\\
			d-0.1d^2+0.01d^3&\text{in GDP 3}
		\end{cases}
	\end{eqnarray*}
	The treatment variable $D$ is generated by
	\begin{eqnarray*}
		D|(X_1,\cdots,X_p)\sim N\left(\sum_{j=1}^{p}\gamma_jX_j,1\right)
	\end{eqnarray*}
	where $(X_1,\cdots,X_p)\sim N(0,\Sigma_p)$ with $\Sigma_p$ be the $p\times p$ variance-covariance matrix whose $(r-c)$ element is $0.5^{2\left(|r-c|+1\right)}$.
	The following four cases of varying sparsity level are considered
	\begin{enumerate}[(i)]
		\item $(\alpha_1,\cdots,\alpha_p)^\intercal=(\gamma_1,\cdots,\gamma_p)^\intercal=(0.5^2,0.5^3,\cdots,0.5^p)$,
		\item $(\alpha_1,\cdots,\alpha_p)^\intercal=(\gamma_1,\cdots,\gamma_p)^\intercal=(0.5^2,0.5^{5/2},\cdots,0.5^{(p+2)}/2)$,
		\item $(\alpha_1,\cdots,\alpha_p)^\intercal=(\gamma_1,\cdots,\gamma_p)^\intercal=(0.5^2,0.5^{7/3},\cdots,0.5^{(p+4)}/3)$,
		\item $(\alpha_1,\cdots,\alpha_p)^\intercal=(\gamma_1,\cdots,\gamma_p)^\intercal=(0.5^2,0.5^{9/4},\cdots,0.5^{(p+6)}/4)$.
	\end{enumerate}
	Across sets of Monte Carlo simulations, we vary DGP$\in$\{DGP1, DGP2, DGP3\}, and the sparsity design $\in$\{(i), (ii), (iii), (iv)\}.
	We set $b(d,x)$ by including powers of $D$ and $X$ up to the third degree, i.e., $b(d,x)=\left(d,x,d^2,(x^2)^\intercal,d^3,(x^3)^\intercal\right)^\intercal$.
	We fix the sample size $n=500$ and the dimension $p=99$ throughout.

	We estimate $\partial_DF_Y(y_\tau|D,X)$ with $\tau\in\{0.1,0.2,\cdots,0.9\}$. $y_{0.1},y_{0.2},\cdots,y_{0.9}$ equal to the 10\%, 20\%, $\cdots$, $90\%$-th quantiles of $Y$ distribution.
	
	Sasaki et al. (2022) propose to estimate $\partial_DF_Y(y_\tau|D,X)$ by
	\begin{eqnarray}
		\widehat{\partial_DF_Y(y_\tau|D,X)}_{\text{Sasaki}}=\partial_D\Lambda\left(b(D,X)\bar{\beta}_\tau\right)=\Lambda'\left(b(D,X)_\tau\bar{\beta}\right)\partial_Db(D,X)\bar{\beta}_\tau\notag
	\end{eqnarray}
	where $\Lambda'(\cdot)$ is the derivative function of logistic function $\Lambda(\cdot)$, $\bar{\beta}_\tau$ is based on the estimation procedure in Sasaki et al. (2022, page 959--960).
	By contrast, we propose to estimate $\partial_DF_Y(y_\tau|D,X)$ by
	\begin{eqnarray}
		\widehat{\partial_DF_Y(y_\tau|D,X)}_{\text{Our}}=\frac{\Lambda\left(b(D+h_n,X)\hat{\beta}_\tau\right)-\Lambda\left(b(D-h_n,X)\hat{\beta}_\tau\right)}{2h_n}\notag
	\end{eqnarray}
	where $h=n^{-1/6}$. $\hat{\beta}_\tau$ is the post-lasso estimator with penalty level and penalty loading described in Step 1 of Section 3.
	This is the special case of our estimator based on Eq. (3.2) by setting $\ell=1$.
	
	From data-generating process the true function form of $\partial_DF(y_\tau|D,X)$ is
	\begin{eqnarray*}
		\partial_DF(y_\tau|D,X)=-\phi\left(y_\tau-g(D)-\sum_{j=1}^{p}\alpha_jX_j\right)g'(D)
	\end{eqnarray*}
	where $\phi$ is the probability density function of standard normal distribution, $g'(\cdot)$ is the derivative function of $g(\cdot)$.
	In each simulation, we calculate $L^2$ distance between estimator and true function by
	\begin{eqnarray*}
		&&\frac{1}{n}\sum_{i=1}^{n}\left(\widehat{\partial_DF_Y(y_\tau|D_i,X_i)}_{\text{Sasaki}}-\partial_DF(y_\tau|D_i,X_i)\right)^2\\
		&&\frac{1}{n}\sum_{i=1}^{n}\left(\widehat{\partial_DF_Y(y_\tau|D_i,X_i)}_{\text{Our}}-\partial_DF(y_\tau|D_i,X_i)\right)^2
	\end{eqnarray*}
	We do 500 iterations to compute mean $L^2$ distance (MeanDist).
	
	Tables B.1--2 summarize the simulation results under the sparsity designs (i), (ii), (iii), and (iv).
	We can conclude that our estimation procedure outperforms the one proposed in Sasaki et al. (2022) in all of data-generating processes, especially at the tails of the unconditional distribution of $Y$.
	\begin{table}[H]
		\centering
		\begin{threeparttable}
			\begin{tabular}{lccrccrcc}	\multicolumn{9}{l}{\small{\textbf{Table B.1 }}\small{Monte Carlo simulation results for the sparsity designs (i) and (ii)}}\\
				\toprule
				\multicolumn{9}{c}{(i) The most sparse design} \\
				& \multicolumn{2}{c}{DGP 1 (i)} &     & \multicolumn{2}{c}{DGP 2 (i)} &     & \multicolumn{2}{c}{DGP 3 (i)} \\
				& \multicolumn{2}{c}{MeanDist} &     & \multicolumn{2}{c}{MeanDist} &     & \multicolumn{2}{c}{MeanDist} \\
				\cmidrule{2-3}\cmidrule{5-6}\cmidrule{8-9}    Quantile & Sasaki et al. & Our paper &     & Sasaki et al. & Our paper &     & Sasaki et al. & Our paper \\
				\midrule
				0.1 & .005 & .004 &     & .008 & .006 &     & .009 & .007 \\
				0.2 & .216 & .003 &     & .247 & .005 &     & .288 & .006 \\
				0.3 & .110 & .013 &     & .107 & .017 &     & .126 & .017 \\
				0.4 & .053 & .009 &     & .060 & .013 &     & .070 & .012 \\
				0.5 & .047 & .008 &     & .048 & .011 &     & .055 & .010 \\
				0.6 & .057 & .009 &     & .058 & .011 &     & .067 & .010 \\
				0.7 & .104 & .013 &     & .087 & .014 &     & .106 & .012 \\
				0.8 & .206 & .020 &     & .144 & .018 &     & .179 & .017 \\
				0.9 & .607 & .041 &     & .367 & .031 &     & .457 & .033 \\
				\multicolumn{9}{c}{(ii) The second most sparse design} \\
				& \multicolumn{2}{c}{DGP 1 (ii)} &     & \multicolumn{2}{c}{DGP 2 (ii)} &     & \multicolumn{2}{c}{DGP 3 (ii)} \\
				0.1 & .005 & .004 &     & .008 & .006 &     & .010 & .007 \\
				0.2 & .245 & .004 &     & .290 & .006 &     & .345 & .007 \\
				0.3 & .116 & .015 &     & .113 & .018 &     & .138 & .019 \\
				0.4 & .064 & .011 &     & .061 & .013 &     & .071 & .013 \\
				0.5 & .050 & .010 &     & .045 & .012 &     & .053 & .011 \\
				0.6 & .062 & .011 &     & .058 & .012 &     & .072 & .011 \\
				0.7 & .114 & .014 &     & .089 & .014 &     & .105 & .013 \\
				0.8 & .221 & .021 &     & .180 & .020 &     & .204 & .019 \\
				0.9 & .614 & .044 &     & .358 & .032 &     & .442 & .035 \\
				\bottomrule
			\end{tabular}%
		\end{threeparttable}
		\label{tab:addlabel}%
	\end{table}%
	\begin{table}[H]
		\centering
		\begin{threeparttable}
			\begin{tabular}{lccrccrcc}	\multicolumn{9}{l}{\small{\textbf{Table B.2 }}\small{Monte Carlo simulation results for the sparsity designs (iii) and (iv)}}\\
				\toprule
				\multicolumn{9}{c}{(iii) The third most sparse design} \\
				& \multicolumn{2}{c}{DGP 1 (iii)} &     & \multicolumn{2}{c}{DGP 2 (iii)} &     & \multicolumn{2}{c}{DGP 3 (iii)} \\
				& \multicolumn{2}{c}{MeanDist} &     & \multicolumn{2}{c}{MeanDist} &     & \multicolumn{2}{c}{MeanDist} \\
				\cmidrule{2-3}\cmidrule{5-6}\cmidrule{8-9}    Quantile & Sasaki et al. & Our paper &     & Sasaki et al. & Our paper &     & Sasaki et al. & Our paper \\
				\midrule
				0.1 & .005 & .004 &     & .009 & .007 &     & .010 & .008 \\
				0.2 & .281 & .004 &     & .295 & .007 &     & .347 & .007 \\
				0.3 & .136 & .016 &     & .143 & .019 &     & .170 & .021 \\
				0.4 & .079 & .012 &     & .072 & .014 &     & .089 & .014 \\
				0.5 & .062 & .011 &     & .055 & .013 &     & .062 & .012 \\
				0.6 & .081 & .011 &     & .070 & .012 &     & .081 & .011 \\
				0.7 & .118 & .014 &     & .099 & .014 &     & .129 & .014 \\
				0.8 & .264 & .024 &     & .189 & .021 &     & .233 & .021 \\
				0.9 & .708 & .045 &     & .437 & .033 &     & .507 & .035 \\
				\multicolumn{9}{c}{(iv) The least sparse design} \\
				& \multicolumn{2}{c}{DGP 1 (iv)} &     & \multicolumn{2}{c}{DGP 2 (iv)} &     & \multicolumn{2}{c}{DGP 3 (iv)} \\
				0.1 & .006 & .005 &     & .009 & .007 &     & .011 & .008 \\
				0.2 & .290 & .005 &     & .326 & .007 &     & .395 & .008 \\
				0.3 & .145 & .017 &     & .153 & .021 &     & .184 & .023 \\
				0.4 & .070 & .012 &     & .080 & .016 &     & .091 & .016 \\
				0.5 & .053 & .010 &     & .061 & .014 &     & .071 & .014 \\
				0.6 & .092 & .013 &     & .075 & .013 &     & .086 & .012 \\
				0.7 & .129 & .016 &     & .118 & .016 &     & .140 & .014 \\
				0.8 & .290 & .025 &     & .219 & .021 &     & .276 & .022 \\
				0.9 & .886 & .050 &     & .505 & .035 &     & .600 & .038 \\
				\bottomrule
			\end{tabular}%
		\end{threeparttable}
		\label{tab:addlabel}%
	\end{table}%

\end{document}